\documentclass{jfm}
\usepackage{graphics}
\usepackage{epstopdf, epsfig}

\usepackage{multiJFMequations}
\usepackage{amsmath,bm}
\usepackage{color}
\newcommand*{\be}{\begin{equation}}
\newcommand*{\ee}{\end{equation}}
\newcommand*{\bse}{\begin{subequations}}
\newcommand*{\ese}{\end{subequations}}
\newcommand*{\bme}{\begin{multiequations}}
\newcommand*{\eme}{\end{multiequations}}
\newcommand*{\se}{\singleequation}
\newcommand*{\de}{\doubleequation}
\newcommand*{\te}{\tripleequation}

\newcommand*{\qie}{\quintequation}

\def\xcheck#1{\overset{\ifx#1f\hspace{.5ex}\fi\lower0.8ex\hbox{\tiny$\vee$}}{#1}}
\def\xbreve#1{\overset{\ifx#1f\hspace{.5ex}\fi\lower0.8ex\hbox{\tiny$\smile$}}{#1}}
\def\xmathring#1{\overset{\ifx#1f\hspace{.5ex}\fi\lower0.8ex\hbox{\tiny$\circ$}}{#1}}

\newcommand*{\mapsfrom}{\mbox{\reflectbox{$\mapsto$}}}

\newcommand*{\XXint}[3]{{\setbox0=\hbox{$#1{#2#3}{\int}$}
\vcenter{\hbox{$#2#3$}}\kern-.5\wd0}}

\newcommand*{\XXnotinfty}[3]{{\setbox0=\hbox{$#1{#2#3}{\to}$}
\vcenter{\hbox{$#2#3$}}\kern-.5\wd0}}

\newcommand*{\ds}{\displaystyle}
\providecommand*{\dfrac}[2]{\ds\frac{#1}{#2}}

\renewcommand*{\Re}{\mbox{Re}}
\renewcommand*{\Im}{\mbox{Im}}

\renewcommand*{\tilde}{\widetilde}
\renewcommand*{\hat}{\widehat}
\renewcommand*{\bar}{\overline}

\newcommand*{\od}[2]{\dfrac{{\mathrm d}{#1}}{{\mathrm d}{#2}}}
\newcommand*{\pd}[2]{\dfrac{\partial{#1}}{\partial{#2}}}

\newcommand*{\erf}{\,{\mbox{erf}}\,}
\newcommand*{\erfc}{\,{\mbox{erfc}}\,}

\newcommand*{\lsum}{{l}}

\renewcommand*{\Lambda}{\varLambda}
\renewcommand*{\Upsilon}{\varUpsilon}
\renewcommand*{\Phi}{\varPhi}
\renewcommand*{\Psi}{\varPsi}
\renewcommand*{\Omega}{\varOmega}
\renewcommand*{\Theta}{\varTheta}
\renewcommand*{\Xi}{\varXi}
\newcommand*{\Omegav}{{\bm{\Omega}}}
\newcommand*{\dR}{{\mathrm d}}\newcommand*{\eR}{{\mathrm e}}\newcommand*{\iR}{{\mathrm i}}
\newcommand*{\CR}{{\mathrm C}}\newcommand*{\IR}{{\mathrm I}}\newcommand*{\JR}{{\mathrm J}}\newcommand*{\SR}{{\mathrm S}}
\newcommand*{\vv}{{\bm{v}}}
\newcommand*{\xS}{{\sf{x}}}\newcommand*{\zS}{{\sf{z}}}
\newcommand*{\CS}{{\sf{C}}}\newcommand*{\DS}{{\sf{D}}}\newcommand*{\ES}{{\sf{E}}}\newcommand*{\FS}{{\sf{F}}}\newcommand*{\GS}{{\sf{G}}}\newcommand*{\PS}{{\sf{P}}}\renewcommand*{\SS}{{\sf{S}}}
\newcommand*{\CC}{{\mathcal C}}\newcommand*{\FC}{{\mathcal F}}\newcommand*{\LC}{{\mathcal L}}
\newcommand*{\pG}{{\mathfrak p}}\newcommand*{\uG}{{\mathfrak u}}\newcommand*{\vG}{{\mathfrak v}}\newcommand*{\wG}{{\mathfrak w}}\newcommand*{\zG}{{\mathfrak z}}
\newcommand*{\EG}{{\mathfrak E}}\newcommand*{\FG}{{\mathfrak F}}\newcommand*{\HG}{{\mathfrak H}}\newcommand*{\UG}{{\mathfrak U}}\newcommand*{\VG}{{\mathfrak V}}\newcommand*{\WG}{{\mathfrak W}}\newcommand*{\ZG}{{\mathfrak Z}}
\newcommand*{\ut}{{\tilde u}}\newcommand*{\vt}{{\tilde v}}

\newcommand*{\chit}{{\tilde \chi}}\newcommand*{\gammat}{{\tilde \gamma}}

\newcommand*{\chimr}{{\mathring \chi}}\newcommand*{\vmr}{{\mathring v}}
\newcommand*{\ubr}{{\breve u}}\newcommand*{\vbr}{{\breve v}}
\newcommand*{\ub}{{\bar u}}\newcommand*{\vb}{{\bar v}}

\newcommand*{\vvb}{{\bar \vv}}
\newcommand*{\chib}{{\bar\chi}}

\newcommand*{\uh}{{\hat u}}\newcommand*{\vh}{{\hat v}}
\newcommand*{\chih}{{\hat \chi}}\newcommand*{\gammah}{{\hat \gamma}}
\newcommand*{\uth}{{\hat \ut}}\newcommand*{\vth}{{\hat \vt}}
\newcommand*{\chith}{{\hat \chit}}

\newcommand*{\ubrh}{{\hat \ubr}}\newcommand*{\vbrh}{{\hat \vbr}}

\newcommand*{\EGh}{{\hat \EG}}\newcommand*{\WGh}{{\hat \WG}}

\newcommand*{\zGh}{{\hat \zG}}\newcommand*{\ZGh}{{\hat \ZG}}

\newcommand*{\VGxw}{{\xcheck \VG}}

\newcommand*{\tE}{{\mbox{\tiny {\itshape E}}}}
\newcommand*{\tG}{{\mbox{\tiny {\itshape G}}}}

\newcommand*{\tIW}{{\mbox{\tiny {\itshape IW}}}}
\newcommand*{\tGH}{{\mbox{\tiny {\itshape MF}}}}
\newcommand*{\tQG}{{\mbox{\tiny {\itshape QG}}}}
\newcommand*{\tAG}{{\mbox{\tiny {\itshape AG}}}}
\newcommand*{\tSS}{{\mbox{\tiny {\itshape S}}}}

\newcommand*{\tDNS}{{\mbox{\tiny {\itshape DNS}}}}
\newcommand*{\tFNS}{{\mbox{\tiny {\itshape FNS}}}}

\renewcommand{\star}{{\dag}}

\newcommand*{\Ro}{{Ro\,}}

\def\today{
\number\day\space
\ifcase\month\or
January\or February\or March\or April\or May\or June\or
July\or August\or September\or October\or November\or December\fi
\space\number\year}

\shorttitle{Spin-down inertial waves. Part I}
\shortauthor{L. Oruba, A.~M. Soward and E. Dormy}

\title{The inertial wave activity during spin-down in a rapidly rotating penny shaped cylinder.\\
Part I The quasi-geostrophic trigger}

\author{L. Oruba \aff{1}$^{,}\dag$
  A. M. Soward \aff{2}$^{,}$\corresp{\email{ludivine.oruba@latmos.ipsl.fr, andrew.soward@ncl.ac.uk, Emmanuel.Dormy@ens.fr}},
 \and E. Dormy\aff{3}$^{,}\dag$}

\affiliation{\aff{1} Laboratoire Atmosph\`eres Milieux Observations Spatiales (LATMOS/IPSL), Sorbonne Universit\'e, UVSQ, CNRS, Paris, FRANCE
\aff{2} School of Mathematics and Statistics, Newcastle University, Newcastle upon Tyne NE1 7RU, UK
  \aff{3} {D\'epartement de Math\'ematiques et Applications, UMR-8553, \'Ecole Normale Sup\'erieure, CNRS, PSL University, 75005 Paris, FRANCE}}

\begin{document}

\maketitle

%\centerline{\today}

\begin{abstract}
  In a previous paper, Oruba, Soward \& Dormy (J.~Fluid Mech., vol.~818, 2017, pp.~205--240) considered the primary quasi-steady geostrophic (QG) motion of a constant density fluid of viscosity $\nu$ that occurs during linear spin-down in a cylindrical container of radius $L$ and height $H$, rotating rapidly (angular velocity $\Omega$) about its axis of symmetry subject to mixed rigid and stress-free boundary conditions for the case $L=H$. Here, Direct Numerical Simulation (DNS) at large $L= 10 H$ and Ekman number $E=\nu/H^2\Omega=10^{-3}$ reveals structured inertial wave activity on the spin-down time-scale. The analytic study, based on $E\ll 1$, builds on the results of Greenspan \& Howard (J.~Fluid Mech., vol.~17, 1963, pp.~385--404) for an infinite plane layer $L\to\infty$. At large but finite distance $r^\dag$ from the symmetry axis, the meridional (QG-)flow, that causes the QG-spin down, is blocked by the lateral boundary $r^\dag=L$, which provides a QG-trigger for inertial waves. The true situation in the unbounded layer is complicated further by the existence of a secondary set of maximum frequency (MF) inertial waves (a manifestation of the transient Ekman layer) identified by Greenspan \& Howard. Their blocking at $r^\dag=L$ provides a secondary MF-trigger for yet more inertial waves that we consider in a sequel (Part II). Here, for the QG-trigger, we solve a linear initial value problem by Laplace transform methods. The ensuing complicated inertial wave structure is explained analytically on approximating our cylindrical geometry at large radius by rectangular Cartesian geometry, valid for $L-r^\star=O(H)$ ($L\gg H$). Other than identifying small scale structure near $r^\star=L$, our main finding is that inertial waves radiated away from the outer boundary (but propagating towards it) reach a distance determined by the group velocity.
\end{abstract}

%%%%%%%%%%%%%%%%%%%%%%%%%%%%%%%%%%%%%
%%%%%%%%%%%%%%%%%%%%%%%%%%%%%%%%%%%%%
%%%%%       SECTION 1
%%%%%%%%%%%%%%%%%%%%%%%%%%%%%%%%%%%%%
%%%%%%%%%%%%%%%%%%%%%%%%%%%%%%%%%%%%%

\section{Introduction\label{Introduction}}

The linear spin--down of a rapidly rotating fluid, when the containing boundary is adjusted by a small amount, is characterised by two distinct transient motions. The primary part which is largely responsible for the spin--down is a quasi--steady geostrophic (QG) flow exterior to any quasi--steady boundary layers. A secondary part is the excitation of inertial waves, which decay either due to boundary layer effects or, if they are on sufficiently short length scale, in the main body of the fluid itself. In our previous paper \citep[][]{OSD17}, we investigated spin-down in a cylindrical container, radius $L$, height $H$, rotating rapidly about its axis of symmetry subject to  mixed rigid and stress--free boundary conditions. There we focused on the aspect ratio $\ell \equiv L/H=1$ and, because our Direct Numerical Solutions (DNS) revealed little inertial wave activity or more precisely the inertial waves decayed very rapidly (for reasons that will become clearer later) and were hardly visible, we only investigated analytically the aforementioned primary QG--flow part. That study was motivated by the possible application to intense nearly axisymmetric vortices, which  develop in geophysical flows, e.g., tornadoes, or hurricanes in the atmosphere, and westward-propagating mesoscale eddies that occur throughout most of the World Ocean \citep{CSS11} as evidenced by the Sea Surface Height variability. The benefits of such modelling by isolated structures is well established \citep[see][and references therein]{PMSM15,ODD17,ODD18}. For those applications the aspect ratio $\ell$ ought to be large, and so our previous choice $\ell=1$ is clearly not the most appropriate. Indeed, later DNS--results for large aspect ratio, specifically $\ell=10$, have revealed considerable persistent inertial wave activity. For that reason we investigate the limit,
\be
\label{ell-large}
\ell\,=\,L/H\, \gg 1\,,
\ee
analytically and compare numerical results based on the asymptotics with the DNS. 

As our work builds upon \cite{OSD17}, we only repeat essential details such as the description of the model and needed results. Our cylindrical container is filled with constant density fluid of viscosity $\nu$ and rotates rigidly with angular velocity $\Omegav$ about its axis of symmetry, the frame, relative to which our analysis is undertaken; the Ekman number is small:
\be
\label{Ek-numb}
E\,=\,\nu \big/\bigl(H^2\Omega\bigr)\,\ll\, 1\,.
\ee
Initially at time $t^\star=0$ the fluid itself rotates rigidly at the slightly larger angular velocity $\Ro\Omegav$, in which the Rossby number $\Ro\!$ is sufficiently small ($\Ro\!\ll E^{1/4}$) for linear theory to apply. Whereas, the nonlinear development of spin-down and spin-up differ \citep[see, e.g.,][and references therein]{Cetal18}, their linear evolution, which we consider, is mathematically equivalent. Relative to cylindrical polar coordinates, $(r^\star,\,\theta^\star,\,z^\star)$, the top boundary ($r^\star<L$, $z^\star=H$) and the sidewall ($r^\star=L$, $0<z^\star<H$) are impermeable and stress-free. The lower boundary ($r^\star<L$, $z^\star=0$) is rigid. For that reason alone the initial state of relative rigid rotation  $\Ro\Omegav$ of the fluid cannot persist and the fluid spins down to the final state of no rotation relative to the container as $t^\star\to\infty$. In order to make our notation relatively compact at an early stage, we use $H$ and $\Omega^{-1}$ as our unit of length and time respectively, and introduce 
\bme
\label{dim-length-time}
\be
\te
r^\star=Hr\,,\qquad z^\star=Hz\,,\qquad\qquad  \Omega t^\star=t\,.
\ee
For our unit of relative velocity $\vv^\star$, we adopt the velocity increment $\Ro L \Omega$ of the initial flow at the outer boundary $r^\star=L$. So, relative to cylindrical components, we set
\be
\vv^\star\,=\,\Ro L \Omega\,\vv\,, \qquad\qquad  \vv\,=\,[u,\,v,\,w]
\ee
\eme
and refer to $[u,v]$ and $w$ as the horizontal and axial components of velocity, respectively.

Relevant to our previous $\ell=1$ study, but of even greater importance to our present $\ell\gg 1$ case, are the aspects of the seminal work of \cite{GH63} that pertain to the unbounded limit $\ell\to \infty$. That study is complicated and many of the key concepts, as they relate to our study are not easily identified. Indeed the most important ideas stem from the even simpler problem of the transient Ekman layer above a flat plate in an otherwise unbounded fluid, which we outline in the following subsection.

\subsection{The transient Ekman layer\label{transient-Ekman-layer}}

The nature of the transient Ekman layer, in the half-space $z>0$ above a rigid boundary $z=0$, is well known \citep[see, e.g.,][]{G68}, but here we provide a summary in order to develop our notation and highlight features upon which we will build. We consider the axisymmetric flow
\bme
\label{sim-variable}
\be
\te
u\,=\,(r/\ell)\,\uG(z,t)\,, \qquad\qquad      v\,=\,(r/\ell)\,\vG(z,t)\,, \qquad\qquad     w\,=\,(1/\ell)\,\wG(z,t)\,,
\ee
\eme
that solves
\bme
\label{sim-eqs}
\be
\partial_t\uG\,-\,2\,(\vG-1)\,=\,E\partial^2_z\uG\,, \qquad\qquad  \partial_t\vG\,+\,2\,\uG\,=\,E\partial^2_z\vG\,
\ee
\eme
subject to $[\,\uG\,,\,\vG\,] =[0,\,1]$ at $t=0$, while subsequently $[\,\uG\,,\,\vG\,] =[0,\,0]$ at $z=0$ and $[\,\uG\,,\,\vG\,] \to [0,\,1]$ as $z\uparrow\infty$ for $t>0$.

We introduce the complex combination
\be
\label{zG}
\zG(z,t)\,=\,{\uG}-\iR\bigl(\vG-1\bigr)
\ee
and consider its Laplace transform (henceforth LT)
\bse
\label{lT-inv-def}
\be
\zGh(z,p)\,=\,\LC_p\bigl\{\zG(z,t)\bigr\}\,\equiv\,\int_0^\infty\zG(z,t)\,\exp(-pt)\,\dR t\,,
\ee
in which the subscript `$p$' to the LT-operator $\LC$ identifies the independent transform variable. The inverse is
\be
\zG(z,t)\,=\,\LC_p^{-1}\Bigl\{\,\zGh(z,p)\Bigr\}\,\equiv\,\dfrac{1}{2\pi\iR}\int_{-\iR \infty}^{\iR \infty}\zGh(z,p)\exp(pt)\,\dR p\,.
\ee
\ese
The LT-solution of the problem posed is
\be
\label{sim-LT-sol}
\zGh(z,p)\,=\,\iR p^{-1}\exp\bigl[-E^{-1/2}(p- 2\iR)^{1/2} z\bigr]\,.
\ee

The $z$-integral
\bse
\label{ZG-int}
\begin{align}
\UG(t)\,-\,\iR\,\VG(t)\,\equiv\,-\,\ZG(t)\,=\,-\,E^{-1/2}\int_0^\infty\zG(z,t)\,\dR z\,=\,&\,-\iR \LC^{-1}\Bigl\{p^{-1}(p- 2\iR)^{-1/2}\Bigr\}\\
=\,&\,\tfrac12(1-\iR)\erf\bigl[(1-\iR)t^{1/2}\bigr]
\end{align}
\ese
\citep[use \S5.3 eq.~(1) of][]{EMOT54I} fixes the horizontal volume flux deficit
\bse
\label{intuvG}
\be
\dfrac{r}{\ell}\int_0^\infty[\,\uG\,,\,\vG-1\,]\,\dR z\,=\,-\,E^{1/2}\,\dfrac{r}{\ell}\,\bigl[\,\UG(t)\,,\,\VG(t)\bigr]
\ee
(the minus sign is motivated by our application (\ref{MF-spin-down-prelim}$a$) in \S\ref{upp-MF}($a$) below), where use of (http://dlmf.nist.gov/7.5.E8) determines
\be
\left[\begin{array}{c}  \! \UG(t) \!\!\\[0.2em]
    \!\! \VG(t) \!\!  \end{array}\right]\,=\,
\left[\begin{array}{c}  \! \SR\bigl(2t^{1/2}/\pi^{1/2}\bigr) \!\!\\[0.2em]
    \!\! \CR\bigl(2t^{1/2}/\pi^{1/2}\bigr) \!\!  \end{array}\right]\,=\,
\int_0^{t} \left[\begin{array}{c}  \!\! \sin (2\tau) \!\!\\[0.2em]
    \!\! \cos(2\tau) \!\!  \end{array}\right]\dfrac{\dR \tau}{\sqrt{\pi \tau}}\,,
\ee
\ese
in which $\CR$, $\SR$ are Fresnel integrals (http://dlmf.nist.gov/7.2.E7,8).  The outflow velocity $E^{1/2}\;\!\WG/\ell$ from the transient Ekman layer is determined by mass continuity:
\bse
\label{sim-outflow}
\be
\WG(t)\,\equiv\,E^{-1/2}\wG_{z\uparrow\infty}\,=\,-2E^{-1/2}\int_0^\infty\uG\,\dR z\,=\,-2\,\Re\bigl\{\ZG\bigr\}\,=\,2\UG(t)
\ee
with Laplace transform
\be
\WGh(p)\,=\,-2\,\Re\Bigl\{\ZGh\Bigr\}\,=\,\dfrac{\iR}{p}\biggl[\dfrac{1}{(p+ 2\iR)^{1/2}}\,-\,\dfrac{1}{(p-2\iR)^{1/2}}\biggr].
\ee
\ese

We now partition the solution $[\,\uG\,,\,\vG\,]$ into two parts; in \S\ref{unbounded-E}, the final steady state $[\,\uG_\tE\,,\,\vG_\tE\,]$ denoted by the subscript `$E\,$'; in \S\ref{unbounded-MF}, the remaining part $[\,\uG_\tGH\,,\,\vG_\tGH\,]$ denoted by the subscript `$MF\,$' for reasons explained following (\ref{transient-EL-sol-z-int}) below.

\subsubsection{The final steady Ekman layer \label{unbounded-E}}

The final steady Ekman layer has horizontal velocity $[u,\,v]=(r/\ell)\bigl[\uG_{\tE},\,\vG_{\!\tE}\bigr]$ determined by $\zG_{\tE}\,={\uG_{\tE}}\,-\,\iR\bigl(\vG_{\tE}-1\bigr)=\iR\exp\bigl[-E^{-1/2}(- 2\iR)^{1/2} z\bigr]$ which is fixed by the residue of the integrand (\ref{sim-LT-sol}) of the inverse-LT integral (\ref{lT-inv-def}$b$)  at the pole $p=0$:
\bse
\label{outflow-sol-E}
\be
\bigl[\,\uG_{\tE}\,,\,\vG_{\tE}-1\,\bigr]\,=\,-\,\bigl[\,\sin\bigl(E^{-1/2}z\bigr)\,,\,\cos\bigl(E^{-1/2}z\bigr)\,\bigr]\exp\bigl(-E^{-1/2}z\bigr).
\ee
The  corresponding boundary layer volume flux deficit determined from (\ref{intuvG}) is
\be
\bigl[\,\UG_\tE\,,\,\VG_\tE\bigr]\,=\,\bigl[\,\UG\,,\,\VG\bigr]_{t\to \infty}\,=\,\bigl[\tfrac12\,,\,\tfrac12 \,]\,.
\ee
\ese
Hence, by (\ref{sim-outflow}$a$), fluid is pumped out with velocity $w=(E^{1/2}/\ell)\;\!\WG_{\!\tE}$, where
\be
\label{outflow-sol-E-WG}
\WG_{\!\tE}\,=\,\WG\big|_{t\to \infty}\,=\,2\;\!\UG_\tE\,=\,1\,.
\ee

\subsubsection{The transient MF-layer\label{unbounded-MF}}

The transient MF-layer structure $(r/\ell)[\,\uG_{\tGH}\,,\,\vG_{\!\tGH}\,]\,=\,(r/\ell)[\,\uG-\uG_{\tE}\,,\,\vG-\vG_{\!\tE}\,]\,$  may be determined from (\ref{lT-inv-def}$b$) and (\ref{sim-LT-sol}) in terms of standard functions of complex argument \citep[see][eqs.~(2.3.4--6)]{G68}. This form is opaque but it is sufficient for us to note that the long time asymptotic behaviour of the inverse-LT integral (\ref{lT-inv-def}$b$) with integrand (\ref{sim-LT-sol}) is determined by the cut contribution in the neighbourhood of the cut-point $p=2\iR$:
\vskip -4mm
\bse
\label{transient-EL-sol}
\begin{align}
\zG_\tGH\,=\,\uG_\tGH\,-\,\iR\,\vG_\tGH\,\approx\,&\,\tfrac12 \exp(2\iR t)\LC^{-1}_q\bigl\{\exp\bigl(-E^{-1/2}q^{1/2}z\bigr)\bigr\}\\[0.2em]
\approx\,&\,\dfrac{\exp(2\iR t)}{2t}\,\dfrac{z}{\sqrt{4\pi E t}}\,\exp\biggl(-\dfrac{z^2}{4Et}\biggr) \hskip 10mm  \mbox{for} \hskip 5mm t\gg 1 \hskip 5mm  
\end{align}
\ese
\citep[use \S5.6 eq.~(1) of][]{EMOT54I}.

On use of (\ref{ZG-int}$b$), the corresponding boundary layer volume flux deficit is given exactly by
\bse
\label{outflow-sol-MF}
\begin{align}
-\UG_\tGH(t)\,+\,\iR\,\VG_\tGH(t)\,\equiv\,\ZG_\tGH(t)\,=\,&\,\tfrac12(1-\iR)\erfc\bigl[(1-\iR)t^{1/2}\bigr]\\
\approx\,&\,\dfrac{\exp(2\iR t)}{\sqrt{4\pi t}} \hskip 28mm  \mbox{for} \hskip 5mm t\gg 1\,,
\hskip 5mm\\[-0.6em]
\intertext{equivalently
\vskip -9mm}
\left[\begin{array}{c}  \! \UG_\tGH(t) \!\!\\[0.2em]
    \!\! \VG_\tGH(t) \!\!  \end{array}\right]\,=\,&\,
-\,\int_t^{\infty} \left[\begin{array}{c}  \!\! \sin (2\tau) \!\!\\[0.2em]
    \!\! \cos(2\tau) \!\!  \end{array}\right]\dfrac{\dR \tau}{\sqrt{\pi \tau}}\,\\
\approx\,&\,\dfrac{1}{\sqrt{4\pi t}}\left[\begin{array}{c}  \!\! -\cos(2t) \!\!\\[0.2em] 
    \!\!  \sin(2t)\!\!  \end{array}\right] \hskip 13mm  \mbox{for} \hskip 5mm t\gg 1\,. \hskip 5mm
\end{align}
\ese
By mass continuity, that fixes the outflow velocity $E^{1/2}\;\!\WG_{\!\tGH}/\ell$, where, by (\ref{sim-outflow}$a$),
\be
\label{outflow-sol-MF-WG}
\WG_{\!\tGH}(t)\,=\,2\;\!\UG_\tGH(t)\,=\,-\,\dfrac{1}{\sqrt{\pi t}}\Bigl[\cos(2t)\,+\,\dfrac{\sin(2t)}{4t}-\cdots\Bigr] \hskip 8mm  \mbox{for} \hskip 5mm t\gg 1\,. \hskip 5mm
\ee

\subsubsection{A late time ($t\gg 1$) summary\label{unbounded-summary}}

For $t\gg 1$, (\ref{transient-EL-sol}$b$) and (\ref{outflow-sol-MF}$b$,$d$) together determine
\bse
\label{transient-EL-sol-z-int}
\be
\left[\begin{array}{c}  \!\! \uG_\tGH \!\!\\[0.2em]
    \!\!  \vG_\tGH \!\!  \end{array}\right]
\,=\,\left[\begin{array}{c}  \!\! \Re\{\zG_\tGH \}\!\!\\[0.2em]
    \!\!-\,\Im\{\zG_\tGH \} \!\!  \end{array}\right]
\approx \,-\,\dfrac{1}{t^{1/2}}\left[\begin{array}{c}  \! \UG_\tGH(t) \!\!\\[0.2em]
    \!\! \VG_\tGH(t) \!\!  \end{array}\right]
\dfrac{z}{\sqrt{4Et}}\,\exp\biggl(-\dfrac{z^2}{4Et}\biggr),
\ee
which, in turn, identifies important characteristics of the oscillating transient Ekman layer. The  horizontal flow (\ref{transient-EL-sol-z-int}$a$) describes a thickening viscous shear layer of width
\be
\Delta(t)=\sqrt{Et}\,, 
\ee
\ese
which emerges from the steady Ekman layer (\ref{outflow-sol-E}$a$) width $\Delta_{\tE}=\Delta(1)=E^{1/2}$ at late time ($t\gg 1$). In concert, the magnitude $\WG_{\!\tGH}\,=\,O(t^{-1/2})$ decays with time, becoming  small compared with $\WG_{\!\tE}\,=\,1$. Since the frequency $\omega$ of inertial waves is bounded by $2$, the value $\omega=2$ appearing in (\ref{outflow-sol-MF}$d$) indicates that the transient flow is composed of inertial waves of maximum frequency; whence our use of the subscript `$MF\,$'. Their temporal algebraic decay implies that the steady Ekman layer forms on the $O(1)$ inertial wave (or rotation) time scale.

\subsection{Spin-down between two unbounded parallel plates\label{unbounded-parallel-plates}}

When the fluid is bounded above by a free boundary at $z=1$, the axial flow $(E^{1/2}/\ell)\;\!\WG$, (\ref{sim-outflow}), in the mainstream outside the bottom boundary layers, is no longer acceptable. Instead the axial velocity is brought to zero at $z=1$ and the realised mainstream flow is composed of two parts:
\begin{itemize}
\item[(i)] QG-motion;
\item[(ii)] quasi (i.e, not purely periodic because of the algebraic decay) inertial waves of maximum frequency, $\omega=2$, which we term MF.
\end{itemize}

Those QG- and MF-parts are carefully identified in the following subsections \S\S\ref{upp-QG} and \ref{upp-MF}, respectively, and attention is drawn to the similarities and any conflicts with the earlier results of \cite{GH63}. 

\subsubsection{The quasi-geostrophic QG-flow\label{upp-QG}}

The primary QG-part $\vvb_{\tQG}^{\,\infty}$ (say, but see (\ref{QG-spin-down}$b$) below) of $\vv$, in the mainstream exterior to the Ekman layer, is a rigid rotation which spins down on the time scale $t_{sd}=E^{-1/2}$ due to blowing of fluid out of the Ekman layer. More precisely the entire velocity associated with this rotation is
\bme
\label{QG-spin-down}
\be
\se
\vvb_{\tQG}\,=\,[\,\ub_{\tQG},\,\vb_{\tQG},\,{\bar w}_{\tQG}\,]\,=\,\biggl[\tfrac12\sigma  E^{1/2}\vb_{\tQG},\,\,\, \vb_{\tQG}, \,\,\,\tfrac12\sigma E^{1/2} (1-z)\dfrac1r\pd{\,}{r}(r\vb_{\tQG})\biggr]\,,
\ee
in which $\vvb_{\tQG}$ is defined by the special choice
\be
\vb_{\tQG}(r,t)\,=\,\vb_{\tQG}^{\,\infty}(r,t)\,\equiv\,\kappa (r/\ell)\,\EG(t)\,, \hskip 14mm  
\left\{\begin{array}{l}  \EG(t)\,=\,\exp(- Q t)\,,\\[0.1em] \hskip 4mm Q\,=\,E^{1/2}\sigma\,,\end{array}\right.
\ee
and where $\kappa$ and $\sigma$ are both close to unity and have expansions
\be
\de
\kappa\,=\,1\,+\,\tfrac14 E^{1/2}\,+\,O(E)\,, \qquad\qquad
\sigma\,=\,1\,+\,\tfrac34 E^{1/2}\,+\,O(E)
\ee
\eme
\citep[see][eqs.~(1.3$a$-$c$)]{OSD17}. Points to note are that:
\begin{itemize}
\item[(i)] The amplitude factor $\kappa$, (\ref{QG-spin-down}$d$), differs from unity because the remaining transient response $\vb_\tGH(r,t)$ considered in \S\ref{upp-MF}($a$) is non-zero at the initial instant $t=0$ (see (\ref{MF-spin-down-initial}$b$));
\item[(ii)] the radial fluid flux in the Ekman layer determines the pumping ${\bar w}_{\tQG}\big|_{z=0}$, which in turn fixes the spin-down rate $Q=\sigma E^{1/2}$ (\ref{QG-spin-down}$c$,$e$).
\end{itemize}

The azimuthal fluid flux deficit $\langle v(r,z,t)\rangle - \vb_{\tQG}(r,t)\,(=(\mu-1)\vb_{\tQG}(r,t)$, say, but see (\ref{bar-angle}$b$) below)), where
\bme
\label{bar-angle}
\be
\se
\langle \bullet\;\! \rangle\,=\,\int_0^1 \bullet\,\, \dR z\,,
\ee
is important for our interpretation of the DNS. For though $\vb_{\tQG}(r,t)$ is well defined in the limit $E\downarrow 0$, it is not easily determined unambiguously from the numerics at finite $E$. Nevertheless, we can readily calculate $\langle v\,  \rangle$ and from it we may extract
\be
\vb_{\tQG}\,=\,\mu^{-1}\langle v \rangle,\qquad\qquad \mbox{where}\qquad\qquad \mu\,=\,1\,-\,\tfrac12 E^{1/2}\,+\,O(E)
\ee
\eme
is the asymptotic prediction encapsulated by eq.~(2.20) of \cite{OSD17}. It not only applies to the particular flow $\vb_{\tQG}^{\,\infty}(r,t)$ (\ref{QG-spin-down}$b$) but also to any flow $\vb_{\tQG}(r,t)$ with arbitrary $r$-dependence, which is dominated by the decay factor $\exp(-E^{1/2}\sigma t)$ while possibly evolving on the longer lateral diffusion time scale, as we will now explain.
  
The main thrust of \cite{OSD17} was to elucidate how the laterally unbounded QG-flow (\ref{QG-spin-down}) is modified by the outer sidewall at $r=\ell$ ($r^\star=L$). There two boundary layers form whose widths $\Delta(t)$ evolve by lateral viscous diffusion according to the rule (\ref{transient-EL-sol-z-int}$b$). One develops into the quasi-steady ageostrophic $E^{1/3}$-Stewartson layer of width $\Delta(t_{\tSS})=\Delta_{\tSS}=E^{1/3}$, which forms on the time-scale $t_{\tSS}=E^{-1/3}$. The other, importantly QG, spreads indefinitely filling the container when $\Delta(t_\ell)=\ell$ at time $t_\ell=\ell^2E^{-1}$. So though (\ref{QG-spin-down}$a$-$c$) provides a valid description of the QG-motion on the spin-down time-scale $t_{sd}=E^{-1/2}$, its radial dependence is more complicated on the longer lateral diffusion time-scale  $t_{\ell}=\ell^2E^{-1}$. The temporal evolution of the QG-flow $\vb_{\tQG}(r,t)$ is sensitive to whether or not the boundary $r=\ell$ is stress-free as in \cite{OSD17} or rigid as in \cite{GH63}. However, here we will filter out any QG-motion and ignore the ageostroghic $E^{1/3}$-Stewartson layer. Subject to those restrictions we will only investigate the remaining wave part. With that proviso our study applies equally to both the stress-free \citep{OSD17} and rigid \cite{GH63} $r=\ell$ boundary cases. The DNS solutions presented here are for the stress-free case, but simulations performed with a no-slip outer wall demonstrated only minor changes to the inertial waves generated. In summary the key times, pertaining to the QG-study of \cite{OSD17}, are ordered as follows
\be
\label{time-scales}
1\ll t_{\tSS}=E^{-1/3}\ll t_{sd}=E^{-1/2}\ll t_{\ell}=\ell^2E^{-1}.
\ee
These times are important to us, as we will report results exterior to all boundary layers for $t>0$. So we need to be aware of any ageostrophic motion that our study cannot explain.

\subsubsection{The inertial wave of maximum frequency \label{upp-MF}}

The secondary MF-part with algebraic decay ($\propto t^{-1/2}$) originates from the oscillatory pumping into and out of the thickening oscillatory shear layer width $\Delta(t)=(Et)^{1/2}$ (\ref{transient-EL-sol-z-int}$b$) adjacent to the lower boundary $z=0$. It spreads reaching the upper boundary $z=1$ at $t=E^{-1}$, when  $\Delta(t)=1$. This situation for MF-waves must be distinguished from that for non-degenerate inertial waves (i.e., non-maximal with frequency less than $2$), which are damped by blowing and suction into an Ekman layer of finite thickness. The distinction becomes blurred for those inertial waves with frequencies close to $2$, for which the Ekman layer width is large. Apparently, this is not an issue here in our study of the inertial waves caused by the so called `QG-trigger' (\ref{triggers}$a$) but causes some concern in our sequel \citep[][henceforth referred to as Part~II]{OSD18}, when we consider the response to the `MF-trigger' (\ref{triggers}$b$), about which we explain in  \S\ref{bounded-parallel-plates} below.

\vskip 3mm
\noindent($a$) {\itshape{MF-mainstream}}
\vskip 2mm

Though, as explained in \S\ref{transient-Ekman-layer}, the MF boundary layer solution is only represented simply in the form (\ref{transient-EL-sol-z-int}$a$) for $t\gg1$, we are not restricted by such considerations for the clearly defined transient MF Ekman blowing $\WG_{\!\tGH}(t)=2\UG_{\!\tGH}(t)$ (\ref{outflow-sol-MF-WG}), for which (\ref{outflow-sol-MF}$c$) is valid at lowest order for all time. So during the interval $0 < t\ll E^{-1}$ and outside, ${\mathrm {max}}\{\Delta(1),\Delta(t)\}\ll z\le 1$, the transient Ekman layer, the mainstream $z$-independent MF-flow exhibits the simple asymptotic form
\bse
\label{MF-spin-down-prelim}
\be
\vvb_{\tGH}\,=\,[\,\ub_\tGH\,,\,\vb_\tGH\,,\,{\bar w}_\tGH\,]\,
=\,\bigl({E^{1/2}}\big/{\ell}\bigr)\,\bigl[\,r \,\UG_{\!\tGH}\,,\,r\,\VGxw_{\!\tGH}\,,\,(1-z)\,\WG_{\!\tGH}\,\bigr]
\ee
and from it we may define a mainstream streamfunction $r\chib_{\tGH}$ as
\be
\chib_{\tGH}\,=\int_z^1\ub_\tGH\,\dR z\,=\,\dfrac{E^{1/2}r}{\ell}(1-z)\UG_{\!\tGH}\,.
\ee
Here the radial volume flux $2\pi r \ub_\tGH$ is equal in magnitude but opposite in sign to that (see (\ref{intuvG}$a$)) carried by the shear layer. This consideration does not apply to $\vb_\tGH$. Instead $\vb_\tGH$ is determined by the azimuthal momentum balance $\partial \vb_\tGH/\partial t -2\ub_\tGH=0$ (see (\ref{sim-eqs}$b$) with E=0), which leads to
\be
\VGxw_{\!\tGH}\,=\,\int_t^\infty\WG_{\!\tGH}\,\dR t\,=\,2\int_t^\infty\UG_{\!\tGH}\,\dR t\,,
\ee
\ese
with the constant of integration chosen to guarantee that $\VGxw_{\!\tGH}\to 0$ as $t\to\infty$. On use of (\ref{outflow-sol-MF}$c$), we may establish
\bse
\label{MF-spin-down-prelim-more}
\be
\od{\,}{t}\bigl(\VGxw_{\!\tGH}-\VG_{\!\tGH}\bigr)\,=\,-\,\dfrac{1}{2\sqrt \pi}\int_t^\infty\dfrac{\cos(2\tau)}{\tau^{3/2}}\,\dR\tau\,,
\ee
from which the asymptotic relation,
\be
\VGxw_{\!\tGH}-\VG_{\!\tGH}\,\approx\,-\,\dfrac{\cos(2 t)}{8\sqrt{\pi}\, t^{3/2}}\,=\,O\bigl(\VG_{\!\tGH}/t\bigr) \hskip 15mm  \mbox{for} \hskip 5mm t\gg 1\,, \hskip 5mm
\ee
\ese
follows. Whereas $\VG_{\!\tGH}(0)=0$, the initial value of $\VGxw_{\!\tGH}$ may be obtained deviously as 
\bse
\label{MF-spin-down-initial}
\be
\VGxw_{\!\tGH}(0)\,=\,\int_0^\infty\bigl[\WG(t)-1\bigr]\,\dR t\,=\,\lim_{p\downarrow 0}\Bigl[\,\WGh(p)\,-\,p^{-1}\Bigr]\,=\,-\,\tfrac14
\ee
(use (\ref{sim-outflow}$b$)). Reassuringly, this leads consistently to the initial rigid rotation value
\be
\vb_\tQG(r,0)+\vb_\tGH(r,0)\, \approx\, (r/\ell)\bigl(\kappa -\tfrac14 E^{1/2}\bigr)\, \approx\,  r/\ell\,,
\ee
\ese
correct to $O(E^{1/2})$  (see (\ref{QG-spin-down}) and particularly (\ref{QG-spin-down}$d$)), before spin-down commences. For large $t$, the estimate (\ref{MF-spin-down-prelim-more}$b$) implies $\VGxw_{\!\tGH}\approx \VG_{\!\tGH}$ so that (\ref{MF-spin-down-prelim}$a$) reduces to
\be
\label{MF-spin-down}
\vvb_{\tGH}\,\approx
\,\bigl({E^{1/2}}\big/{\ell}\bigr)\,\bigl[\,r \,\UG_{\!\tGH}\,,\,+\,r\,\VG_{\!\tGH}\,,\,2(1-z)\,\UG_{\!\tGH}\,\bigr]\hskip 10mm  \mbox{for} \hskip 5mm t\gg 1\,. \hskip 5mm
\ee

The large-time value of $\vvb_\tQG+\vvb_\tGH$ determined by (\ref{QG-spin-down}) and (\ref{MF-spin-down}) agrees with the results Eqs.~(3.18), (3.19) of \cite{GH63} in their $R^{-1}\,(\mbox{our }E)=0$ limit. Those equations are, however, followed by an inappropriate approximation that results in their erroneous Eqs.~(3.20), which describe some spurious Ekman damping of the MF-flow. This error is rectified in their un-numbered equations at the bottom of the page, where approximations are given that are consistent with  our (\ref{QG-spin-down})  and  (\ref{MF-spin-down}). Indeed the true decay rates of those QG and MF-velocities are important:
\begin{itemize}
\item[(i)] The  primary QG-part (\ref{QG-spin-down}) decays exponentially $\propto \exp(-E^{1/2}\sigma t)$;
\item[(ii)] the  secondary MF-part (\ref{MF-spin-down}) decays algebraically $\propto t^{-1/2}$.
\end{itemize}
At large time, their relative magnitudes are
\bme
\label{QG-versus-GH}
\be
\biggl|\dfrac{\vb_\tGH}{\vb_\tQG}\biggr|\,=\,O\Bigl((E/t)^{1/2}\eR^{\sigma E^{1/2}t}\Bigr),\qquad\qquad
\biggl|\dfrac{\ub_\tGH}{\ub_\tQG}\biggr|\,=\,O\Bigl(t^{-1/2}\eR^{\sigma E^{1/2}t}\Bigr),
\ee
\eme
The factor $E^{1/2}$ in the estimate of the ratio $\bigl|\vb_\tGH\big/\vb_\tQG\bigr|$ suggests that the MF-wave may remain insignificant on the spin-down time. However, the absence of that factor  $E^{1/2}$ in the ratio $\bigl|\ub_\tGH\big/\ub_\tQG\bigr|$ for the smaller radial velocities is interesting, because it suggests that, on the Ekman layer formation time-scale $t=O(1)\ll t_{sd}$, the $\ub_\tGH$ and $\ub_\tQG$ contributions may be of comparable size.

\vskip 3mm

\noindent($b$) {\itshape {MF-boundary layer}}
\vskip 2mm

According to the transient Ekman layer solution (\ref{transient-EL-sol-z-int}), for $t\gg 1$, the mainstream MF-flow is linked to an expanding shear layer adjacent to the $z=0$ boundary of width $\Delta(t)=\sqrt{Et}$, which emerges from the Ekman layer on the rotation time $t=O(1)$ ($\Delta(1)=\Delta_E=E^{1/2}$). It remains thin compared to the plate separation provided $\Delta(t)\ll 1$. So for
\be
\label{MF-range}
                  1 \,\ll\, t\,\ll\, E^{-1}\, 
\ee
the MF-wave flow can be separated into a $z$-independent mainstream flow, $\bigl[\ub_\tGH,\vb_\tGH\bigr]$
%(see
(\ref{MF-spin-down}),
%),
which, when combined with (\ref{transient-EL-sol-z-int}$a$), determines the complete MF-flow
\bse
\label{MF-combined}
\be
\left[\begin{array}{c}  \!\! u_\tGH \!\!\\[0.2em]
    \!\!  v_\tGH \!\!  \end{array}\right]
\approx\,E^{1/2}\,\dfrac{r}{\ell}\,\left[\begin{array}{c}  \!\! \UG_{\!\tGH} \!\!\\[0.2em]
    \!\! \VG_{\!\tGH}\!\!  \end{array}\right]\biggl[1-\dfrac{z}{2Et}\,\exp\biggl(-\dfrac{z^2}{4Et}\biggr)\biggr].
\ee
The corresponding streamfunction $r\chi_\tGH$ is given by
\be
\chi_\tGH(r,z,t)\,=\,-\int_0^z u_\tGH\,\dR z\,\approx\,E^{1/2}\,\dfrac{r}{\ell}\,\UG_{\!\tGH}\, \biggl[1-z-\exp\biggl(-\dfrac{z^2}{4Et}\biggr)\biggr].
\ee
\ese
Provided that $Et\ll 1$, the approximate value of total radial mass flux $-\chi_\tGH(r,1,t)$ given by (\ref{MF-combined}$b$) is exponentially small, a value which is adequate to approximate the true boundary condition  $\chi_\tGH(r,1,t)=0$. The MF boundary layer contribution for $z=O\bigl(\Delta(t)\bigr)$ is identified by the factor $\exp\bigl(-z^2/(4Et)\bigr)$. To meet the mass flux condition, the boundary layer velocity in (\ref{MF-combined}$a$) is larger than the mainstream part by a factor of order $1/\Delta(t)=(Et)^{-1/2}$. From that point of view, the vanishing of the boundary layer contribution on $z=0$ is exactly what is needed to approximate the true boundary condition $\bigl[u_\tGH\,,\,v_\tGH\bigr]=0$ at lowest order.

The value of the $z$-average $[\langle v_\tGH \rangle\,,\,\langle v_\tGH\rangle]$ of the complete MF-flow is pertinent to our comparisons with the DNS in \S\ref{numerics}. For $t\gg 1$, we may integrate (\ref{MF-combined}$a$) and, as in (\ref{MF-combined}$b$), readily deduce that the $z$-average vanishes. For $t=\!O(1)$, we need to proceed more \\\vskip -4.5mm
\noindent
cautiously combining the mainstream contribution $(E^{1/2} r/\ell)\,[\UG_{\!\tGH}\,,\,\VGxw_{\!\tGH}]$ from (\ref{MF-spin-down-prelim}$a$) with the integrated boundary layer contribution $-\,(E^{1/2}r/\ell)\,[\UG_{\!\tGH}\,,\,\VG_{\!\tGH}\bigr]$ determined by (\ref{intuvG}) and (\ref{outflow-sol-MF}). The radial volume flux so determined vanishes,  while in the azimuthal direction the combined sum gives
 \be
 \label{MF-v-z-average}
\langle v_\tGH\rangle\,=\,(E^{1/2}r/\ell)\,\bigl(\VGxw_{\!\tGH}-\VG_{\!\tGH}\bigr).
\ee
Its right-hand side, determined by (\ref{MF-spin-down-prelim-more}), enables us to make the leading order estimates
\vskip -3mm
\bme
\label{GH-waves-series-means}
\be
\langle v_{\tGH}\rangle\,=\,\left\{
\begin{array}{llll}
O(\vb_\tGH)\hskip 5mm &\mbox{for} &  \hskip 5mm &\!\!t=O(1)\,,\\[0.4em]
O(t^{-1}\vb_\tGH)\hskip 5mm  &\mbox{for} & \hskip 5mm  1\ll\!\!&\!\! t\ll E^{-1}\,.
\end{array}  \right.
\ee
\eme
This means that, though initially $\langle v_{\tGH}\rangle$ is comparable to $\vb_\tGH$, its relative size decreases $\propto t^{-1}$ and is small $O(E^{1/2})$ on the spin-down time $t_{sd}=E^{-1/2}$. Despite the clarification gleaned from the boundary layer approach, we employ the alternative MF-harmonic expansion obtained by \cite{GH63} and outlined in appendix~\ref{MF-harmonic} for our comparisons with the DNS.
  
\subsection{Spin-down between two parallel plates bounded at $r^\star=L$\label{bounded-parallel-plates}}

The inclusion of a lateral boundary at $r=\ell$ complicates matters. In our previous study \citep{OSD17} of the quasi-steady part of the spin-down our primary concern was the evolution of the laterally diffusing QG-layer from that outer boundary on the long $t_{\ell}=\ell^2E^{-1}$ time scale. However, even in the unbounded case discussed in \S\ref{unbounded-parallel-plates}, inertial waves are excited by the initial impulse, albeit limited to the degenerate MF-type identified by \cite{GH63}. Now it is well known that a myriad of inertial waves exist in our circular cylinder geometry as elucidated for example by \cite{KB95} and \cite{ZL08} \citep[see also][]{ZL17}. Though, the inertial waves triggered by the initial impulse in the bounded cylinder geometry are evidently axisymmetric, the realised mode selection in the closed cylinder remains complicated and is the objective of our present study.

Indeed, evidence from the unbounded case, namely the algebraic $t^{-1/2}$ decay of the MF-modes, suggests that inertial wave generation is a minor effect. This point of view was supported by the DNS results of \cite{OSD17}, which showed little evidence of any significant inertial wave generation for the case $\ell=1$. However more recent DNS results for large-aspect ratio (shallow) containers, namely $\ell=10$ have revealed significant inertial wave activity on the spin-down time $t_{sd}=E^{-1/2}$, as manifest particularly by the contours of $\chi=\,$const.~in figures~\ref{fig1}, \ref{fig2} (below) at various times (panels ($a$), ($d$), ($g$)). For that reason, our intention here is to study analytically the limit $\ell \gg 1$, but we will comment briefly on the relative absence of  wave activity for $\ell=1$ in our concluding \S\ref{conclusion}. 

Our asymptotic approach is based on the premise, that the  \cite{GH63} infinite plane layer solution gives a first approximation to the large $\ell$ bounded problem. However, the main weakness of that solution is its serious failure to meet the impermeable boundary condition $u=0$ at $r=\ell$. Specifically, the $z$-independent part $\ub(\ell,t)$ of the radial velocity has two parts:
\bse
\label{triggers}
\begin{align}
\qquad   \ub_{\tQG}(\ell,t)\,=\,&\,\tfrac12\sigma\kappa E^{1/2}\,\EG(t) & \mbox{(QG-trigger)}&
\intertext{(see (\ref{QG-spin-down}$a$--$e$)) and}
\ub_{\tGH}(\ell,t)\,=\,&\,\tfrac12 E^{1/2}\,\WG_{\tGH}(t) &  \mbox{(MF-trigger)}&
\end{align}
\ese
(see (\ref{MF-spin-down-prelim}$a$,$b$)). Our objective is to elucidate the corrections to the Greenspan \& Howard solution which are ``triggered'' by demanding that the radial velocity correction is $-\,\ub(\ell,t)$ at $r=\ell$.

When $t=O(1)$, the two contributions $-\,\ub_{\tQG}(\ell,t)$ and $-\,\ub_{\tGH}(\ell,t)$ are of comparable size, but later up until the spin-down time is reached, $1\ll t \le O(E^{-1/2})$, (\ref{QG-versus-GH}$b$) gives the estimate $\bigl|\ub_{\tGH}(\ell,t\bigr)\big|/\bigl|\ub_{\tQG}(\ell,t)|=O\bigl(t^{-1/2}\bigr)\ll 1$ suggesting that the QG-triggered motion dominates. The idea that $-\,\ub_{\tQG}(\ell,t)$ provided the trigger for the expanding QG-shear layer at $r=\ell$ was the {\itshape modus operandi} for our study of the QG-evolution \citep{OSD17}. For our relatively large $\ell$, the shallow cylinder also acts as a wave guide for inertial waves triggered at $r=\ell$. In this paper we again consider the QG-trigger  $-\,\ub_{\tQG}(\ell,t)$ but investigate the ensuing inertial wave activity instead. We expect any inertial wave generation by the MF-trigger $-\,\ub_{\tGH}(\ell,t)$ to be of lesser importance and relegate its investigation to our sequel, Part~II. Here we study only the response to the simpler QG-trigger, which identifies and highlights the key physical processes. Nevertheless, as conspicuous features of the guided wave are caused by the initial impulse (i.e., on the early Ekman layer formation time $t=O(1)$, when the two triggers are of comparable size), some essential differences are found in Part~II on retaining the MF-trigger, which lead to far better agreement with  the DNS.

\subsection{Outline\label{Outline}}

In \S\ref{mathematical-problem} we formulate the mathematical problem for the triggered wave motion, $E^{1/2}\vv^{\rm {wave}}=\vv-\vvb_{\tQG}$ (\ref{c-of-v}), and in \S\ref{Fourier-series} simplify using a Fourier series in $z$. In \S\ref{LT-solution} we include viscosity and solve by the LT-method leading to a Fourier-Bessel series in $r$ (see \S\ref{r-Fourier-Bessel-series} and appendix~\ref{Fourier-Bessel}). Though damping by internal friction leads to decay rates that are generally small compared to the Ekman suction decay rate $d^E=O(E^{1/2})$ (\ref{Ek-decay-rate}), a mode with length scale $\delta$ decays at a rate $d^\delta=O(\delta^{-2}E)$, which is faster for sufficiently short length scale modes $\delta\ll E^{1/4}$. Very short length scale modes are generated close to $r=\ell$ and are quickly destroyed near that boundary at the relatively moderate value $E=10^{-3}$ used in the DNS. In \S\ref{Ekman-layer-damping}, we add the additional effect of Ekman suction which leads to the decay rate
\bme
\label{Ek-decay-rate}
\be
d^E\,=\,\tfrac12 E^{1/2}\sigma_+\sigma_-\bigl(\sigma_+^3+\sigma_-^3\bigr), \hskip 10mm \mbox{where}\hskip 10mm
\sigma_\pm=\sqrt{1\pm \omega/2}
\ee
\eme
\citep[see][]{ZL08,KB95}, which is larger than $d^\delta$ when $\delta\gg E^{1/4}$.
The QG-limit $\omega\to 0$ leads to $d^E\to Q$ in agreement with (\ref{QG-spin-down}$b$,$c$,$e$), while for the MF-limit  $|\omega|\uparrow2$ leads to $d^E\downarrow 0$. Then the only damping mechanism is internal friction confined to the expanding shear layer, given by (\ref{transient-EL-sol-z-int}) but included in (\ref {MF-combined}), adjacent to the lower boundary. 

In \S\ref{numerics} we note that the entire inertial wave motion $E^{1/2}[u_\tIW,v_\tIW]$, namely the sum of the triggered inertial waves $E^{1/2}[u^{\rm {wave}},v^{\rm {wave}}]$ studied in \S\ref{mathematical-problem} and the basic state MF-waves $[u_\tGH,v_\tGH]$ described here in \S\ref{upp-MF}, may be obtained asymptotically from the full solution by removing the QG-part as explained in \S\ref{filtered-velocity} (see (\ref{iw-v}) and (\ref{iw-u})). By use of the same recipe, we may extract from the DNS, at small but finite $E$, our so called filtered-DNS, or simply FNS, $[u_\tFNS,v_\tFNS]$ (see (\ref{filter-v}) and (\ref{filter-u})). Our prime objective, the comparison of $[u_\tFNS,v_\tFNS]$ with the analytic results for $E^{1/2}[u_\tIW,v_\tIW]$ undertaken in \S\ref{iw-filter-comparison}, is only applicable outside all quasi-steady boundary layers; they include both the Ekman layer width $\Delta_E=E^{1/2}$ on the rigid $z=0$ boundary and the Stewartson sidewall layer width $\Delta_s=E^{1/3}$ abutting the boundary $r=\ell$. Note, however, that the expanding MF shear layer width $\Delta(t)=\sqrt{Et}$ above the $z=0$ boundary is correctly accounted for by the representation  (\ref{MF-combined}), when $t\gg 1$.

In order to obtain a sharper picture of the various detailed structures identified in \S\ref{numerics}, we neglect viscous damping and set $E=0$ in \S\ref{No-damping}. This allows us to see clearly small scale features that are heavily damped at $E=10^{-3}$ used in the DNS. To understand the complex (but also elegant) wave patterns that emerge, we further restrict our domain of interest in \S\ref{Cartesian-limit} to the large $r$-limit $\ell-r=O(1)$ ($\ell\gg 1$), for which a rectangular Cartesian approximation is applicable. Two distinct solution techniques are employed. Firstly, due to the omission of viscosity, the simplicity of the top $z=1$ and bottom $z=0$ boundaries permits our use in \S\ref{images} of the method of images, essentially a convenient device for handling wave reflection at $z=1$ and $0$. Not surprisingly this clarifies the detailed nature of the inertial waves in the vicinity of $r=\ell$. Further away, wave interference leaves simpler cell forms with dimensions of the gap width unity. So secondly in \S\ref{IFm}, we consider the $(r,\,t)$-evolution of individual $m$-modes of the $z$-Fourier series (\ref{FS}). For given $m$, we use the method of stationary phase (essentially a group velocity consideration) in \S\ref{X-very-large} to identify the dominant structure  at given $(r,\,t)$. The method also shows that the wave becomes evanescent (see also \S\ref{vartheta-very-large*}) beyond a certain distance $x_c(t)=\ell-r_c(t)$ from the outer boundary. The realised distance (\ref{front-line}) is inversely proportional to $m$, demonstrating the importance of the smallest $m=1$ mode and explaining why detailed structure, associated with larger $m$, is only to be found for small $\ell-r$ or at any rate $\ell-r=O(1)$. We end with a few concluding remarks in \S\ref{conclusion}.

%%%%%%%%%%%%%%%%%%%%%%%%%%%%%%%%%%%%%
%%%%%%%%%%%%%%%%%%%%%%%%%%%%%%%%%%%%%
%%%%%       SECTION 2
%%%%%%%%%%%%%%%%%%%%%%%%%%%%%%%%%%%%%
%%%%%%%%%%%%%%%%%%%%%%%%%%%%%%%%%%%%%

\section{The mathematical problem\label{mathematical-problem}}

As already explained our objective is to investigate the inertial wave motion, velocity $E^{1/2}\vv^{\rm {wave}}$, which is excited by the initial impulse caused by the failure of $\vvb_{\tQG}^{\,\infty}$ (\ref{QG-spin-down}$a$,$b$) to meet the boundary condition $u=0$ at $r=\ell$. This failure leads in part to a QG-correction in the form of a shear layer, width $\Delta(t)=\sqrt{Et}$, expanding from $r=\ell$. We denote the entire QG-velocity by $ \vvb_{\tQG}$ (\ref{QG-spin-down}$a$), but not, of course, limited to the special rigid rotation case (\ref{QG-spin-down}$b$). Together they determine
\be
\label{c-of-v}
\vv\,=\,\vvb_{\tQG}\,+\,E^{1/2}\vv^{\rm {wave}}
\ee
in the mainstream exterior to the Ekman layer adjacent to $z=0$  and ageostrophic $E^{1/3}$-sidewall shear layers adjacent to $r=\ell$. From this perspective, the boundary condition at $r=\ell$  may be expressed as
\be
\label{ell-bc}
\bigl(\ub_{\tQG}\,-\,\ub_{\tQG}^{\,\infty}\bigr)\,+\,\,E^{1/2}u^{\rm {wave}}\,=\,-\,\ub_{\tQG}^{\,\infty}\,.
\ee
Here the difference $\ub_{\tQG}\,-\,\ub_{\tQG}^{\,\infty}$ simply recovers the  expanding QG-shear layer with boundary condition $\ub_{\tQG}=\ub_{\tQG}^{\,\infty}$ studied by \cite{OSD17}. So, in what follows, we simply suppose that the inertial waves are triggered by the remaining balance
\be
\label{boundary-condits-prelim}
u^{\rm {wave}}\,=\,-\,E^{-1/2}\ub_{\tQG}^{\,\infty}\,=\,-\,\tfrac12\kappa\sigma\!\;\EG(t) \hskip 20mm \mbox{at} \hskip 6mm r=\ell
\ee
(see (\ref{QG-spin-down}$a$-$e$), also (\ref{triggers}$a$), and (\ref{boundary-condits}$b$) below).

The above description totally ignores the MF-contribution $\vv_\tGH$ (\ref{MF-combined}) equivalently (\ref{GH-waves}). Of course, that is the basis upon which we have adopted the boundary condition (\ref{boundary-condits-prelim}). Nevertheless, when we attempt to make contact with the DNS we will need to include the MF-waves. Throughout this section we will simply solve for the inertial waves $\vv^{\rm {wave}}$ triggered by (\ref{boundary-condits-prelim}) and to simplify the notation drop the superscript `wave' and write $\vv=[u,\,v,\,w]\,(\mapsfrom\,\vv^{\rm {wave}}$).

The inertial wave problem is: Solve
\bme
\label{gov-eqs}
\begin{align}
\pd{v}{t}\,+\,2\,u\,&=\,E\bigl(\nabla^2-r^{-2}\bigr)v\,,&    u\,&=\,-\,\pd{\chi}{z}\,,\\[0.3em]
\pd{\gamma}{t}\, -\,2\,\pd{v}z\,&=\,E\bigl(\nabla^2-r^{-2}\bigr)\gamma\,,&
\gamma\,&=\,-\, \bigl(\nabla^2-r^{-2}\bigr)\chi
\end{align}
\eme
subject to the initial ($t=0$) conditions 
\bme
\label{initial-condits}
\be
v = 0,\qquad\qquad\qquad     \gamma=0,
\ee
\eme
and for $t\ge 0$ the boundary conditions 
\bse
\label{boundary-condits}
\begin{align}
r\chi\,&=\,0& \mbox{at} &&r\,=&\,0\, & (0&<z\le 1)\,,\\ 
r\chi\,&=\,\tfrac12 \ell\kappa\sigma (z-1)\!\;\EG(t)& \mbox{at} &&r\,=&\,\ell\, & (0&<z\le 1)\,,\\ 
\chi\,&=\,0&  \mbox{at}&& z\,=&\,0,\,1 & (0&<r<\ell),
\end{align}
\ese
where (\ref{boundary-condits}$b$) corresponds to (\ref{boundary-condits-prelim}) and for our modal expansions (\ref{FS}$a$) below we need
\be
\label{z-minus-one}
\tfrac12 (z-1)\, = \,-\,\sum_{m=1}^\infty\dfrac{(-1)^m}{m\pi}\sin(m\pi (z-1))\hskip 15mm      (0<z\le 1)\,.
\ee
Some care is needed in the interpretation and implementation of the boundary conditions (\ref{boundary-condits}), which strictly apply to the inviscid $E=0$ problem and are insufficient for the viscous ($E\not=0$) equations (\ref{gov-eqs}). For the moment, in this semi-inviscid spirit we only address interior viscous dissipation and will later incorporate the effects of Ekman boundary layers. Indeed we ignore any viscous sidewall layers at $r=\ell$ completely, as the discussion between (\ref{se-chi}) and (\ref{LT-se-chi-bc}) below emphasises. The reason for this
%rather
cavalier approach is two-fold:
\begin{itemize}
\item[(i)] We are not interested in any quasi-steady shear layers. That was remit of \cite{OSD17};
\item[(ii)] our primary concern is to identify the inertial wave generation. Their damping is a secondary {\itshape bookkeeping} exercise  needed to identify what is realised at finite $E$ so that comparisons can be made with the DNS.
\end{itemize}

\subsection{The $z$-Fourier series \label{Fourier-series}}

We seek a  $z$-Fourier series solution
\bse
\label{FS}
\begin{align}
\left[\begin{array}{c}  \!\! \chi \!\!\\[0.2em]
      \!\!  \gamma \!\!  \end{array}\right]\,=\,&\,
-\,\kappa\sigma\sum_{m=1}^\infty\dfrac{(-1)^m}{m\pi}
\left[\begin{array}{c}  \!\! \chit_m \!\!\\[0.2em]
      \!\!  \gamma_m \!\!  \end{array}\right]
\sin\bigl(m\pi (z-1)\bigr)\,,\,
\intertext{chosen so that $\chi(r,z,t)$ satisfies the top and bottom boundary conditions (\ref{boundary-condits}$c$), together with}
\left[\begin{array}{c}  \!\! u  \!\!\\[0.2em]
      \!\! v \!\!  \end{array}\right]  
\,=\,&-\,\kappa\sigma\sum_{m=1}^\infty\dfrac{(-1)^m}{m\pi}
\left[\begin{array}{c}  \!\! \ut_m \!\!\\[0.2em]
      \!\! \vt_m \!\!  \end{array}\right]
\cos\bigl(m\pi (z-1)\bigr)\,.
\end{align}  
A further property of $v$, due to its assumed form (\ref{FS}$b$), is
\be
\langle v  \rangle\,=\,0\,, \hskip 5mm \mbox{strictly} \hskip 5mm O(E^{1/2})
\ee
\ese
when the consequences of the Ekman layer are taken into account.
The series (\ref{FS}$a$,$b$) satisfy (\ref{gov-eqs}) when $\chit_m(r,t)$ and $\vt_m(r,t)$ are governed by
\bme
\label{F-Series-eqs}
\begin{align}
    \pd{\vt_m}{t}\,+\,2\ut_m\,&=\,E\DS_m\vt_m\,,\qquad\qquad  \ut_m\,=\,-\,m\pi \chit_m\,,\\[0.3em]
\pd{\gammat_m}{t}\, +\,2m\pi \vt_m\,&=\,E\DS_m\gammat_m\,,\qquad\qquad \gammat_m\,=\,-\,\DS_m \chit_m\,,
\end{align}
where
\be\se
\DS_m\,\bullet\,=\,\dfrac1r\pd{\,}r\biggl(r\pd{\,\bullet}{r}\biggr)-\biggl(\dfrac1{r^2}+(m\pi)^2\biggr)\,\bullet\,.
\ee
\eme
They are to be solved subject to the initial ($t=0$) conditions 
\bme
\label{LT-init-condit}
\te
\be
\vt_m = 0,\qquad\qquad
\gammat_m =0 \qquad \Longrightarrow \qquad \chit_m(r,0)\,=\,\IR_1(m\pi r)\big/\IR_1(m\pi \ell)
\ee
\eme
(see (\ref{initial-condits}) together with (\ref{F-Series-eqs}$d$) and (\ref{LT-bdry-condit}) below), and for $t\ge 0$ the boundary conditions 
\bse
\label{LT-bdry-condit}
\begin{align}
r\chit_m\,&=\,0& \mbox{at} &&r\,=&\,0\, & (0&<z<1)\,,\\
r\chit_m\,&=\,\ell\!\;\EG(t)& \mbox{at} &&r\,=&\,\ell\, & (0&<z<1)
\end{align}
\ese
(see (\ref{boundary-condits}$a$,$b$)). For our LT-solution of this initial value problem in the following \S\ref{LT-solution}, it is useful to note that $\chit_m(r,0)$ (\ref{LT-init-condit}$c$) can be represented, via the use of the Fourier-Bessel series (\ref{Fourier-Bessel-end}) with $q=\iR$ (giving $\JR_1\big(\iR m\pi r\bigr) =\iR\,\IR_1\big(m\pi r\bigr)$), in the form
\bme
\label{Fourier-Bessel-series}
\be
\se
\dfrac{\IR_1(m\pi r)}{\IR_1(m\pi\ell)}\,=\,-\,\sum_{n=1}^\infty\FG_{mn}\,\dfrac{\JR_1(j_nr/\ell)}{j_n\JR_0(j_n)} \hskip 10mm \mbox{on} \hskip 10mm  0\,\le\, r\,<\,\ell\,,
\ee
where $j_n$ denotes the $n^{\rm{th}}$ zero ($>0$) of $\JR_1(x)$, and
\begin{align}
\FG_{mn}\,=\,&\,\HG^2_{mn} \big/2\,,  &    \HG_{mn}\,=\,&\,q_{mn}\omega_{mn}\,,\\
\omega_{mn}\,=\,&\,{2}\Big/\!\bigl(q_{mn}^2+1\bigr)^{1/2}, &  q_{mn}\,=\,&\,j_n/(m\pi\ell)\,.
\end{align}
\eme
Obviously the series (\ref{Fourier-Bessel-series}$a$) only holds on $0\le r<\ell$ as each term vanishes at $r=\ell$.

\subsection{The Laplace transform (LT-)solution\label{LT-solution}}

The Laplace transform  of the governing equations (\ref{F-Series-eqs}) and initial conditions (\ref{LT-init-condit}) determine
\bse
\label{LT-eqs}
\begin{align}
p \vth_m\,-\,2m\pi\chith_m\,&=\,E\DS_m\vth_m\,,\\[0.3em]
p\DS_m\chith_m -\,2m\pi \vth_m\,&=\,E\DS_m^2\chith_m\,,
\end{align}
where
\be
\bigl[\,\chith_m\,,\,\vth_m\,\,\bigr](z,p)\,=\, \LC_p \bigl\{\bigl[\,\chit_m\,,\,\vt_m\,\bigr]\bigr\}
\ee
\ese
and $\LC_p$ is defined by (\ref{lT-inv-def}). Elimination of $\vth_m$ leads to a single equation for $\chith_m$:
\be
\label{se-chi}
\bigl(p-E\DS_m\bigr)^2\DS_m\chith_m\,-\,4(m\pi)^2\chith_m\,=\,0\,.
\ee

As already stressed, we ignore viscous boundary layers and solve (\ref{se-chi}) on the basis that, when $E=0$, it is second order in $r$ for which the end-point boundary conditions
\bme
\label{LT-se-chi-bc}
\be
r\chith_m = 0 \hskip 5mm \mbox{at} \quad r=0 \hskip 10mm\mbox{and} \hskip 10mm
\left\{\begin{array}{l} r\chith_m =\ell\EGh(p)  \hskip 10mm \mbox{at} \quad r=\ell\,,\\[0.1em]
\hskip 0mm  \EGh(p)=(p+Q)^{-1},\end{array}\right.
\ee
\eme
namely the Laplace transforms of (\ref{LT-bdry-condit}) and (\ref{QG-spin-down}$c$), suffice. The pertinent solution is
\bse
\label{LT-se-chi-sol}
\be
\left[\begin{array}{c}  \!\! \chith_m \!\!\\[0.2em]
    \!\!  \vth_m \!\!  \end{array}\right]\,=
\left[\begin{array}{c}  \!\! 1 \!\!\\[0.3em]
    \!\!  2m\pi/\pG \!\!  \end{array}\right]\,\EGh(p)\,\dfrac{\JR_1\bigl(m\pi qr\bigr)}{\JR_1\big(m\pi q\ell\bigr)}\,.
\ee
Here, since $\chith_m$ satisfies
\be
\DS_m\chith_m =\,-\,\bigl(q^2+1\bigr)(m\pi)^2\chith_m\,,
\ee
\ese
it follows from (\ref{se-chi}) that $p$ and $q$ are related by the ``dispersion relation''
\bme
\label{LT-se-chi-sol-disp-rel}
\begin{align}
\pG^2\,=\,&\,-\,4\big/\bigl(q^2+1\bigr)& \mbox{in which} && \pG\,=\,&\,p\,+\,E(q^2+1)(m\pi)^2\\
\intertext{(the definition of $\pG$), equivalently}
q^2+1\,=\,&\,-\,4/\pG^2&\mbox{and}\hskip 7mm && p\,=&\,\,\pG\,+\,E(2m\pi)^2/\pG^2\,,
\end{align}
from which we obtain the useful result
\be
\se
\pG\od{p}{\pG}\,=\,\pG\,-\,2E\,\dfrac{(2m\pi)^2}{\pG^2}\,.
\ee
\eme

The inverse-LT (\ref{lT-inv-def}$b$) of $\chith_m$, defined by (\ref{LT-se-chi-sol}$a$), is
\be
\label{se-chi-sol}
\chit_m\,=\,\LC_p^{-1}\,\biggl\{\EGh(p)\,\dfrac{\JR_1\bigl(m\pi qr\bigr)}{\JR_1\big(m\pi q\ell\bigr)}\biggr\}.
\ee
The initial condition (\ref{LT-init-condit}c) is recovered on expanding the integrand of the inverse-LT integral (\ref{se-chi-sol}) about the limit $p\to \infty$, for which $q\to \iR$ (see (\ref{LT-se-chi-sol-disp-rel}$a$,$b$)) and $p\EGh(p)\to 1$. For $t>0$ the inverse of (\ref{LT-se-chi-sol}$a$) has two parts,
\be
\label{wave+AG}
\bigl[\chit_m\,,\,\vt_m\bigr]\,=\,\bigl[\chit^\daleth_m\,,\,\vt^\daleth_m\bigr]\,+\,\bigl[\chit^{\tAG}_{m}\,,\,\vt^{\tAG}_m\bigr]\,.
\ee
The former inertial wave part $\bigl[\chit^\daleth_m\,,\,\vt^\daleth_m\bigr]$ stems from the residues (denoted by Res$\{\,\}$) at the set $\daleth$ of poles $p\,(\not=0)$ linked to the zeros of $\JR_1\big(m\pi q\ell\bigr)$; for $t>0$, they have the property $\chit^\daleth_m(\ell,t)=0$ (see (\ref{LT-se-chi-v-sol-disp-rel}) below). The latter ageostrophic-part $\bigl[\chit^{\tAG}_{m}\,,\,\vt^{\tAG}_m\bigr]$ stems from the residues at the poles of $\EGh(p)=(p+Q)^{-1}$ and $\pG^{-1}$. When internal friction is included ($d_{mn}\not=0$), this AG-part determines a Stewartson $E^{1/3}$-layer and alone meets the boundary condition $\chit^{\tAG}_m(\ell,t)=\EG(t)$ (see (\ref{LT-se-chi-bc}$b$)). However, since we have not applied any stress related boundary conditions, the flow so determined is unphysical and we consider it no further. Hence, the wave part of the velocity $\vv^{\rm {wave}}$ alluded to in (\ref{c-of-v}) is simply $\vv^\daleth$, valid on the entire range $0\le r \le \ell$.

\subsection{The $r$-Fourier-Bessel series\label{r-Fourier-Bessel-series}}

The residues at the poles $p\in\daleth$ are rendered more illuminating by use of the Fourier-Bessel series (\ref{Fourier-Bessel-end}) with $q^2=-1-4/\pG^2$ (\ref{LT-se-chi-sol-disp-rel}$a$), $q_{mn}^2=-1+4/\omega^2_{mn}$ (see (\ref{Fourier-Bessel-series}$d$)) giving $2q_{mn}^2\big/\bigl(q_{mn}^2-q^2\bigr)=\FG_{mn}\pG^2\big/\bigl(\pG^2+\omega^2_{mn}\bigr)$ on use of (\ref{Fourier-Bessel-series}$b$,$c$). It enables us to express the residue sum for $\bigl[\chit^\daleth_m\,,\,\vt^\daleth_m\bigr]$ derived from (\ref{LT-se-chi-sol}$a$) in the form
\begin{align}
\left[\begin{array}{c}  \!\! \chit^\daleth_m \!\!\\[0.3em]
    \!\!  \vt^\daleth_m \!\!  \end{array}\right]\,=\,&\,-\,\sum_{n=1}^\infty\dfrac{\JR_1(j_nr/\ell)}{j_n\JR_0(j_n)}\,
\FG_{mn}\,\,\underset{p\in\daleth}{\mathrm{Res}}\left\{\left[\begin{array}{c}  \!\! \pG \!\!\\[0.3em]
    \!\!  2m\pi \!\!  \end{array}\right]\,\dfrac{\pG\exp(pt)}{\pG^2+\omega_{mn}^2}\,\dfrac{1}{p+Q}\right\}\nonumber\\
=\,&\,-\,\sum_{n=1}^\infty\dfrac{\JR_1(j_nr/\ell)}{j_n\JR_0(j_n)}
\,\FG_{mn}\left(\underset{p\in\daleth+\!\!}{\mathrm{Res}}\left\{\left[\begin{array}{c}  \!\! \pG/2 \!\!\\[0.3em]
    \!\!  m\pi \!\!  \end{array}\right]\,\dfrac{\exp(pt)}{\pG-\iR\omega_{mn}}\,\dfrac{1}{p+Q}\right\}\,+\,\mbox{c.c.}\right).
\label{LT-se-chi-v-sol-disp-rel}
\end{align}
Here we have identified the half-set $\daleth+$ of poles
\bme
\be
\pG\,=\,\pG_{mn}\,=\,\iR\omega_{mn}   \hskip 10mm  \Longleftrightarrow \hskip15mm p\,=\,p_{mn}\,=\,\iR\omega_{mn}\,-\,d_{mn}
\ee
having $\omega_{mn}=\Im\{\pG_{mn}\}>0$, which when combined with their complex conjugates (denoted by c.c.) form the complete set $\daleth$. On use of (\ref{LT-se-chi-sol-disp-rel}$d$,$e$) we determine
\be
d_{mn}\,=\,\dfrac{E(2m\pi)^2}{\omega_{mn}^2} \hskip 10mm  \Longrightarrow \hskip 8mm
\biggl[\pG\od{p}{\pG}\biggr]_{\pG=\iR\omega_{mn}}=\,\iR\omega_{mn}\,+\,{2d_{mn}}\,.
\ee
\eme

Evaluation of the residues in (\ref{LT-se-chi-v-sol-disp-rel})  yields
\bse
\label{solution-preliminary}
\begin{align}
\left[\begin{array}{c}  \!\! \chit^\daleth_m \!\!\\[0.3em]
    \!\!  \vt^\daleth_m \!\!  \end{array}\right]\,=\,&\,\sum_{n=1}^\infty \dfrac{\JR_1(j_nr/\ell)}{2\JR_0(j_n)}
\left[\begin{array}{c}  \!\! \chimr^\daleth_{mn} \!\!\\[0.3em]
    \!\!  \vmr_{mn} \!\!  \end{array}\right]\exp({-\lambda_{mn}t})\,,
\intertext{in which}
\left[\begin{array}{c}  \!\! \chimr^\daleth_{mn} \!\!\\[0.3em]
    \!\!  \vmr^\daleth_{mn} \!\!  \end{array}\right]\,=\,&\,
\left[\begin{array}{c}  \!\! -\,\FG_{mn}/j_n \!\!\\[0.3em]
    \!\! \iR\HG_{mn}/\ell \!\!  \end{array}\right]\dfrac{\iR\omega_{mn}+2d_{mn}}{\iR\omega_{mn}-d_{mn}+Q}\exp(\iR\omega_{mn}t)\,+\,\mbox{c.c.}\,,
\end{align}
\ese
where $\FG_{mn}$ and $\HG_{mn}$ are defined by (\ref{Fourier-Bessel-series}$b$,$c$). Written explicitly, (\ref{solution-preliminary}) is
\bme
\label{solution}
\se
\begin{align}
\!\!\chit^\daleth_m=\,&\,-\sum_{n=1}^\infty\FG_{mn}\,\dfrac{\JR_1(j_nr/\ell)}{j_n\JR_0(j_n)}
\bigl(\CS^\EG_{mn}\cos\phi_{mn} +\SS^\EG_{mn}\sin\phi_{mn}\bigr)\exp({-\lambda_{mn}t})\,,\\
\!\!\vt^\daleth_m=\,&\,-\sum_{n=1}^\infty\HG_{mn}\,\dfrac{\JR_1(j_nr/\ell)}{\ell\JR_0(j_n)}\bigl(\CS^\EG_{mn}\sin\phi_{mn} -\SS^\EG_{mn}\cos\phi_{mn}\bigr)\exp({-\lambda_{mn}t})\,,
\end{align}
where
\be
\de
\CS^\EG_{mn}\,=\,1\,-\,\dfrac{(3d_{mn}-Q)(d_{mn}-Q)}{\omega^2_{mn}+(d_{mn}-Q)^2}\,,  \hskip10mm    \SS^\EG_{mn}\,=\,\dfrac{(3d_{mn}-Q)\omega_{mn}}{\omega^2_{mn}+(d_{mn}-Q)^2}\,,
\ee
and
\be
\de
\phi_{mn}(t)=\,\omega_{mn}t\,, \hskip 25mm  \lambda_{mn}\,=\,d_{mn}\,.\hskip 8mm
\ee
\eme

At the initial instant $t=0$ all $(m,n)$-harmonics are synchronised because of their vanishing phase $\phi_{mn}(0)=0$. So, though the Fourier-Bessel series  (\ref{solution-preliminary}$a$) for the pole-contribution ($\chit^\daleth_m$) is valid on  $0\le r\le\ell$ vanishing at $r=\ell$, the Fourier-Bessel series  (\ref{Fourier-Bessel-series}$a$) for the initial value of the combined sum ($\chit^\daleth_m +\chit^{\tAG}_m$) (see (\ref{LT-init-condit}$c$)) only holds on $0\le r<\ell$ with a discontinuity at $r=\ell$. This subtle complication is illustrated further by the undamped case ($d_{mn}=0$, $Q=0$), for which the initial values $\vt_m(r,0) = 0$, $\chit_m(r,0)=\IR_1(m\pi r)\big/\IR_1(m\pi \ell)$ (see (\ref{LT-init-condit}$a$,$c$)) are recovered from (\ref{solution}) with $\CS^\EG_{mn}=1$ and $\SS^\EG_{mn}=0$ on  $0\le r<\ell$. The initial discontinuity at $r=\ell$ is real, but may be resolved in the damped case, for which $\CS^\EG_{mn}\not=1$ and $\SS^\EG_{mn}\not=0$. Then small difference with the true initial values can be accounted for by the quasi-steady ageostrophic-part $\bigl[\chit^{\tAG}_{m}\,,\,\vt^{\tAG}_m\bigr]$ that we have discarded.

\subsection{Ekman layer damping\label{Ekman-layer-damping}}

As we remarked in our introduction \S\ref{bounded-parallel-plates}, the Ekman layer provides a relatively strong damping of inertial waves, quantified by the decay rate $d^E$ (\ref{Ek-decay-rate}$a$). The basis of that formula is eq.~(4.5) of \cite{ZL08}, which consists of three sets of terms. The first corresponds to our internal friction decay rate $d_{mn}$. The second proportional to their $\Gamma^{-1}$  (our $\ell^{-1}$) corresponds to decay caused by the end wall boundaries, which is negligible in our large aspect $\ell\gg 1$ limit. Indeed that friction is absent for our stress-free outer boundary. The third, namely the remaining pair of terms, identifies the decay rate $d_E$. For that, we halve the Zhang \& Liao result because we only have an Ekman layer on $z=0$ and no layer on $z=1$. The correction $\omega^E$ to the frequency is not given by Zhang \& Liao, but can be deduced as $\omega^E=\tfrac12 E^{1/2}\sigma_+\sigma_-\bigl(\sigma_+^3-\sigma_-^3\bigr)$ ($\sigma_\pm$ defined by  (\ref{Ek-decay-rate}$b$)) from the formula (2.12) of \cite{KB95}. It follows that, to accommodate Ekman layer dissipation, we should add the complex growth rate
\bse
\label{EL-damping}
\be
p^{E\pm}_{mn}\,=\,-\,d^E_{mn}\,\pm \iR\omega^E_{mn}\,,
\ee
\vskip -2mm
\noindent where
\begin{align}  
d^E_{mn}\,=\,&\,\tfrac12 E^{1/2}\bigl(1-(\omega_{mn}/2)^2\bigr)^{1/2}\bigl[(1+\omega_{mn}/2)^{3/2}+(1-\omega_{mn}/2)^{3/2}\bigr]\,,\\
\omega^E_{mn}  \,=\,&\,\tfrac12 E^{1/2}\bigl(1-(\omega_{mn}/2)^2\bigr)^{1/2}\bigl[(1+\omega_{mn}/2)^{3/2}-(1-\omega_{mn}/2)^{3/2}\bigr]\,.
\end{align}
\ese
With this dissipation added, the formula (\ref{solution}$a$,$b$) continues to hold, but with (\ref{solution}$e$,$f$) replaced by
\bme
\label{EL-solution-parameters}
\be
\phi_{mn}(t)\,=\,\bigl(\omega_{mn}+\omega^E_{mn}\bigr)t+\epsilon^E_{mn}\,, \qquad\qquad \lambda_{mn}\,=\,d_{mn}+d^E_{mn}\,.
\ee
\eme
Here we have included small phase corrections $\epsilon^E_{mn}$, which are not determined by the earlier results to which we have appealed. Being small we do not expect them to be important and so have not attempted to quantify them. In all our numerical evaluations we have simply set $\epsilon^E_{mn}=0$.

%%%%%%%%%%%%%%%%%%%%%%%%%%%%%%%%%%%%%
%%%%%%%%%%%%%%%%%%%%%%%%%%%%%%%%%%%%%
%%%%%       SECTION 3
%%%%%%%%%%%%%%%%%%%%%%%%%%%%%%%%%%%%%
%%%%%%%%%%%%%%%%%%%%%%%%%%%%%%%%%%%%%

\section{Comparison with the DNS\label{numerics}}

To solve the entire spin-down problem, we performed DNS of the full governing equations (\ref{gov-eqs}) subject to the initial conditions
\be
\label{ic}
{v/r}\,=\,1\,,\qquad\qquad {r\chi}\,=\,0 \qquad\qquad\mbox{everywhere at} \qquad t=0\,,
\ee
and boundary conditions
\bse
\label{bc}
\begin{align}
{r \chi}\,=\,\pd{{(v/r)}}{r}\,=\,\pd{{w}}{r}\,&=\,0& \mbox{at} &&r\,=&\,0\,\,\,\mbox{ and }\,\,\ell & (0&<z<1)\,,\\
{r \chi}\,=\,\pd{{(r \chi)}}{z}\,=\,{v/r}\,&=\,0&  \mbox{at}&& z\,=&\,0 & (0&<r<\ell)\,,\\
\qquad{r \chi}\,=\,\pd{^2{(r \chi)}}{z^2}\,=\,\pd{{(v/r)}}{z}\,&\,=\,\,0&  \mbox{at} &&z\,=&\,1 &(0&<r<\ell)\,.\qquad
\end{align}
\ese

We solved (\ref{gov-eqs}) using second-order finite differences in space, and an implicit second-order backward differentiation (BDF2) in time. We used a stretched grid, staggered in the $z$-direction. The simulations were performed with a spatial resolution of $3000 \times 500$, a convergence study confirmed that this resolution is sufficient at the Ekman number considered here.

In \S\ref{mathematical-problem} we considered, from an asymptotic point of view, the inertial wave response $E^{1/2}\vv^{\rm {wave}}$ outside the Ekman layer (see  (\ref{c-of-v})) to the QG-trigger subject to the reduced set of initial and boundary conditions (\ref{initial-condits}) and (\ref{boundary-condits}). The superscript `wave', dropped in \S\ref{mathematical-problem}, is reinstated throughout this section. On excluding the side-wall layers, we have $E^{1/2}\vv^{\rm {wave}}=E^{1/2}\vv^\daleth$. Those considerations ignored the MF-wave contribution $\vv_{\tGH}$, which needs to be added to $E^{1/2}\vv^{\rm {wave}}$ to construct the complete inertial wave (IW-)structure $E^{1/2}\vv_\tIW$:
\bme
\label{IW-def}
\be
\vv\,=\,\vvb_{\tQG}\,+\,E^{1/2}\vv_{\tIW}\,, \hskip 20mm  E^{1/2}\vv_{\tIW}\,=\,\vv_{\tGH}\,+\,E^{1/2}\vv^{\rm {wave}}\,.
\ee
\eme
Our goal is to compare the \S\ref{mathematical-problem} results with the DNS identified by the subscript `$DNS\,$' and illustrated in panels ($a$), ($d$), ($g$) of figures~\ref{fig1}--\ref{fig4} (below). Care must be taken with the scale factor $E^{1/2}$ introduced in $E^{1/2}\vv_{\tIW}$, $E^{1/2}\vv^{\rm {wave}}$ and evident in the relations (\ref{iw-v})--(\ref{filter-u}) (below). Once these inter-relations have been set up in the following \S\ref{filtered-velocity}, we adopt the scaling $E^{-1/2}\vv$ (as in $\vv_{\tIW}$, $\vv^{\rm {wave}}$) for our reference velocity unit in our description of the figures in \S\ref{iw-filter-comparison}.

\subsection{The filtered DNS-velocity $\vv_{\tFNS}$\label{filtered-velocity}}

The most dominant feature of the spin-down, exterior to the bottom Ekman layer, is the $z$-independent azimuthal QG-flow $\vb_\tQG$, which is larger by a factor of at least $O\bigl(E^{-1/2}\bigr)$ than almost all other contributions to the complete flow description. So, to make comparison with results based on our \S\ref{mathematical-problem} theory for $E^{1/2}v^{\rm {wave}}$, we need to remove $\vb_\tQG$ from $v$. As $\vb_\tQG$ is not easily identifiable from the numerics, we determine it indirectly from the $z$--average $\langle v \rangle$ of $v$. To this end, we note that, on ignoring all wave motion, (\ref{bar-angle}$b$) indicates that $\langle v \rangle =\mu\vb_{\tQG}+O(E)$, a result that even holds in the expanding QG-shear layer adjacent to the outer boundary $r=\ell$. Interestingly, for $t\gg 1$, though the MF-waves have $\vb_{\tGH}=O\bigl((E/t)^{1/2}\bigr)$ (see (\ref{MF-spin-down}) with (\ref{outflow-sol-MF}$d$)), their $z$--average $\langle \vb_{\tGH}\rangle$ is smaller by a factor $O\bigl(t^{-1}\bigr)$: $\langle v_{\tGH}\rangle=O\bigl(E^{1/2}t^{-3/2}\bigr)$ (see (\ref{GH-waves-series-means}$b$)). Furthermore the inertial waves $E^{1/2}v^{\rm {wave}}$ in their assumed form (\ref{FS}$b$) have zero $z$--average. That assumption was based on neglect of their associated Ekman layer. In practice, these Ekman layers carry an azimuthal flux smaller by a factor $O\bigl(E^{1/2}\bigr)$ so that $E^{1/2}\langle v^{\rm {wave}} \rangle =O(E)$. This fortuitous estimate indicates that the (IW-)contribution $E^{1/2}v_\tIW$ (\ref{IW-def}$b$), outside the Ekman layer, is related to the full solution by
\be
\label{iw-v}
v_\tIW
\,=\,E^{-1/2}v_{\tGH}\,+\,v^{\rm {wave}}
\,=\,E^{-1/2}\bigl(v\,-\mu^{-1}\langle v \rangle\bigr) +\,O(E^{1/2})
\ee
on the spin-down time $t=O\bigl(E^{-1/2}\bigr)$. Owing to the relative simplicity of the MF-harmonic expansions (\ref{GH-waves}$a$,$b$) of appendix~\ref{MF-harmonic}, we used them to determine $\chi_{\tGH}$, $v_{\tGH}$. However, we did compare the results with a quantitatively improved version of the blended mainstream and boundary layer form (\ref{MF-combined}) and found no discernible differences to graph plotting accuracy.

We also assume that the quasi-steady $z$-dependent correction to $\vb_\tQG$ is relatively small $O(E|\vb_\tQG|)$ \citep[see][eq.~(2.11$a$)]{OSD17} so that its presence on the right-hand side of (\ref{iw-v}) does not corrupt the recipe for the IW-part $v_\tIW$, at any rate to the order of accuracy needed. Importantly, we may evaluate $v-\mu^{-1}\langle v \rangle$ directly from the DNS-results and refer to
\be
\label{filter-v}
v_{\tFNS}\,=\,E^{-1/2}\bigl(v_{\tDNS}\,-\mu^{-1}\langle v_{\tDNS} \rangle\bigr)
\ee
as the ``filtered DNS'' or simply FNS. In figures~\ref{fig3},~\ref{fig4} (below), we portray $v_{\tFNS}$ in the FNS-panels ($b$), ($e$), ($h$), derived from $E^{-1/2}v_{\tDNS}$ illustrated in the DNS-panels ($a$), ($d$), ($g$), while $v_\tIW$ is shown in the IW-panels ($c$), ($f$), ($i$).

\begin{figure}
\centerline{}
\vskip 3mm
\centerline{
\includegraphics*[width=1.0 \textwidth]{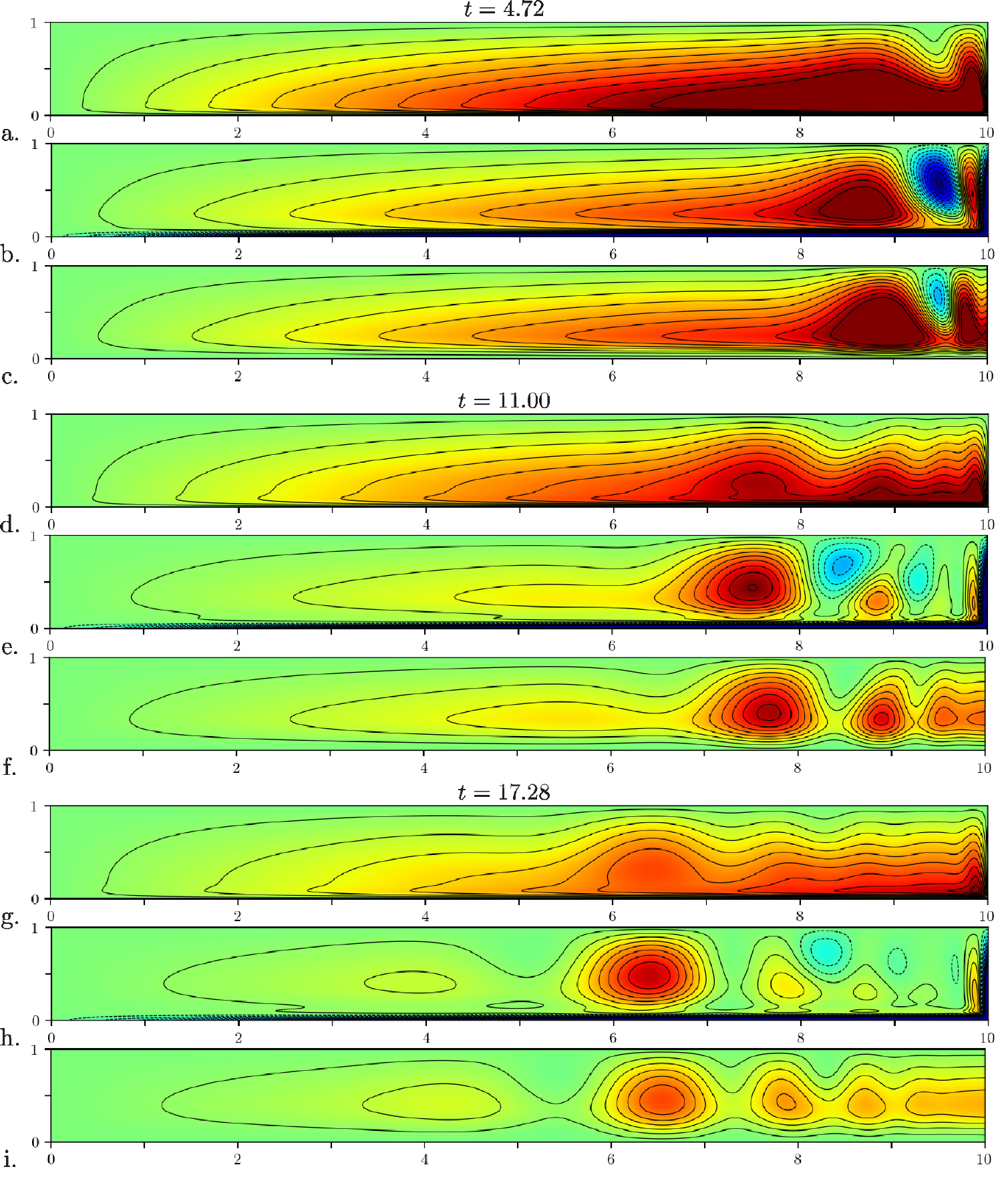}
}
\caption{(Colour online) The case $E=10^{-3}$, $\chi$-contours at three distinct instants $t=N\pi/2$ ($N=3,\,7,\,11$) when $E^{-1/2}\chib_{\tGH}$ is maximised: ($a$)--($c$), ($d$)--($f$), ($g$)--($i$) correspond to $t=4.72$, $11.00$, $17.28$ respectively. ($a$), ($d$), ($g$) show $E^{-1/2}\chi_{\tDNS}$ (colour scale from $-3$ to $3$); ($b$), ($e$), ($h$) and ($c$), ($f$), ($i$) show  $\chi_{\tFNS}$ and $\chi_\tIW$ respectively (colour scale from $-1$ to $1$).
%%  \hskip 50mm  -----------------------------------------------------------------------------------------------------------------------------
}
\label{fig1}
\end{figure}

\begin{figure}
\centerline{}
\vskip 3mm
\centerline{
\includegraphics*[width=1.0 \textwidth]{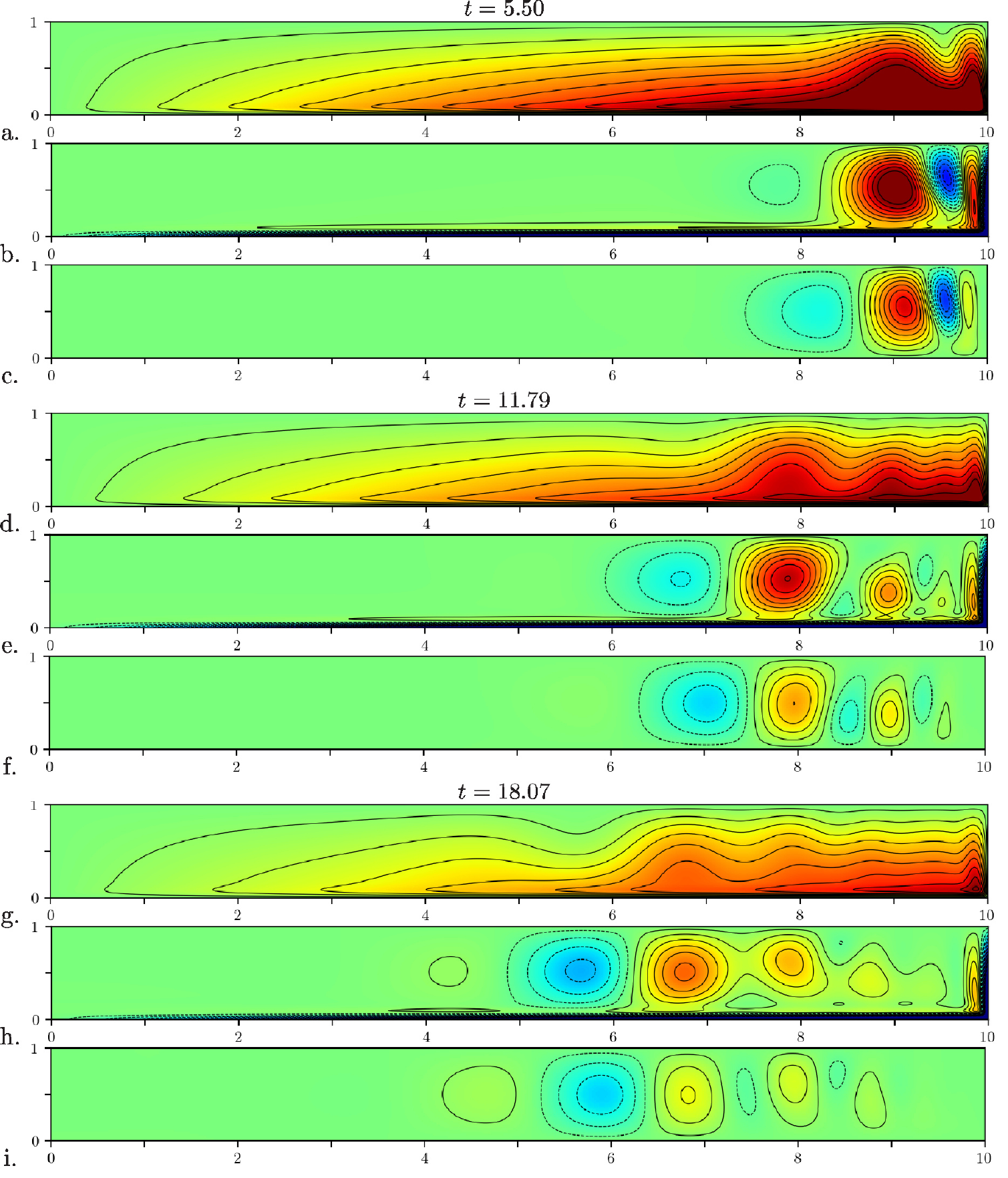} 
}
\caption{(Colour online) As in figure~\ref{fig1} but now at three distinct instants $t=(N+\tfrac12)\pi/2$ ($N=3,\,7,\,11$) at which $E^{-1/2}\chib_{\tGH}=0$. ($a$)--($c$), ($d$)--($f$), ($g$)--($i$) correspond to $t=5.50$, $11.79$, $18.07$ respectively.}
\label{fig2}
\end{figure}

\begin{figure}
\centerline{}
\vskip 3mm
\centerline{
\includegraphics*[width=1.0 \textwidth]{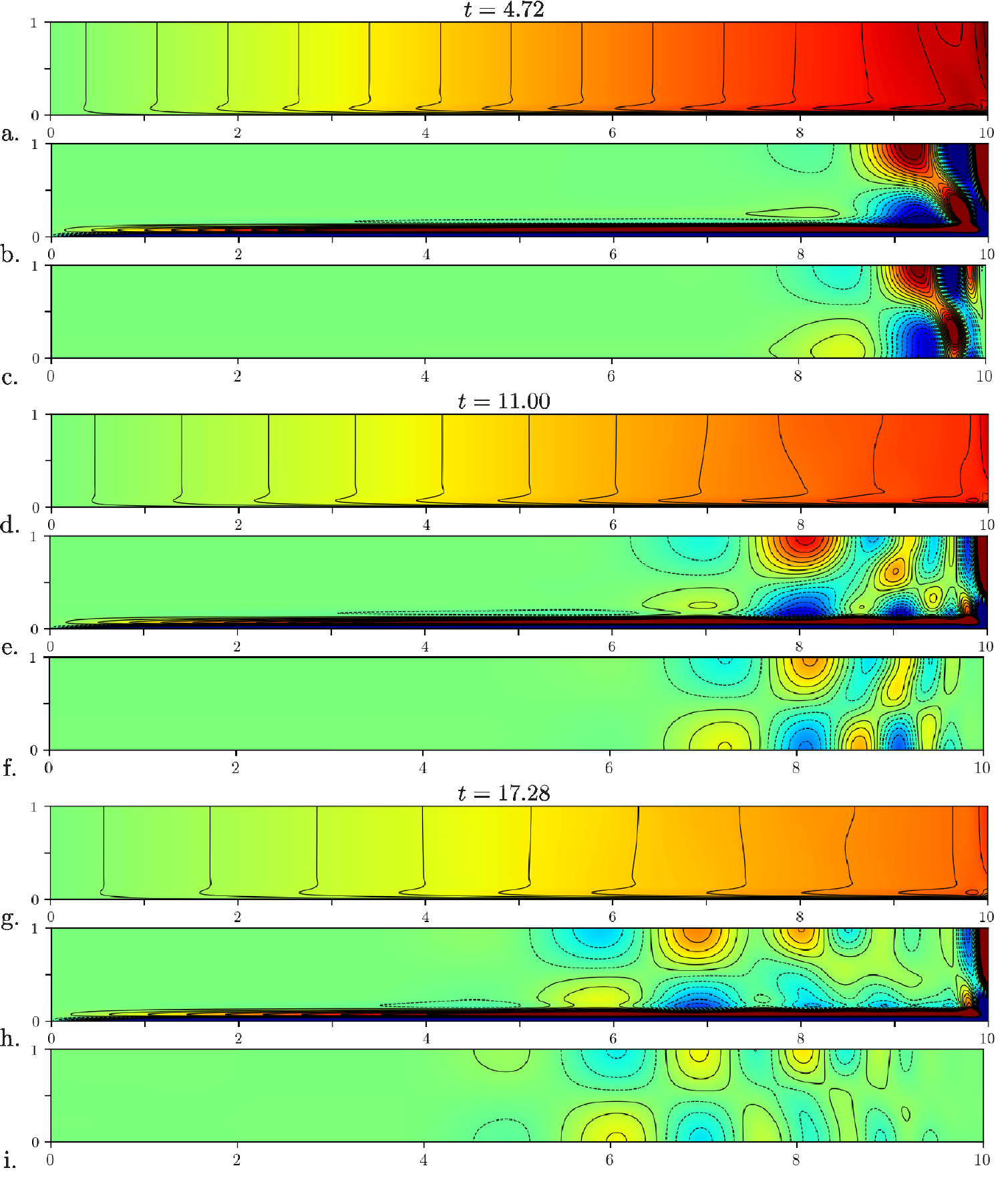}  
}
\caption{(Colour online) As in figure~\ref{fig1} but now $v$-contours for the same instants, at which $E^{-1/2}\vb_{\tGH}=0$ (equivalent to $E^{-1/2}\chib_{\tGH}$ maximised). ($a$), ($d$), ($g$) show $E^{-1/2}v_{\tDNS}$ (colour scale from $-300$ to $300$); ($b$), ($e$), ($h$) and ($c$), ($f$), ($i$) show  $v_{\tFNS}$ and $v_\tIW$ respectively (colour scale from $-5$ to $5$).
}
\label{fig3}
\end{figure}

\begin{figure}
\centerline{}
\vskip 3mm
\centerline{
\includegraphics*[width=1.0 \textwidth]{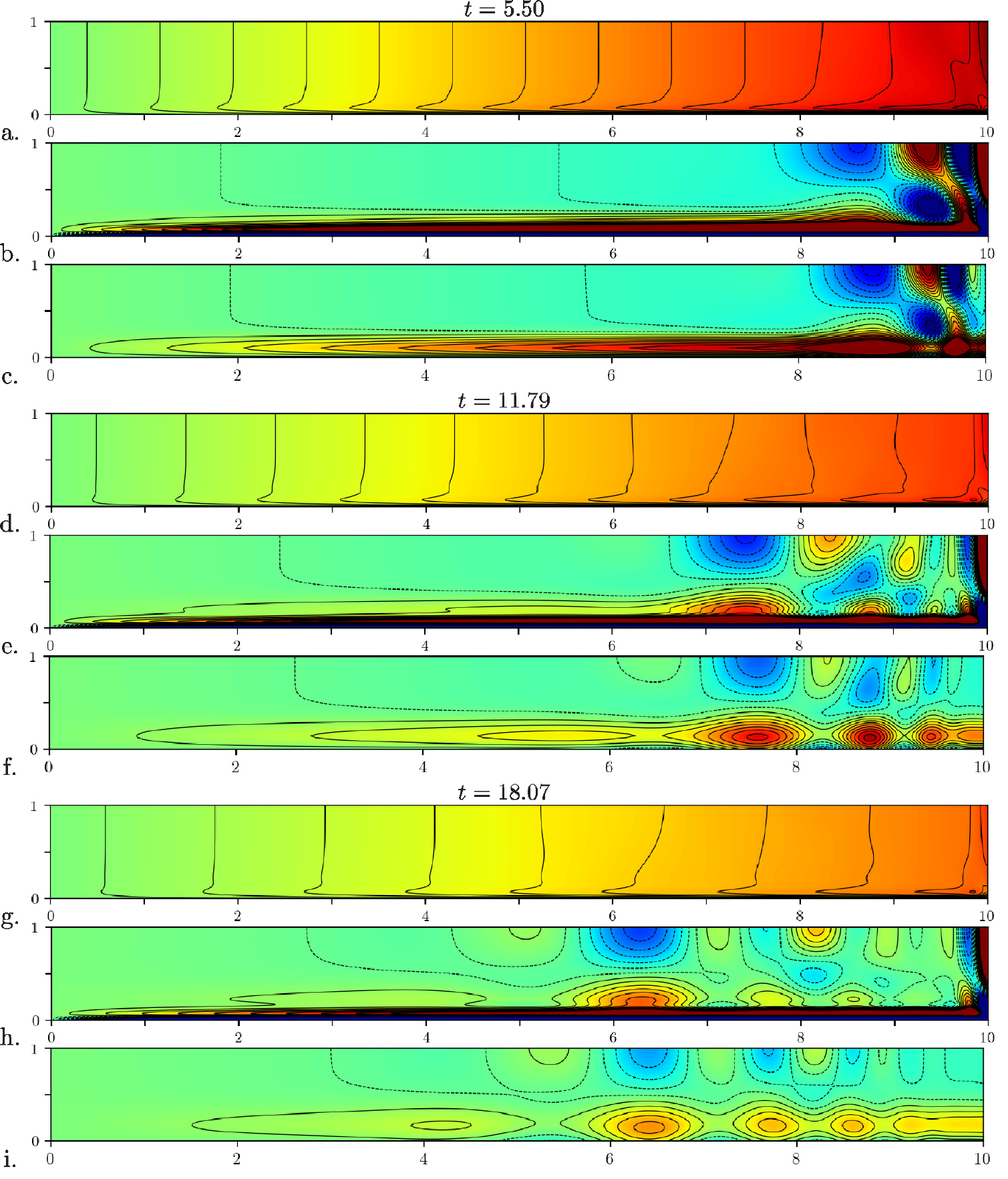}  
}
\caption{(Colour online) As in figure~\ref{fig2} but now $v$-contours for the same instants, at which $E^{-1/2}\vb_{\tGH}$ is maximised (equivalent to $E^{-1/2}\chib_{\tGH}=0$). Panel description as in figure~\ref{fig3}.
}
\label{fig4}
\end{figure}

All contributions to the radial flow $u$ are $O(E^{1/2})$. Nevertheless, just as for $v$, we need to first identify the QG-part $\ub_\tQG=\tfrac12\sigma E^{1/2}\vb_\tQG=\tfrac12(\sigma/\mu) E^{1/2}\langle v_{\tQG}\rangle$ (see (\ref{QG-spin-down}$a$)) and note that the IW-contribution, outside the Ekman layer, is
\bse
\label{iw-u}
\begin{align}
u_\tIW\,=\,E^{-1/2}u_{\tGH}\,+\,u^{\rm {wave}}\,=\,&\,E^{-1/2}(u\,-\,\ub_\tQG)\, +\,O(E^{1/2})\nonumber\\
\,=\,&\,E^{-1/2} u\,-\,\tfrac12(\sigma/\mu)\langle v \rangle\, +\,O(E^{1/2}) 
\end{align}
on the spin-down time $t=O\bigl(E^{-1/2}\bigr)$. Exactly as before in  our consideration of $\vb_\tQG$, we neglect the small quasi-steady $z$-dependent correction $O(E|\ub_\tQG|)$ to $\ub_\tQG$ \citep[see][eq.~(2.11$b$)]{OSD17} on the right-hand side of (\ref{iw-u}$a$). On defining the mainstream streamfunction $r\chi$ as $\chi=\int_z^1u\,\dR z$, we may extract the IW-part via the recipe
\be
\chi_\tIW\,=\,E^{-1/2}\chi_{\tGH} \,+\,\chi^{\rm {wave}}\,=\,E^{-1/2}\chi\,-\,\tfrac12(\sigma/\mu)(1-z)\langle v \rangle\, +\,O(E^{1/2})\,.
\ee
\ese
Guided by the results (\ref{iw-u}$a$,$b$), we define the radial FNS-velocity and streamfunction by
\bse
\label{filter-u}
\begin{align}
  u_{\tFNS}\,=\,&\,E^{-1/2}u_{\tDNS}\,-\,\tfrac12(\sigma/\mu)\langle v_{\tDNS} \rangle\,,\\
\chi_{\tFNS}\,=\,&\,E^{-1/2}\chi_{\tDNS}\,-\,\tfrac12(\sigma/\mu)(1-z)\langle v_{\tDNS} \rangle\,.
\end{align}
\ese
In figures~\ref{fig1},~\ref{fig2} (below), we portray $E^{-1/2}\chi_{\tDNS}$ in the DNS-panels ($a$), ($d$), ($g$), $\chi_{\tFNS}$ in the FNS-panels ($b$), ($e$), ($h$) and $\chi_{\tIW}$ in the IW-panels ($c$), ($f$), ($i$).

%%%%%%%%%%%%%%%%%%%%%%%%%%%%%%
%%%%%%%%%% FIGURE 1 %%%%%%%%%%
%%%%%%%%%%%%%%%%%%%%%%%%%%%%%%

%%%%%%%%%%%%%%%%%%%%%%%%%%%%%%
%%%%%%%%%%%%%%%%%%%%%%%%%%%%%%

%%%%%%%%%%%%%%%%%%%%%%%%%%%%%%
%%%%%%%%%% FIGURE 2 %%%%%%%%%%
%%%%%%%%%%%%%%%%%%%%%%%%%%%%%%

%%%%%%%%%%%%%%%%%%%%%%%%%%%%%%
%%%%%%%%%%%%%%%%%%%%%%%%%%%%%%

%%%%%%%%%%%%%%%%%%%%%%%%%%%%%%
%%%%%%%%%% FIGURE 3 %%%%%%%%%%
%%%%%%%%%%%%%%%%%%%%%%%%%%%%%%

%%%%%%%%%%%%%%%%%%%%%%%%%%%%%%
%%%%%%%%%%%%%%%%%%%%%%%%%%%%%%

%%%%%%%%%%%%%%%%%%%%%%%%%%%%%%
%%%%%%%%%% FIGURE 4 %%%%%%%%%%
%%%%%%%%%%%%%%%%%%%%%%%%%%%%%%

%%%%%%%%%%%%%%%%%%%%%%%%%%%%%%
%%%%%%%%%%%%%%%%%%%%%%%%%%%%%%

%%%%%%%%%%%%%%%%%%%%%%%%%%%%%%
%%%%%%%%%% FIGURE 5 %%%%%%%%%%
%%%%%%%%%%%%%%%%%%%%%%%%%%%%%%

\begin{figure}
\centerline{}
\vskip 3mm
\centerline{
\includegraphics*[width=1.0 \textwidth]{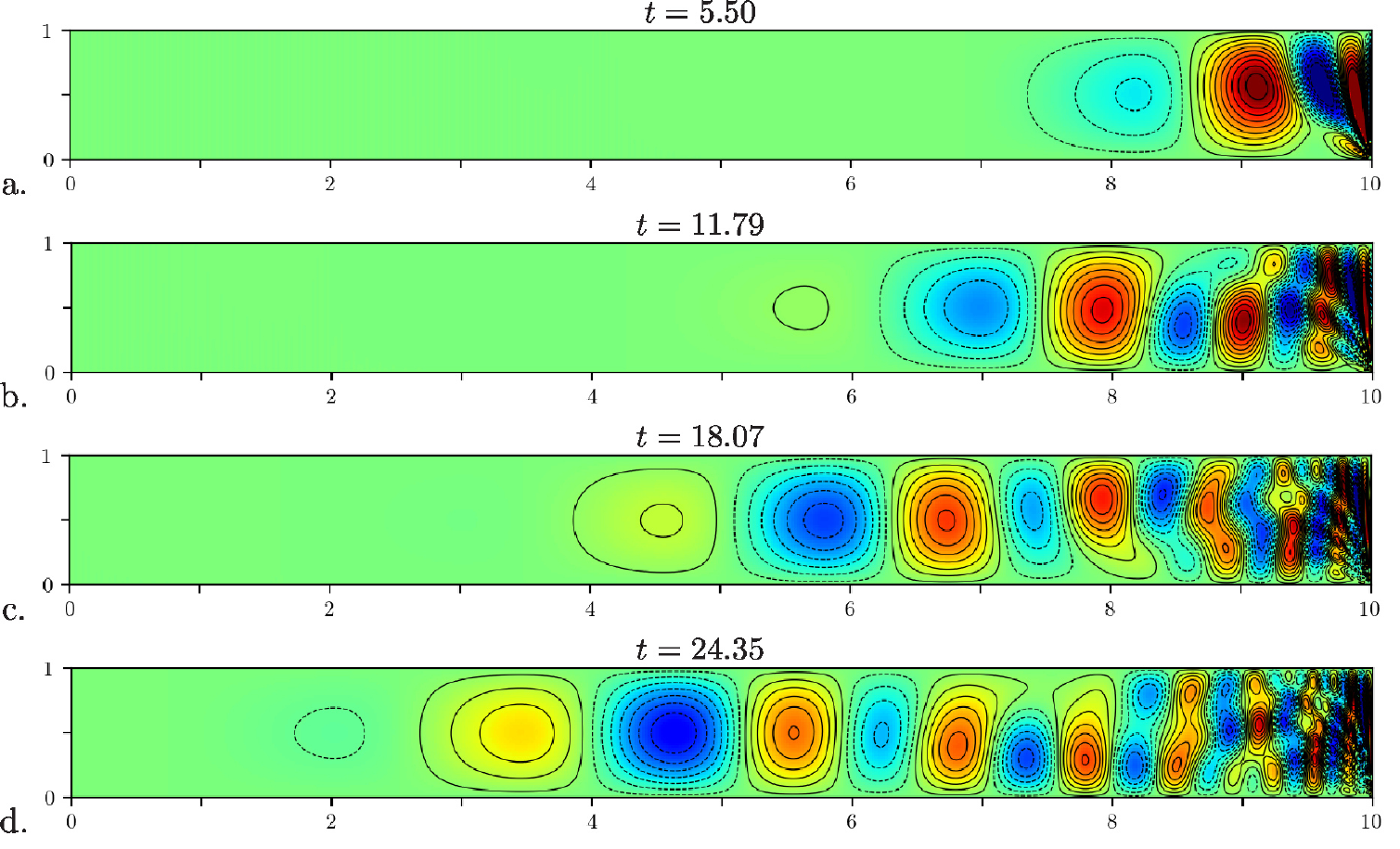}  
}
\caption{(Colour online) $\chi^{\rm{wave}}$-contours in the $E\downarrow 0$ limit at sequential times ${}$$\qquad t=(N+\tfrac12)\pi/2$. ($a$) $N=3$, ($b$) $N=7$, ($c$) $N=11$, cf., figures~\ref{fig2}($c$),($f$),($i$) respectively.\hfill\break ($d$) $N=15$.
}
\label{fig7}
\end{figure}

%%%%%%%%%%%%%%%%%%%%%%%%%%%%%%
%%%%%%%%%%%%%%%%%%%%%%%%%%%%%%

%%%%%%%%%%%%%%%%%%%%%%%%%%%%%%
%%%%%%%%%% FIGURE 6 %%%%%%%%%%
%%%%%%%%%%%%%%%%%%%%%%%%%%%%%%
\begin{figure}
\centerline{}
\vskip 3mm
\centerline{
\includegraphics*[width=1.0 \textwidth]{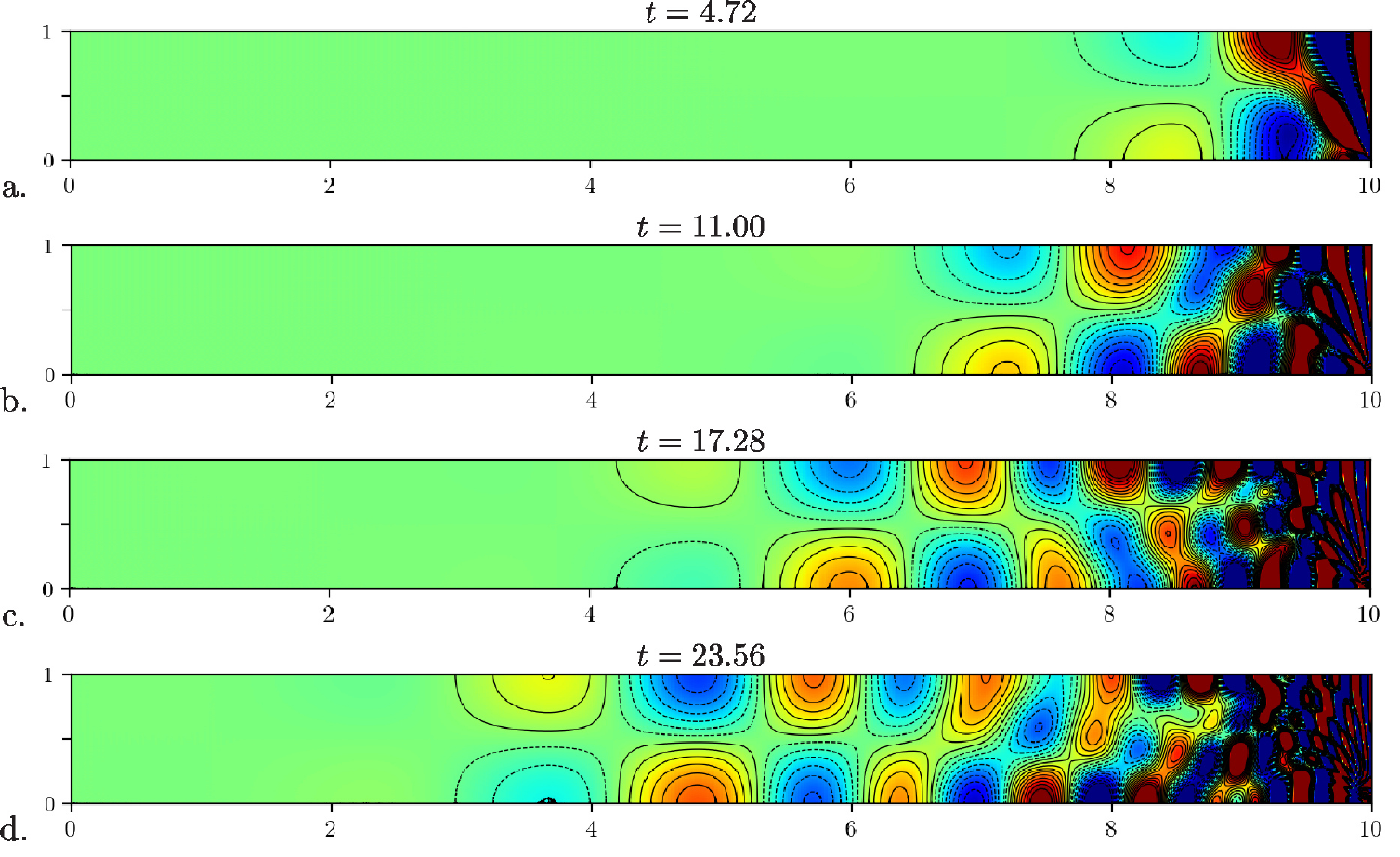}  
}
\caption{(Colour online) $v^{\rm{wave}}$-contours in the $E\downarrow 0$ limit at sequential times $t=N\pi/2$.\hfill\break
($a$) $N=3$, ($b$) $N=7$, ($c$) $N=11$, cf., figures~\ref{fig3}($c$),($f$),($i$) respectively. ($d$) $N=15$.
}
\label{fig8}
\end{figure}

%%%%%%%%%%%%%%%%%%%%%%%%%%%%%%
%%%%%%%%%%%%%%%%%%%%%%%%%%%%%%

\subsection{The inertial wave $\vv_\tIW$ comparison with $\vv_{\tFNS}$\label{iw-filter-comparison}}

In the FNS- and IW-panels of figures~\ref{fig1}--\ref{fig4}, contours are scaled consistently as in (\ref{iw-v})--(\ref{filter-u}) so that amplitude comparisons are readily discernible. The full DNS-results  however exhibit a wider  amplitude range, because they contain, in addition to the IW-contribution, the generally large QG-part. It is therefore impractical to employ the same scaling on the DNS-panels as used on the FNS- and IW-panels. The DNS-panels are, however, important as they illustrate the entire spin-down process and provide a visual measure of the relevance of the IW-contribution. This is particularly pertinent to $E^{-1/2}v_{\tDNS}$ which is $O\bigl(E^{-1/2}\bigr)$ larger than both $v_{\tFNS}$ and $v_\tIW$.

The results portrayed in figures~\ref{fig1}--\ref{fig4} all concern $E=10^{-3}$. The lower Ekman layer has width $E^{1/2}\doteqdot 0.03$, which is perhaps most readily identifiable in the azimuthal velocity $E^{-1/2}v$ contour plots of figures~\ref{fig3},~\ref{fig4}. The well known Ekman spiral is evident in the DNS-panels, whereas on the blown up scale of the FNS-panels it blurs and appears as thin black shaded layer. There is also a persistent ageostrophic $E^{1/3}$-sidewall layer at $r=\ell$ of width $0.1$. Our FNS- and IW-results are only meaningful in the regions exterior to those quasi-static boundary layers. Note that neither the Ekman layer nor the sidewall layer appear on the IW-panels as they are not part of either of the constituents  ($\chi_{\tGH}$, $v_{\tGH}$) or ($\chi^{\rm {wave}}$, $v^{\rm {wave}}$)  that together compose the IW-solution.

The time range of our plots starts at $t=4.72>1$ (i.e., large compared to the spin-down time) in figures~\ref{fig1},~\ref{fig3}, panels ($a$)--($c$) and ends at $t=18.07<10^3$ (i.e., short compared to the MF boundary layer (width $\Delta(t)=\sqrt{Et}$ (\ref{transient-EL-sol-z-int}$b$)) diffusion time~$E^{-1}$ needed to fill $0<z<1$) in figures~\ref{fig2},~\ref{fig4}, panels ($g$)--($i$). Essentially the results apply on the spin-down time $t_{sd}=E^{-1/2}\doteqdot 30$. The actual times chosen on figure~\ref{fig1}~(\ref{fig3}) are $t=N\pi/2$ ($N=3,\,7,\,11$) at which the MF-wave contribution $E^{-1/2}\chib_{\tGH}\propto \cos(2t)$ given by (\ref{MF-combined}$c$) is maximised (for $E^{-1/2}\vb_{\tGH}\propto \sin(2t)=0$, see (\ref{MF-combined}$a$)). The times $t=(N+\tfrac12)\pi/2$ ($N=3,\,7,\,11$) used on figure~\ref{fig2}~(\ref{fig4}) are when  $E^{-1/2}\chib_{\tGH}\propto \cos(2t)=0$ ($E^{-1/2}\vb_{\tGH}\propto \sin(2t)$ is maximised). The idea is that at times when $E^{-1/2}\chib_{\tGH}=0$ ($E^{-1/2}\vb_{\tGH}=0$), the FNS- and IW-panels for $E^{-1/2}\chi$ ($E^{-1/2}v$) on figure~\ref{fig2}~(\ref{fig3}) simply describe $\chi^{\rm {wave}}$ ($v^{\rm {wave}}$). However at times, when $E^{-1/2}\chib_{\tGH}$ ($E^{-1/2}\vb_{\tGH}$) are maximised, the FNS- and IW-panels for $E^{-1/2}\chi$ ($E^{-1/2}v$) on figure~\ref{fig1}~(\ref{fig4}), through comparison with figure~\ref{fig2}~(\ref{fig3}), identify the role of the $E^{-1/2}\chib_{\tGH}$ ($E^{-1/2}\vb_{\tGH}$) contribution. Perhaps the most striking characteristic of this comparison is that $E^{-1/2}\chib_{\tGH}$ ($E^{-1/2}\vb_{\tGH}$) identified in figure~\ref{fig1}~(\ref{fig4}) is non-zero throughout the entire domain just as predicted by (\ref{MF-combined}$c$ ($a$)). By contrast  $\chi^{\rm {wave}}$ ($v^{\rm {wave}}$) identified in figure~\ref{fig2}~(\ref{fig3}) is only non-zero for a limited radial extent from the outer boundary $r=\ell$. In the following \S\ref{No-damping} we ignore all damping and in \S\ref{IFm} explain this phenomenon. There is also much detailed structure in a subdomain close to $r=\ell$, which we explain in \S\ref{images}.

As explained in \S\ref{upp-MF}($b$) and noted earlier in this subsection in the context of time scales, the MF-wave possesses a spreading boundary layer width $\Delta(t)=\sqrt{Et}$ adjacent to the lower boundary $z=0$ quantified by (\ref{MF-combined}). This layer is most clearly evident in the IW-panels of figure~\ref{fig4} (sufficiently far to the left for $v^{\rm {wave}}$ to be negligible), for which the Ekman layer is absent. It is also evident in the FNS-panels, where it extends beyond the prominent Ekman layer. These features can also be identified, but less obviously, in the corresponding panels of figure~\ref{fig1} (sufficiently far to the left for $u^{\rm {wave}}$ to be negligible).

The values of $\chi^{\rm {wave}}$, $v^{\rm {wave}}$ used for our IW-plots are given by the $z$-Fourier series (\ref{FS}$a$,$b$) using $\chit_m$, $\vt_m$ determined by (\ref{solution}$a$,$b$). Since we found that the slow decay of the QG-flow has virtually no effect on the result, we set
\be
\label{Q=0}
Q\,=\,E^{1/2}\sigma\,=\,0
\ee
in (\ref{solution}$c$,$d$), which define the parameters $\CS^\EG_{mn}$, $\SS^\EG_{mn}$ (\ref{solution}$a$,$b$). As the formulae (\ref{solution}$c$,$d$) for the phase $\phi_{mn}(t)$ and decay rate $\lambda_{mn}$ account only for the damping by internal friction, we used instead (\ref{EL-solution-parameters}$a$,$b$) which also takes into account Ekman damping. As explained at the end of \S~\ref{Ekman-layer-damping}, we make the simplifying assumption $\epsilon^E_{mn}=0$, because we have no theory to predict its value. The slight change of phase linked to $\epsilon^E_{mn}$ is of lesser importance to us than the change of magnitude $d^E_{mn}$ of mode quenching and frequency increment $\omega^E_{mn}$, which both lead to secular changes. To assess whether or not our damping predictions are reasonable, we need to compare the FNS- IW-panels for $\chi_{\tFNS}$ and $\chi_\tIW$ on figure~\ref{fig2} (noting that $E^{-1/2}\chib_{\tGH}\approx 0$) and $v_{\tFNS}$ and $v_\tIW$ on figure~\ref{fig3} (noting that $E^{-1/2}\vb_{\tGH}\approx 0$). Our theoretical model, though generally good, appears to slightly over estimate damping on the shorter length scales. This appears to be a shortcoming of our choice of the QG-trigger $\ub_{\tQG}(\ell,t)=\tfrac12\sigma\kappa E^{1/2}\,\EG(t)$ (\ref{triggers}$a$). When we employ the MF-trigger $\ub_{\tGH}(\ell,t)=\tfrac12 E^{1/2}\,\WG_{\tGH}(t)$ (\ref{triggers}$b$) in our sequel Part~II, the comparison is much improved.

%%%%%%%%%%%%%%%%%%%%%%%%%%%%%%%%%%%%%
%%%%%%%%%%%%%%%%%%%%%%%%%%%%%%%%%%%%%
%%%%%       SECTION 4
%%%%%%%%%%%%%%%%%%%%%%%%%%%%%%%%%%%%%
%%%%%%%%%%%%%%%%%%%%%%%%%%%%%%%%%%%%%

\section{No damping $E\downarrow 0$\label{No-damping}}

With dissipation included the $z$-Fourier series representations (\ref{FS}) for $\chi$ and $v$ (the superscript `wave' is again dropped), possessing $r$-Fourier-Bessel series coefficients (\ref{solution}$a$,$b$) with parameter values (\ref{solution}$c$,$d$) and (\ref{EL-solution-parameters}$a$,$b$), determine results, which compare well with the DNS for $E=10^{-3}$, as figures~\ref{fig1}--\ref{fig4} in \S\ref{numerics} illustrate. Nevertheless, at that moderately small $E$, motion on small scales suffers considerable dissipation and decays rapidly. As the full DNS at smaller $E$ is computationally intensive, we rely entirely on our analytic results, which we now describe and analyse in the limit  $E\downarrow 0$. By this device, some detailed structures visible in figures~\ref{fig1}--\ref{fig4} are identified more clearly. 

\subsection{Formulation\label{Formulation}}

On setting $E=0$ in (\ref{gov-eqs}$a$--$d$), these governing equations together with the initial conditions (\ref{initial-condits}$a$,$b$) and boundary condition $\chi=0$ at $z=1$ (\ref{boundary-condits}$c$) determine
\bme
\label{gov-eqs-ints}
\be\te
\chi\,=\int_z^1 u\,\dR z\,, \hskip 8mm v\,=\,-\,2\!\int_0^t u\,\dR t\,\,=\,2\!\int_0^t\pd{\chi}{z}\,\dR t\,,\hskip 8mm  \gamma\,=\,2\!\int_0^t\pd{v}{z}\,\dR t\,.
\ee
\eme
We continue to consider the representations (\ref{FS}) and (\ref{solution}$a$,$b$), in which $\FG_{mn}=q_{mn}^2\omega_{mn}^2/2$, $\HG_{mn}=q_{mn}\omega_{mn}$ (\ref{Fourier-Bessel-series}$b$,$c$) and $\phi_{mn}=\omega_{mn}t$ (\ref{solution}$e$), but set $E^{1/2}\sigma=Q=0$, $\kappa=\sigma=1$, $d_{mn}=0$ so that the coefficients become $\CS^\EG_{mn}=1$, $\SS^\EG_{mn}=0$ (see (\ref{solution}$c$,$d$)) and $\lambda_{mn}=0$ (see (\ref{solution}$f$)). In this way, (\ref{solution}$a$,$b$) yield
\bse
\label{sol-tilde}
\begin{align}
\chit_m\,=\,\chit^\daleth_m\,=\,&\,-\,\sum_{n=1}^\infty\dfrac{q_{mn}^2\omega_{mn}^2}{2}\,\dfrac{\JR_1(j_nr/\ell)}{j_n\JR_0(j_n)}\,\cos(\omega_{mn}t)\,,\\
\left[\begin{array}{c}  \!\! \ut_m \!\!\\[0.2em]
      \!\!  \vt_m \!\!  \end{array}\right]
=\,\left[\begin{array}{c}  \!\! \ut^\daleth_m \!\!\\[0.2em]
      \!\!  \vt^\daleth_m \!\!  \end{array}\right]
=\,&\,\sum_{n=1}^\infty\dfrac{q_{mn}\omega_{mn}^2}{2}\,\dfrac{\JR_1(j_nr/\ell)}{\ell\JR_0(j_n)}
\left[\begin{array}{c} \!\!\!\cos(\omega_{mn}t) \!\!\! \\[0.3em]
    \!\!\! -(2/\omega_{mn})\sin(\omega_{mn}t) \!\!\! \end{array}\right].
\end{align}
\ese

Some solutions $\chi^{\rm{wave}}=\chi$ and $v^{\rm{wave}}=v$, realised by substitution of (\ref{sol-tilde}) into (\ref{FS}), are illustrated in figures~\ref{fig7}~and~\ref{fig8} respectively. The times $t= 5.50$, $11.79$, $18.07$ adopted on the first three panels ($a$)-($c$) of figures~\ref{fig7} correspond to the prescription $t=(N+\tfrac12)\pi/2$ ($N=3,\,7,\,11$)  adopted in figure~\ref{fig2}, for which $E^{-1/2}\chib_{\tGH}= 0$. By this choice, we see how the small scale structure of $\chi^{\rm{wave}}$ visible in figures~\ref{fig7}($a$)-($c$) particularly near the outer boundary $r=\ell$  is largely eliminated by dissipation in the contour plots of $\chi_{\tIW}$ in figures~\ref{fig2}($c$),($f$),($i$). Likewise, at times $t=4.72$, $11.00$, $17.28$, i.e., $t=N\pi/2$ ($N=3,\,7,\,11$)  a similar comparison of figures~\ref{fig8}($a$)-($c$) with figures~\ref{fig3}($c$),($f$),($i$), for which $E^{-1/2}\vb_{\tGH}= 0$, can be made. Results for the individual Fourier modes
\be
\label{Fourier-m-modes}
\chit_m^{\,\rm {wave}}(r,z,t)\,=\,-\,(m\pi)^{-1}\chit_m\sin(m\pi z)
\ee
(see  (\ref{FS}$a$)) with  $m=1,\,2$ are illustrated in figures~\ref{fig5}~and~\ref{fig6} below.

\subsection{The Cartesian limit, $\ell\gg 1$, $\ell-r=O(1)$\label{Cartesian-limit}}

The wave-solutions are best understood by their behaviour at large $r$. So throughout this section and the following \S\S\ref{images},~\ref{IFm} we restrict attention to
\be 
\label{l-large}
x\,=\,\ell\,-\,r\,=\,O(1)  \hskip15mm   (\ell\gg 1)\,,
\ee
for which two key approximations follow:
\bse
\label{Bessel-large-arg}
\begin{align}
  \dfrac{\JR_1(j_nr/\ell)}{\JR_0(j_n)}\,\approx &\,-\,\sin (n\pi x/\ell)\,,\qquad j_n\,\approx\, n\pi \qquad\mbox{for} \qquad n\gg 1\,,\\
  \dfrac{\IR_1(m\pi r)}{\IR_1(m\pi \ell)}\,\approx &\,\exp(-m\pi x)\,.
\end{align}
\ese
Henceforth we will adopt $x$ rather than $r$ as our independent variable, but it must be remembered that $x$ measures distance in the opposite direction to $r$ (see (\ref{l-large})).

The essential idea is that for $x=O(1)$, the $r=0$ axis is unimportant. So, with $\ell\gg 1$, we may regard $n/\ell$ as a continuous rather than discrete variable and approximate the sum $\sum_{n=1}^\infty\bullet_{n}$ in (\ref{sol-tilde}) by the integral $\int_{n=0}^\infty\bullet_{n}\dR n$ instead. In this way, from (\ref{Fourier-Bessel-series}$d$,$e$) and (\ref{Bessel-large-arg}$a$), we obtain
\bme
\label{continuous-variables}
\begin{align}  
n\,\approx\,j_n/\pi\,=\,&\,(\ell/\pi)k\,,   &   \dR n\,\approx\,&\,(\ell/\pi)\,\dR k\,,\\
q\,=\,q_{mn}\,=\,&\,k/(m\pi)\,,   &   \omega\,=\, \omega_{mn}\,=\,&\,2m\pi\big/\sqrt{k^2+(m\pi)^2}\,. 
\end{align}
\eme
Accordingly, (\ref{sol-tilde}$b$) is approximated by
\bme
\label{sol-tilde-int-u-v}
\be
\left[\begin{array}{c}  \!\! \ut_m \!\!\\[0.2em]
      \!\!  \vt_m \!\!  \end{array}\right]
\approx\,-\,\dfrac{2}{\pi}\int_0^\infty\dfrac{\omega^2 k \sin(k x)}{4m\pi}
\left[\begin{array}{c} \!\!\!\cos(\omega t) \!\!\! \\[0.3em]
    \!\!\! -(2/\omega)\sin(\omega t) \!\!\! \end{array}\right]\,\dR k\,, \hskip 10mm \chit_m\,=\,-\,\dfrac{\ut_m}{m\pi}\,.
\ee
\eme
On noting that $\LC_p\{\exp(\iR \omega t)\}=(p+\iR\omega)\big/(p^2+\omega^2)$, the Fourier sums (\ref{FS}$b$) for $u$ and $v$, based on (\ref{sol-tilde-int-u-v}), have Laplace transforms
\bme
\label{sol-tilde-int-LT-prelim}
\be\se
\left[\begin{array}{c}  \!\!\uh  \!\!\\[0.2em]
    \!\! \vh \!\!  \end{array}\right]
\,\approx\,\dfrac{2}{\pi p}\left[\begin{array}{c}  \!\! 1  \!\!\\[0.2em] \!\! -2/p  \!\!  \end{array}\right]
\sum_{m=1}^\infty\,\biggl[\int_0^\infty\dfrac{k\sin(k x)}{k^2+(m\pi s)^2}\,\dR k\biggr]\, \cos(m\pi z)\,,
\ee
where, on setting $E=0$ in (\ref{LT-se-chi-sol-disp-rel}) to obtain
\be
s\,=\,-\,\iR q \,=\,(p^2+4)^{1/2}\big/p  \hskip 10mm \Longleftrightarrow \hskip 10mm {4}/{p^2}\,=\,s^2-1\,,
\ee
\eme
we have noted  from (\ref{continuous-variables}$d$) and (\ref{sol-tilde-int-LT-prelim}$c$) that ${4}/{\omega^2}+{4}/{p^2}= k^2/(m\pi)^2+s^2$. Evaluation of the integral in (\ref{sol-tilde-int-LT-prelim}$a$) \citep[use \S2.2 eq.~(15) of][]{EMOT54I} yields
\bme
\label{sol-tilde-int-LT}
\se
\begin{align}
\!
\left[\begin{array}{c}  \!\!\uh  \!\!\\[0.2em]
    \!\! \vh \!\!  \end{array}\right]
=\,&\,\dfrac{1}{p}\left[\begin{array}{c}  \!\! 1  \!\!\\[0.2em]
    \!\! -2/p  \!\!  \end{array}\right]\sum_{m=1}^\infty\,\exp(-m\pi s\,x)\, \cos(m\pi z)\,\\
\,=\,&\,\,\dfrac{1}{2p}\left[\begin{array}{c}  \!\! 1  \!\!\\[0.2em]
    \!\! -2/p  \!\!  \end{array}\right]
\biggl[- \,1 \,+ \,\dfrac{\sinh(s\pi x)}{\cosh(s\pi x)\,-\,\cos(\pi z)}\biggr]\,.  \hskip 0mm
\end{align}
Application of the formula $\chih=\int_z^1 \uh\,\dR z$  determines
\be\de
\chih\,\approx\,-\dfrac{1}{\pi p}\biggl[-\,\dfrac{\pi z}{2}\,+\,\tan^{-1}\biggl(\dfrac{\tan(\pi z/2)}{\tanh(s \pi x/2)}\biggr)\biggr], \hskip 15mm
\gammah\,\approx\,\dfrac{2}{p}\pd{\vh}{z}\,. \hskip 8mm
\ee
\eme

In order to invert the Laplace transforms, we need to note that $s=(p^2+4)^{1/2}/p\to 1$ as $|p|\to \infty$, i.e., $s$ is defined by a cut connecting $p=-2\iR$ to $p=2\iR$ along the $\Im\{p\}$-axis and by analytic continuation elsewhere. This consideration is essential to guarantee that we take the correct sign of the square root of $(p^2+4)^{1/2}$. Indeed this property may be used to extract the initial values of $\chi$ and $u$, which are determined by the form of $\uh$ and $\chih$ in the limit $|p|\to \infty$. In that limit, evaluation of the inverse-LTs is achieved by simply setting $s=1$ and then evaluating the residues of (\ref{sol-tilde-int-LT}$b$,$c$) at the only remaining singularity, the pole at $p=0$ ($s=1$), so determining
\bse
\label{u-chi-initial}
\begin{align}
 u(x,z,0)\,\approx\,& \, \dfrac{1}{2}\biggl[-1\,+\,\dfrac{\sinh(\pi x)}{\cosh(\pi x) -\cos (\pi z)}\biggr]\,,\\
\chi(x,z,0)\,\approx\,& -\,\dfrac{1}{\pi}\,\biggl[-\,\dfrac{\pi z}{2}\,+\,\tan^{-1}\biggl(\dfrac{\tan(\pi z/2)}{\tanh(\pi x/2)}\biggr)\biggr]\,.
\end{align}
\ese
Initially $v$ and $\gamma$ are zero, but for $t\ll 1$ (\ref{gov-eqs-ints}$b$,$c$) determine
\bme
\label{v-gamma-small-t}
\be
v\,\approx\,-2t u(x,z,0), \hskip 15mm  \gamma\,\approx\,-2t^2 \pd{u}{z}(x,z,0)\,.
\ee
\eme

Despite the apparent simplicity of the Laplace transforms $\uh$, $\vh$ and $\chih$ given by (\ref{sol-tilde-int-LT}$b$,$c$), their direct inversion is not straightforward. That is partly due to the essential singularity of $\tanh(s\pi x)$ at $p=0$ which leads to some apparently suspicious results following LT-inversion. For example, the form (\ref{sol-tilde-int-LT}) hints at a pole at $p=0$, where none exists in the primitive form (\ref{sol-tilde-int-LT-prelim}$a$) (recall that $p^2s^2\to 4$ as $p\to 0$). An alternative approach is suggested by the formula
\be
\label{Poisson-sum}
-1\,+\,\dfrac{\sinh(s\pi x)}{\cosh(s\pi x) -\cos (\pi z)}\,=-1\,+\,\dfrac{1}{\pi}\sum_{\lsum=-\infty}^\infty\dfrac{2s x}{(z-2\lsum)^2+s^2x^2}
\ee
\citep[][\S1.445, eq.~(9)]{GR07}, which we substitute into (\ref{sol-tilde-int-LT}$b$). Due to the invariance of the sum (\ref{Poisson-sum}) under the shift $z\mapsto z+2$, there is only one independent solution $[\ubr,\vbr]$ linked to $\lsum=0$. We refer to the others, $[\ubr_\lsum,\vbr_\lsum]$ for $\lsum\not=0$, as the ``image system''.

The various LT-representations suggest two distinct strategies for their inversion. In \S\ref{images}, we adopt the ``method of images'', based on (\ref{sol-tilde-int-LT}$b$) and (\ref{Poisson-sum}), to explain detailed features of the solution particularly evident at small $x$. In \S\ref{IFm} we study the evolution of the individual  $z$-Fourier $m$-modes (\ref{sol-tilde-int-u-v}). The smallest, $m=1$, identifies the dominant structure at large~$x$.

%%%%%%%%%%%%%%%%%%%%%%%%%%%%%%%%%%%%%
%%%%%%%%%%%%%%%%%%%%%%%%%%%%%%%%%%%%%
%%%%%       SECTION 5
%%%%%%%%%%%%%%%%%%%%%%%%%%%%%%%%%%%%%
%%%%%%%%%%%%%%%%%%%%%%%%%%%%%%%%%%%%%

\section{$E\downarrow 0$:\, The ``method of images''\label{images}}

The inverse-LT of (\ref{sol-tilde-int-LT}$b$) with (\ref{Poisson-sum}) takes the form
\bme
\label{Poisson-sum-representation}
\be
\left[\begin{array}{c}  \!\! u  \!\!\\[0.2em]
    \!\! v \!\!  \end{array}\right]
\,\approx\,\left[\begin{array}{c}  \!\! -\tfrac12  \!\!\\[0.2em]
    \!\! t \!\!  \end{array}\right]\,+\,
\sum_{\lsum=-\infty}^\infty\left[\begin{array}{c}  \!\! \ubr_\lsum  \!\!\\[0.2em]
    \!\! \vbr_\lsum \!\!  \end{array}\right],  \hskip10mm
\left[\begin{array}{c}  \!\!\ubr_\lsum  \!\!\\[0.2em]
    \!\! \vbr_\lsum \!\!  \end{array}\right](x,z,t)\,
=\,\left[\begin{array}{c}  \!\!\ubr  \!\!\\[0.2em]
    \!\! \vbr \!\!  \end{array}\right](x,z-2\lsum,t)\,.\hskip4mm
\ee
\eme
In \S\ref{elementarysolution} we consider only the primary $\lsum=0$ mode $[\,\ubr\,,\,\vbr\,](x,z,t)$, which describes motion throughout the half-plane $x>0$, $-\infty<z<\infty$ due to a sink at $(x,z)={\bf 0}$, or more precisely $\ubr(0,z,t)=\delta(z)$, where $\delta(z)$ is the Dirac $\delta$-function. In \S\ref{fullsolution}, we compose the complete solution $[u,\,v](x,z,t)$, defined by (\ref{Poisson-sum-representation}$a$) formed upon superimposing the flows due to the image sinks at $(x,z)=(0,\pm 2\lsum)$, whose net outflow is compensated by the additional uniform flow contribution $u=-\tfrac12$. In turn, the corresponding contribution $v=t$ follows from (\ref{gov-eqs-ints}$b$).

\subsection{The primary $\lsum=0$ mode \label{elementarysolution}}

According to (\ref{sol-tilde-int-LT}$b$) and (\ref{Poisson-sum}), the Laplace transform of the primary mode is
\bme
\label{Poisson-sum-LT}
\be\se
\left[\begin{array}{c}  \! \ubrh  \!\\[0.2em]
    \! \vbrh \!  \end{array}\right](x,z,p)
=\,\dfrac{1}{\pi\varpi}\dfrac{(p^2+4)^{1/2}\xS}{p^2+4\xS^2}\left[\begin{array}{c}  \!\! 1  \!\!\\[0.2em]
    \!\! -2/p  \!\!  \end{array}\right]
\ee
in which we have introduced the unit vector
\be
[\xS\,,\,\zS]\,=\,[x\,,\,z]/\varpi\,, \hskip 20mm \varpi\,=\,\sqrt{x^2+z^2}\,.
\ee
\eme

In view of our remarks in the penultimate paragraph of \S\ref{Cartesian-limit}, the pole at $p=0$ determines an unexpected steady geostrophic flow $[\ubr_\tG,\,\vbr_\tG]$:
\bme
\label{G}
\be
\ubr_\tG\,=\,0\,, \hskip 20mm       \vbr_\tG\,=\,-\,(\pi x)^{-1}\,.
\ee
\eme
When, however, we consider the full solution in the following \S\ref{fullsolution},  we see that this unwelcome contribution is eliminated under accumulation with the image flows. Indeed the entire flow evolves indefinitely with no identifiable non-oscillatory part.

The inverse-LT of (\ref{Poisson-sum-LT}$a$) at $z=0$ (i.e. $\xS=1$) is
\bme
\label{Poisson-sum-sol-small-t-zero-z}
\be
\ubr(x,0,t)\,=\,\dfrac{1}{\pi x}\JR_0(2t)\,, \hskip 10mm  \vbr(x,0,t)\,-\,\vbr_\tG(x)
\,=\,\dfrac{2}{\pi x}\int_t^\infty\JR_0(2\tau)\dR \tau\,.
\ee
\eme
Elsewhere (indeed $\forall z$) it is
\bme
\label{Poisson-sum-sol-small-t}
\be\se
\left[\begin{array}{c}  \!\!\ubr  \!\!\\[0.2em]
    \!\! \vbr \!\!  \end{array}\right]=
\,\dfrac{1}{\pi\varpi}\biggl\{\left[\begin{array}{c}  \!\!\xS\cos(2\xS t)  \!\!\\[0.2em]
    \!\! -\sin(2\xS t) \!\!  \end{array}\right]+\,\int_0^{t}\dfrac{\JR_1(2\tau)}{\xS\tau}\,
\left[\begin{array}{c}  \!\!\xS\ES_i\bigl(2\xS(t-\tau)\bigr)  \!\!\\[0.2em]
    \!\! \ES_r\bigl(2\xS(t-\tau)\bigr)\!\!  \end{array}\right] \,\dR\tau\biggr\}
\ee
in which
\be
\ES(\varphi)\,=\,-1\,+\,\exp(\iR \varphi)\,,\hskip 10mm
\left[\begin{array}{c}  \!\!\ES_i(\varphi)  \!\!\\[0.2em]
    \!\! \ES_r(\varphi) \!\!  \end{array}\right]=
\left[\begin{array}{c}  \!\! \sin\varphi \!\!\\[0.2em]
    \!\! -1\,+\,\cos\varphi \!\!  \end{array}\right].
\ee
\eme
On use of  $\LC_p\{t^{-1}\JR_1(2t)\}=2\big/\bigl[(p^2+4)^{1/2}+p\bigr]$ \citep[see \S4.14 eq.~(5) of][]{EMOT54I}, it is readily verified that the Laplace transform of (\ref{Poisson-sum-sol-small-t}$a$) is (\ref{Poisson-sum-LT}$a$). In view of the unlikely relevance of $2\big/\bigl[(p^2+4)^{1/2}+p\bigr]$ to (\ref{Poisson-sum-LT}$a$), the direct derivation (without hindsight) of (\ref{Poisson-sum-sol-small-t}$a$) was not obvious to us. The primitive form (\ref{Poisson-sum-sol-small-t}$a$) is useful for $t=O(1)$. However, the identity $\int_0^\infty\tau^{-1}\JR_1(2\tau)\ES(2\xS\tau)\dR\tau=-1+|\zS|+\iR\xS$ \citep[use \S1.12 eq.~(4) and \S2.12 eq.~(5) of][]{EMOT54I} permits the alternative representation
\bse
\label{Poisson-sum-sol-large-t}
\be
[\,\ubr\, ,\, \vbr\,  ]\,=\, [\,\ubr\, ,\, \vbr\,]_{ms}\,+\,[\,\ubr\, ,\, \vbr\,  ]_{bl}\,,
\ee
useful for $t\gg 1$, where
\begin{align}
  \left[\begin{array}{c}  \!\!\ubr  \!\!\\[0.2em]
      \!\! \vbr\!\!  \end{array}\right]_{ms}
\,-\,\left[\begin{array}{c}  \!\! 0  \!\!\\[0.2em]
      \!\! \vbr_\tG\!\!  \end{array}\right]
=\,&\,
\dfrac{|\zS|}{\pi x}\left[\begin{array}{c}  \!\!\xS\sin(2\xS t)  \!\!\\[0.2em]
    \!\!  \cos(2\xS t) \!\!  \end{array}\right]
=\,\dfrac{|\zS|}{\pi x}\left[\begin{array}{c}  \!\!\xS \ES_i(2\xS t)  \!\!\\[0.2em]
    \!\! 1+ \ES_r(2\xS t) \!\!  \end{array}\right],
\\[0.5em]
\left[\begin{array}{c}  \!\!\ubr \!\!\\[0.2em]
\!\! \vbr\!\!  \end{array}\right]_{bl}
=\,&\,
-\,\dfrac{1}{\pi x}\int_{t}^\infty\dfrac{\JR_1(2\tau)}{\tau}\,\left[\begin{array}{c}  \!\!\xS\ES_i\bigl(2\xS(t-\tau)\bigr)  \!\!\\[0.2em]
    \!\! \ES_r\bigl(2\xS(t-\tau)\bigr)\!\!  \end{array}\right]
\,\dR\tau\,.
\end{align}
\ese

At late time, the primary pole-contribution ($p=\pm \iR 2\xS$ of (\ref{Poisson-sum-LT})), $[\ubr,\vbr]_{ms}$, defines mainstream motion, while the secondary cut-contribution (cut points $p=\pm 2\iR$), $[\ubr,\vbr]_{bl}$, identifies an ever thinning boundary layer, width $\Delta_{bl}=xt^{-1/2}$ (see appendix \ref{esoteric}), which slowly evaporates. Though both  $[\ubr,\vbr]_{ms}-[0,\vbr_\tG]$ and $[\ubr,\vbr]_{bl}$ are of comparable size within the boundary layer, they are both relatively small near $z=0$ and so dominated by the image system when superimposed. The detailed character of $[\ubr,\vbr]_{bl}$ is interesting but, being technical, is relegated to appendix~\ref{esoteric}. It suffices to remark here that the factor $\tau^{-1}$ in the integrand of (\ref{Poisson-sum-sol-large-t}$c$) provides the key to obtaining our tight large-$t$ estimates on the size of  $[\ubr,\vbr]_{bl}$, which in turn confirm (see our following \S\ref{fullsolution}) that (\ref{Poisson-sum-sol-large-t}$a$-$c$) is indeed an appropriate and useful {\itshape ms}-{\itshape bl} decomposition.

\subsection{The full solution\label{fullsolution}}

A dominant feature of $[\ubr,\vbr]_{ms}$ (\ref{Poisson-sum-sol-large-t}$b$) is the lines $2\xS t$ of constant phase linked to $\exp(\iR 2\xS t)$ that emanate from the corner $(x,z)={\bf 0}$. The nodes $\xS_n=n\pi/(2t)$ $(n=1,2,\cdots)$ for $\ubr$ lie on a fan that contracts with time as illustrated on figures~\ref{fig7},~\ref{fig8}. Also visible are the waves reflected at $z=1$. They correspond to the $\lsum=1$ image fan  emanating from the image sink $(x,z)=(0,2)$ and are particularly evident in panels~($a$)--($c$). Owing to the intensity of the reflections including further interference from other images, the last panel~($d$) (longest time) is ``busy'' and a little confused.

Though much of what is visible in figures~\ref{fig7},~\ref{fig8} may be understood in terms of the primary mode $[\ubr,\vbr]_{ms}$, the complete description, at least within the asymptotic approximations (\ref{continuous-variables}), (\ref{sol-tilde-int-u-v}) for $x=O(1)$ and $\ell\gg 1$, is given by the sum
%solution
(\ref{Poisson-sum-representation}$a$). As already remarked $[\ubr,\vbr]_{bl}$ is small for $t\gg 1$. So we omit its contribution to the sum (\ref{Poisson-sum-representation}$a$) and define what remains,
\be 
\label{ms-m}
\left[\begin{array}{c}  \!\! u  \!\!\\[0.2em]
    \!\! v \!\!  \end{array}\right]_{ms}
\,=\,\left[\begin{array}{c}  \!\! -\tfrac12  \!\!\\[0.2em]
    \!\! t \!\!  \end{array}\right]\,+\,
\sum_{\lsum=-\infty}^\infty\left[\begin{array}{c}  \!\! \ubr_\lsum  \!\!\\[0.2em]
    \!\! \vbr_\lsum \!\!  \end{array}\right]_{ms},  
\ee
as the mainstream solution.

A disconcerting feature of (\ref{ms-m}) is the presence of the divergent contribution $t$ to $v_{ms}$. To test the worth of the approximation (\ref{ms-m}), which ignores the $[\ubr_\lsum,\vbr_\lsum]_{bl}$ contributions, we consider the $z$-average of that mainstream solution. Since $[\ubr,\vbr](x,z,t)$ is symmetric in $z$, we note that $[\langle\ubr\rangle,\langle\vbr\rangle](x,t)=\tfrac12\int_{-1}^1[\ubr,\vbr]\dR z$ with the implication $[\langle\ubr_\lsum\rangle,\langle\vbr_\lsum\rangle](x,t)=\tfrac12\int_{-1-2\lsum}^{1+2\lsum}[\ubr,\vbr]\dR z$. This property permits us to express the integral of the infinite sum in (\ref{ms-m}) as a single infinite integral:
\bse
\label{mean-values-large-t}
\begin{align}
\left[\begin{array}{c}  \!\!\langle u_{ms}\rangle\!\!\\[0.2em]
      \!\! \langle v_{ms}\rangle\!\!  \end{array}\right]
=\,&\left[\begin{array}{c}  \!\!-\tfrac12\!\! \\[0.2em]
    \!\!t\!\! \end{array}\right]+\dfrac{1}{2\pi x}
\int_{-\infty}^\infty\left[\begin{array}{c}  \!\!\xS|\zS|\ES_i(2\xS t)  \!\!\\[0.2em]
\!\!-1 +|\zS|+|\zS| \ES_r(2\xS t) \!\!  \end{array}\right]
\dR z\,,\nonumber\\
=\,&\left[\begin{array}{c}  \!\!-\tfrac12\!\! \\[0.2em]
    \!\!t\!\! \end{array}\right]+\dfrac{1}{2\pi}
\int_{-1}^1\left[\begin{array}{c}  \!\!\xS\ES_i(2\xS t)  \!\!\\[0.2em]
\!\!-|\zS|^{-1}+1+ \ES_r(2\xS t) \!\!  \end{array}\right]
\dfrac{\dR \xS}{\xS^2}\,,
\end{align}
in which we have used (\ref{Poisson-sum-sol-large-t}$b$) with $\vbr_\tG=-\,(\pi x)^{-1}$ and noted that $x^{-1}\dR z=-\xS^{-2}|\zS|^{-1} \dR \xS$.
After manipulations, that use  $\int_0^\infty\bigl[\phi^{-1}\sin\phi,\,\phi^{-2}(1-\cos\phi)\bigr]\,\dR \phi=[\pi/2,\pi/2]$, (\ref{mean-values-large-t}$a$) may be recast as
\be
\left[\begin{array}{c}  \!\!\langle u_{ms}\rangle\!\!\\[0.2em]
      \!\! \langle v_{ms}\rangle\!\!  \end{array}\right]
=\,-\,\dfrac{1}{\pi}
\int_{2t}^\infty\left[\begin{array}{c} \!\!\!\phi^{-1}\sin\phi\!\!\! \\[0.2em]
    \!\!\!2t\phi^{-2}\cos\phi\!\!\! \end{array}\right]\dR \phi\,.
\ee
\ese
For $t\gg 1$ it behaves like
\be
\label{mean-values-large-t-more}
[\,\langle u_{ms}\rangle\, ,\, \langle v_{ms}\rangle\,  ]\,\approx\,(2\pi t)^{-1}[\,-\,\cos(2t)\, ,\,\sin(2t) \,  ]\,+\,O\bigl(t^{-2}\bigr).
\ee
Reassuringly, the diverging contribution $t$ to $v$ in (\ref{mean-values-large-t}$a$) is eliminated by the summation and its mean value (\ref{mean-values-large-t-more}), being $O(t^{-1})$, decays. By implication, since $[\langle u\rangle,\langle v\rangle]={\bf 0}$, the $z$-average of the remaining  $\left[u,v\right]_{bl}=\sum_{\lsum=-\infty}^\infty\left[\ubr_\lsum,\vbr_\lsum\right]_{bl}$, namely
\be
\label{mean-values-large-t-yet-more}
[\,\langle u_{bl}\rangle\, ,\, \langle v_{bl}\rangle\,  ]\,=\,-\,[\,\langle u_{ms}\rangle\, ,\, \langle v_{ms}\rangle\,  ]\,,
\ee
decays at the same rate. Interestingly, the results (\ref{mean-values-large-t-more}),  (\ref{mean-values-large-t-yet-more}) and (\ref{bl-mean-values-large-t}$c$) show, correct to leading order, that $[\langle u_{bl}\rangle,\langle v_{bl}\rangle]\approx [\langle \ubr_{bl}\rangle,\langle \vbr_{bl}\rangle]$, i.e., only the primary $\lsum=0$ mode of the infinite sum contributes. The result reinforces the view that, for $t\gg 1$, we may ignore entirely the $\lsum\not=0$ contributions $\left[\ubr_\lsum,\vbr_\lsum\right]_{bl}$ and make the approximation $\left[u,v\right]_{bl}\approx\left[\ubr,\vbr\right]_{bl}$.

To estimate whether or not $\left[\ubr,\vbr\right]_{bl}$ is significant near $z=0$, we consider briefly the value of $[u,v]_{ms}$ there. Substitution of (\ref{Poisson-sum-sol-large-t}$b$) into (\ref{ms-m}), and noting that  $\vbr_{ms}(x,0,t)=\vbr_\tG(x)=-\,(\pi x)^{-1}$, determines
\bme
\label{uv-ms-0}
\be\se
\left[\begin{array}{c}  \!\! u _{ms} \!\!\\[0.2em]
    \!\! v_{ms}-\vbr_\tG \!\!  \end{array}\right]_{z=0}
\,=\,\left[\begin{array}{c}  \!\! -\tfrac12  \!\!\\[0.2em]
    \!\! t \!\!  \end{array}\right]\,+\,
\dfrac{2}{\pi x}\sum_{\lsum=1}^\infty\left[\begin{array}{c}  \!\!\xS_\lsum|\zS_\lsum|\sin(2\xS_\lsum t)  \!\!\\[0.2em]
    \!\!-1+|\zS_\lsum| \cos(2\xS_\lsum t) \!\!  \end{array}\right],
\ee
where
\be
  [\xS_\lsum,\,\zS_\lsum]\,=\,[x,\,2\lsum]/\varpi_\lsum\,, \hskip 20mm \varpi_\lsum\,=\,\sqrt{x^2+(2\lsum)^2}
\ee
and, as before, we have appealed to the symmetry in $z$. By arguments similar to those used to derive (\ref{mean-values-large-t}$b$), we may show that (\ref{uv-ms-0}$a$) may be re-expressed as
\be\se
\left[\begin{array}{c}  \!\! u _{ms} \!\!\\[0.2em]
    \!\! v_{ms}-\vbr_\tG \!\!  \end{array}\right]_{z=0}
\,=\,\dfrac{2}{\pi x}\Biggl[\sum_{\lsum=1}^\infty-\int_0^\infty\dR \lsum\biggr]
\left[\begin{array}{c}  \!\!\xS_\lsum|\zS_\lsum|\sin(2\xS_\lsum t)  \!\!\\[0.2em]
    \!\!-1+|\zS_\lsum| \cos(2\xS_\lsum t) \!\!  \end{array}\right]\Biggr\}
+\left[\begin{array}{c}  \!\!\langle u_{ms}\rangle\!\!\\[0.2em]
      \!\! \langle v_{ms}\rangle\!\!  \end{array}\right],
\ee
\eme
by which the integral-sum difference avoids secular behaviour with good convergence for $\lsum\gg x$, where $[\xS_\lsum,\zS_\lsum]\to[0,1]$. In fact, only terms with $\lsum=O(x)$ contribute to the integral-sum difference which therefore must be no larger than $O(x)$ and when multiplied by $2/(\pi x)$ is $O(1)$. Of course, the complete flow velocity $[u,v]=[u,v]_{ms}+[u,v]_{bl}$ on $z=0$ is obtained by adding together all $\lsum$ contributions. So $[u,v]_{ms}(x,0,t)$ is determined by (\ref{uv-ms-0}$d$),  while $[u,v]_{bl}(x,0,t)$ is approximated by the dominant $\lsum=0$ contribution $[\ubr,\vbr]_{bl}(x,0,t)$. Since $[\ubr,\vbr]_{ms}(x,0,t) =[0,\vbr_\tG]$, it follows that $[\ubr,\vbr]_{bl}(x,0,t) =[\ubr,\vbr-\vbr_\tG]_{z=0}$, in which (\ref{Poisson-sum-sol-small-t-zero-z}) determines
\begin{align}
\left[\begin{array}{c}  \!\! u- u_{ms} \!\!\\[0.2em]
    \!\! v- v_{ms} \!\!  \end{array}\right]_{z=0}
\,\approx\,\left[\begin{array}{c}  \!\! \ubr_{bl} \!\!\\[0.2em]
    \!\! \vbr_{bl} \!\!  \end{array}\right]_{z=0}
\,=\,&\,\dfrac{1}{\pi x}\left[\begin{array}{c}  \!\! \JR_0(2t) \!\!\\[0.2em]
    \!\! 2\int_t^\infty \JR_0(2\tau) \dR \tau\!\!  \end{array}\right]\nonumber\\
\,\approx\,&\,\dfrac{1}{\pi x\sqrt{\pi t}}\left[\begin{array}{c}  \!\! \cos(2t-\pi/4) \!\!\\[0.2em]
    \!\! -\,\sin(2t-\pi/4)\!\!  \end{array}\right]  \hskip 10mm  \mbox{for} \hskip 5mm t\gg 1\,.\hskip 3mm
\label{uv-ms-0-SD}
\end{align}
In summary the relative sizes of the terms identified by (\ref{uv-ms-0-SD}) are
\be
\label{uv-OofM}
%\be\te
\left[\begin{array}{c}  \!\! u_{ms}  \!\!\\[0.2em]
    \!\! v_{ms}  \!\!  \end{array}\right]_{z=0}=\,O(1) \hskip 2mm \gg \hskip 2mm
\left[\begin{array}{c}  \!\! \ubr_{bl}  \!\!\\[0.2em]
    \!\! \vbr_{bl}  \!\!  \end{array}\right]_{z=0}=\,O(x^{-1}t^{-1/2}) \hskip 2mm \gg \hskip 2mm
\left[\begin{array}{c}  \!\! \langle u_{ms}\rangle \!\!\\[0.2em]
    \!\! \langle v_{ms}\rangle  \!\!  \end{array}\right]=\,O(t^{-1})
\ee
for $\Delta_{bl}=xt^{-1/2}\ll 1$.

The most striking feature of our complete solution (\ref{Poisson-sum-representation}) and (\ref{Poisson-sum-sol-large-t}) is the myriad of inertial waves generated. The presence of the secular contribution $[-\tfrac12, t]$ to $[u,v]$ in  (\ref{Poisson-sum-representation}) emphasises the resonant nature of the waves in the infinite sum that by necessity compensates it. So though the representation (\ref{uv-ms-0}) on $z=0$ is complicated, it clearly indicates how the compensation is achieved but it is unfortunate that the sum-integral difference in (\ref{uv-ms-0}$d$) is difficult to evaluate. That said, the true strength of the method of images lies in its explanation of the fan-like structures visible in figures~\ref{fig7}~and~\ref{fig8} for small-$x$. The method is not suited to explain the cell-structure visible at moderate-$x$ nor for that matter the absence of motion at large-$x$. These are matters resolved in the following \S\ref{IFm} by consideration of individual $z$-Fourier $m$-modes.

%%%%%%%%%%%%%%%%%%%%%%%%%%%%%%%%%%%%%
%%%%%%%%%%%%%%%%%%%%%%%%%%%%%%%%%%%%%
%%%%%       SECTION 6
%%%%%%%%%%%%%%%%%%%%%%%%%%%%%%%%%%%%%
%%%%%%%%%%%%%%%%%%%%%%%%%%%%%%%%%%%%%

\section{$E\downarrow 0$:\, Individual $z$-Fourier $m$-modes\label{IFm}}

Except close to $r=\ell$, motion is dominated by the $m=1$ mode of the $z$-Fourier series (\ref{FS}$a$,$b$), and so here we focus attention on the individual $m$-modes $[\ut_m,\vt_m]$ given by (\ref{sol-tilde-int-u-v}). Noting (\ref{sol-tilde-int-LT}$a$) their LT-solution is
\bme
\label{chitm-LT}
\be\se
-\,(m\pi)^{-1}\bigl[\,\uth_m\,,\,\vth_m\,\bigr]\,=\,\bigl[\,1\,,\,-2/p \,\bigr]\,\chith_m\,,
\ee
where
\be
\chith_m(x,p)\,=\,p^{-1}\exp(-m\pi s\,x) \hskip 10mm \mbox{with} \hskip 5mm s(p)=(p^2+4)^{1/2}/p
\ee
as before in (\ref{sol-tilde-int-LT-prelim}$b$). We also find it convenient to connect the cuts from $p=\pm\iR$, rather than letting them extend to $-\infty$, and to deform the LT-contour of integration into a circuit $\CC$ containing the cut and the essential singularity at $p=0$:
\be
\chit_m(x,t)\,=\,\dfrac{1}{2\pi\iR}\oint_\CC\dfrac{\exp(\xi(p))}{p}\,\dR p \hskip 10mm\mbox{with} \hskip 5mm \xi\,=\,-m\pi s(p)x+pt\,.
\ee
\eme

To simplify our analysis, we restrict attention to $\chit_m(x,t)$, which we evaluate in two distinct asymptotic limits namely $\vartheta\gg 1$ in \S\ref{vartheta-very-large*} and  $X \gg 1$ in \S\ref{X-very-large}, where
\bme
\label{vartheta-X*}
\begin{align}
  \vartheta\,= \,&\,\sqrt{{m\pi x}/{(2t)}}\,,   &     X\,=\,&\,\sqrt{2m\pi xt}\\
\hskip -10mm  \Longleftrightarrow \hskip 12mm    \vartheta X\,= \,&\,m\pi x\,,   &     X/\vartheta\,=\,&2t\,.\hskip 25mm
\end{align}
\eme

\subsection{$\vartheta\gg 1$: Large $|p|$ asymptotics \label{vartheta-very-large*}}

The essential idea in considering the limit
\be
\label{vartheta-large*}
\vartheta\gg 1 \hskip 10mm \Longleftrightarrow \hskip 10mm  m\pi x\gg 2t
\ee
is that the contour path $\CC$ of the LT-inversion integral (\ref{chitm-LT}$d$) may be chosen advantageously to be restricted to $|p|\gg 1$, on which $s\approx 1+2p^{-2}$.  This approximation leads to the similarity solution
\bme
\label{chith-large-vartheta-LT*}
\be
\chit_m\,\approx \,\exp(-m\pi x)\,\FC(\Pi), \hskip 10mm  \mbox{where} \hskip 10mm
\Pi\,=\,X\big/\vartheta^{1/3}\,=\,m\pi x\big/\vartheta^{4/3}
\ee
and, upon setting $p=2\vartheta^{2/3}P$,
\be\se
\FC(\Pi)\,=\,\dfrac{1}{2\pi\iR}\oint_\CC\dfrac{\exp\bigl[ \bigl(-\,\tfrac12\:\! P^{-2}+P\bigr)\Pi\bigr]}{P}\,\dR P\,.
\ee
After expressing the exponential by its power series, the residues at the poles determine the entire function
\be\se
\FC(\Pi)\,=\,1\,+\,\sum_{k=1}^\infty(-1)^k\dfrac {(\Pi^3/2)^k}{(2k)!\,k!}
\ee
\eme 	
(c.f., the power series expansion (http://dlmf.nist.gov/10.2.E2) for the Bessel function~$\JR_0$). For large $\Pi$, a steepest descent evaluation of (\ref{chith-large-vartheta-LT*}$c$) over the saddle points at $P=1\pm \iR\sqrt{3}/2$ yields the dominant contribution
\be
\label{chith-large-vartheta*}
\FC(\Pi)\,\approx\,\biggr(\dfrac{2}{3\pi\Pi}\biggl)^{\!1/2}\,\exp\biggl(\dfrac{3}{4}\Pi\biggr)\sin\biggl(\dfrac{3^{3/2}}{4}\Pi-\dfrac{2\pi}{3}\biggr) \hskip 15mm \mbox{for} \hskip 5mm \Pi\gg 1\,.
\ee

The initial $t=0$ solution valid for all $x>0$ is given by
\be
\label{chitm_x=0*}
\chit_m\,\approx \,\exp(-m\pi x)\,\FC(0)\,=\,\exp(-m\pi x)\,.
\ee
From this point of view, we may regard the factor $\FC(\Pi)$ in (\ref{chith-large-vartheta-LT*}$a$) as an amplitude modulation of the primary structure $\exp(-m\pi x)$. Substitution of the large $\Pi$ asymptotic result (\ref{chith-large-vartheta*}) into (\ref{chith-large-vartheta-LT*}$a$)  determines
\bse
\label{saddle-point*}
\be
\chit_m\,\approx\,\biggr(\dfrac{2}{3\pi\Pi}\biggl)^{\!1/2}\,\exp\biggl[-m\pi x\biggl(1-\dfrac{3}{4\vartheta^{4/3}}\biggr)\biggr]\sin\biggl(\dfrac{3^{3/2}}{4}\Pi-\dfrac{2\pi}{3}\biggr),
\ee
provided that
\be
\Pi\,=\,(2t)^{2/3}(m\pi x)^{1/3}\,\gg \, 1 \hskip 10mm \Longleftrightarrow \hskip 10mm  m\pi x\,\gg\, (2t)^{-2}\,.
\ee
So, in view of (\ref{vartheta-large*}), the asymptotic result (\ref{saddle-point*}$a$) only applies, at fixed $x$,
%  $m\pi x\,(\gg 1)$,
for a limited period of time:
\be
m\pi x \,\gg \,2t\, \gg \,(m\pi x)^{-1/2}\,,
\ee
\ese
a domain that only exists for $m\pi x\gg 1$. So though, the amplitude modulation $\FC(\Pi)$ increases with $\Pi$, its influence in relative importance decreases, as identified by the factor $\vartheta^{-4/3}$ in the exponential of (\ref{saddle-point*}$a$). At early time, $t\ll 1$, the wave-like behaviour, evident in (\ref{saddle-point*}$a$) for $\Pi\gg 1$, encroaches inwards from infinity back to $m\pi x = O(t^{-2})$. The main point, however, is that the full solution (\ref{chith-large-vartheta-LT*}) decays exponentially whenever $m\pi x \gg 2t$, i.e., $\vartheta\gg 1$. 

\subsection{$X\gg 1$: Large $t$ asymptotics\label{X-very-large}}

The analysis of the previous \S\ref{vartheta-very-large*}, points at the importance of the case $\vartheta=O(1)$, which, when $t\gg 1$, may again be studied by steepest descent methods. For $\vartheta>\vartheta_c$ (see (\ref{saddle-point-critical}$a$) below), we note the existence of evanescent solutions similar in character to (\ref{saddle-point*}$a$). However, the emphasis in this section is on the completely wave-like solutions that occur for $\vartheta<\vartheta_c$.

To help guide our asymptotics, we express the exponent $\xi$ (\ref{chitm-LT}$e$) as
\bme
\label{chim-inverse-LT}
\be
\xi\,=\,\varXi X  \hskip 15mm \mbox{with} \hskip 10mm \varXi(\vartheta,p)\,=\,-\,\vartheta \,s(p)\,+\,\tfrac12  \vartheta^{-1} p 
\ee
\eme
and evaluate the integral (\ref{chitm-LT}$d$) asymptotically, in the limit
\be
\label{very-large-X}
X\gg 1\,.
\ee
The dominant contributions stem from the saddle points located, where the $p$-derivative 
\be
\label{saddle-point-Xi}
\varXi_{,p}\,\equiv\,\od{\varXi}{p}\,=\,\dfrac{4\vartheta}{p^3s}\,+\,\dfrac{1}{2\vartheta}
\ee
vanishes. The relevant saddle points occur at purely imaginary locations defined parametrically by
\bme
\label{saddle-point-varphi}
\be
p\,=\,p_g\,=\,2\iR\,\bigl(1-\varphi^2\bigr)^{1/2}\,,  \hskip 10mm s\,=\,s_g\,=\,2\varphi/p_g
\ee
together with their complex conjugates, all chosen to satisfy (\ref{chitm-LT}$c$). The condition $\varXi_{,p}=0$ implies that $p_g^3s_g=-(2\vartheta)^2$, a requirement which is met when $\varphi$ is one of the two real positive roots $\varphi_+$ and $\varphi_-$ of
\be\se
\varphi^3-\varphi+\vartheta^2\,=\,0\,.
\ee
We order them, $0<\varphi_-<\varphi_+<1$,  such that they define
\be
p_g\,=\,p_\pm\,=\,\iR\omega_\pm \,,  \hskip 10mm    \omega_\pm\,=\,2(1-\varphi_\pm^2\bigr)^{1/2}\,,
\hskip 10mm  2\,>\,\omega_-\,>\,\omega_+\,>\,0\,.
\ee
\eme

At $\vartheta=0$, we have $\omega_+=0$ and $\omega_-=2$. On increasing $\vartheta$, $\omega_+$ increases and $\omega_-$ decreases until they coalesce, when $\vartheta$ reaches
\bme
\label{saddle-point-critical}
\be\se
\vartheta\,=\,\vartheta_c\,=\,2^{1/2}3^{-3/4}\,,
\ee
at which
\be
\varphi\,=\,\varphi_c\,=\,3^{-1/2} \,,  \hskip 20mm  \omega\,=\,\omega_c\,=\,2^{3/2}3^{-1/2}\,.
\ee
\eme
The roots only form a real pair for $ 0<\vartheta<\vartheta_c$. By contrast, for $ \vartheta>\vartheta_c$ a complex root with $\Re\{p_g\}>0$ identifies the saddle which yields the dominant contribution to the integral, similar to that described by the results of \S\ref{vartheta-very-large*} when $\vartheta\gg 1$.

Restricting attention to the half-plane $\Im\{p\}\ge 0$, we note that $\Re\{\varXi\}=0$ along the imaginary $p$-axis $p=\iR \omega$ on the range $0\le \omega<2$. At neighbouring points $p=\iR \omega+\delta p$, where $\delta p$ is real and small, the corresponding real increment of $\varXi$ is
\bme
\label{increment}
\be
\delta\varXi \,\approx\varXi_{,p}\,\delta p\,, \hskip 10mm \mbox{where}\hskip 10mm
\varXi_{,p}\,=\,-\,\dfrac{\vartheta}{\omega^2\sqrt{4-\omega^2}}\,+\, \dfrac{1}{2\vartheta}\,.
\ee
\eme
In view of the divergent nature of the first term on the right-hand side of (\ref{increment}$b$) near both $\omega=0$ and $\omega=2$, we have $\delta\varXi<0$ for $\delta p>0$ both at $\omega=0$ and as $\omega\uparrow 2$. However, since $\varXi_{,p}$ vanishes at both $\omega=\omega_\pm$, we have $\varXi_{,p}<0$ on $0\le \omega<\omega_+$ and  $\omega_-<\omega<2$,  while $\varXi_{,p}>0$ on $\omega_+<\omega<\omega_-$. The steepest descent contour remains in the domains, identified by $\delta\varXi<0$. Essentially it starts on the real $p$ axis in the half-plane $\Re\{p\}>0$. It crosses into the half-plane $\Re\{p\}<0$ at the saddle point $p=\iR\omega_+$ and remains there until crosses the imaginary axis again at the saddle point $p=\iR\omega_-$. It continues in the half-plane $\Re\{p\}>0$ until it passes around the cut-point $p=2\iR$ and returns back into the half-plane $\Re\{p\}<0$. Though this saddle point approach is mathematically clear and straightforward, the physical content of the results is best understood via its reformulation as an equivalent stationary phase problem in the following subsection.

\subsection{A stationary phase formulation\label{stationary-phase}}

In \S\ref{vartheta-very-large*}, the circuit $\CC$ for the inverse-LT was limited to large $|p|$. Now, instead, we shrink $\CC$ right down to two lines either side of the imaginary $p$-axis connecting the cut-points $ p=\pm 2\iR$. We therefore set $p=\iR\omega$, $m\pi s=-\iR k$ and $\xi=\iR(kx+\omega t) $, and change the integration variable from $p$ to $s$, noting also that $-(m\pi)^{-1}\dR p/\dR s=\dR \omega/\dR k=1/(\omega^3k)$. Then on taking considerable care with the signs of $k$ and $\omega$ (real) on each of the four sections of $\CC$ (recall that the cut-point $p=2\iR$, and the essential singularity at $p=0$ are now at $k=0$ and $k=\infty$, respectively), we may express (\ref{chitm-LT}$d$) in the form
\be
\label{sol-tilde-int-chi}
\chit_m\,=\,\dfrac{1}{\pi}\int_0^\infty\dfrac{k\sin(k x+\omega t) + k\sin(k x-\omega t)}{k^2+(m\pi)^2}\,\dR k
\ee
equivalent to (\ref{sol-tilde-int-u-v}), where $\omega=2m\pi\big/\sqrt{k^2+(m\pi)^2}$ as defined in (\ref{continuous-variables}$d$). 

On the basis that $k>0$ and $\omega>0$, the waves with phase $k x+\omega t$ ($k x-\omega t$) travel outwards (inwards) in the direction of $x$ decreasing (increasing). The integrals in the complex $p$-plane from which the $k x+\omega t$ ($k x-\omega t$) contribution originates stem from the sections of  $\CC$ with $\Re\{p\}>(<)\,0$. Only the first set of waves with phase $k x+\omega t$ have points of stationary phase, which correspond to the saddle points $p=\iR \omega_\pm$ identified in \S\ref{X-very-large} above, and so we limit our attention to them. 

In order to take advantage of the $\vartheta$, $X$ formalism (\ref{chim-inverse-LT})--(\ref{saddle-point-critical}) of the \S\ref{X-very-large} steepest descent problem, we introduce the new variables $K$ and $\Upsilon$ defined by
\bme
\label{K-Upsilon*}
\be\te
{k}/(m\pi)\,=\,{K}/\vartheta\,, \hskip 10mm   \omega\,=\,2\vartheta \Upsilon \hskip 8mm \Longrightarrow \hskip 8mm
kx\,+\,\omega t\,=\,(K+ \Upsilon)X\,.
\ee
\eme
Following the parallel formalisms, the phase velocity $c_p=\omega/k\,(>0)$ is given by
\be
\label{phase-velocity}
m\pi c_p\,=\,\bigg\{\begin{array}{lll}
\!\!m \pi \omega/k\,,  \hskip 3mm       & \mbox{where}    \hskip 5mm &   \omega\,=\,2m\pi\big/\sqrt{k^2+(m\pi)^2}\,,\\[0.4em]
\!\!2\Theta^2\Upsilon/K\,,   \hskip 3mm &  \mbox{where}   \hskip 5mm & \Upsilon\,=\,1\big/\sqrt{K^2+\vartheta^2}\,.
\end{array}
\ee
The group velocity $c_g=\partial\omega/ \partial k \,(<0$, see below) is given by
\be
\label{group-velocity}
m\pi c_g\,=\,\bigg\{\begin{array}{lll}
\!\!m \pi \omega^{\prime}\,,  \hskip 8mm       & \mbox{where}    \hskip 5mm &   \omega^{\prime}\,=\,-\,\omega^3 k\big/(2m\pi)^2\,,\hskip 7mm\\[0.4em]
\!\!2\Theta^2\dot{\Upsilon}\,,   \hskip 8mm &  \mbox{where}   \hskip 5mm & \hskip 1mm \dot{\Upsilon}\,=\,-\,\Upsilon^3K\,,
\end{array}
\ee
where the prime and dot denote the partial derivatives with respect to $k$ and $K$ respectively. Being negative the group velocity is directed inwards ($x$ increasing, $r$ decreasing), i.e, opposite to the phase velocity. On further differentiation we obtain
\be
\label{group-velocity-derivative}
(m\pi)^2  c^{\prime}_g\,=\,\bigg\{\begin{array}{lll}
\!\!(m\pi)^2 \omega^{\prime\prime}\,, \hskip 3mm    &   \mbox{where}    \hskip 5mm&  \omega^{\prime\prime}\,=\,\bigl[1-3\bigl(\omega k/(2m\pi)\bigr)^2\bigr]\omega^{\prime}\big/k\,,\hskip 7mm\\[0.4em]
\!\!2\vartheta^3 \ddot{\Upsilon}\,,  \hskip 3mm      & \mbox{where}    \hskip 5mm &   \ddot{\Upsilon}\,=\,\bigl[1-3(K\Upsilon)^2\bigr]\dot{\Upsilon}\big/K  \,.
\end{array}
\ee

The points $k=k_\pm$  ($k_-<k_+$ with $\omega^{\prime}<0\,\Longrightarrow\,\omega_->\omega_+$) of stationary phase occur, where the $k$-derivative of the phase $kx+\omega t$ vanishes:
\be
\label{st-ph}
m\pi(kx+\omega t)^{\prime}\big|_\pm\,=\,\biggl\{\begin{array}{l}
m\pi(x+c_{g\pm}  t)   \\[0.4em]
(1+{\dot{\Upsilon}}_\pm)\vartheta X
\end{array}\biggr\}\,=\,0 \hskip 5mm\Longrightarrow\hskip 5mm
\bigg\{\begin{array}{l}
c_{g\pm}\, =\,-\, x/t\,,   \\[0.4em]
\,\,{\dot{\Upsilon}}_\pm\,=\,-\,1\,.
\end{array}
\ee
Substitution of ${\dot{\Upsilon}}_\pm=-1$ into (\ref{group-velocity}) determines $\Upsilon_\pm^3K_\pm=1$ which together with (\ref{phase-velocity}) yields $K_\pm^{2/3}=K_\pm^2+\theta^2$. This shows that
\bme
\label{st-ph-1}
\be\te
K_\pm\,=\,\varphi_\pm^{3/2}\,,    \hskip10mm     \Upsilon_\pm\,=\,\varphi_\pm^{-1/2}\,,     \hskip10mm
{\ddot{\Upsilon}}_\pm\,=\,\bigl(3\varphi_\pm^2-1\bigr)\big/\varphi_\pm^{3/2}\,\gtrless\, 0\,,
\ee
\eme
where $\varphi_\pm$ are the positive roots  ($0<\varphi_-<\varphi_+<2$) of (\ref{saddle-point-varphi}$c$).
  
Using the results (\ref{phase-velocity}), (\ref{group-velocity}) and (\ref{st-ph}), the prefactor to $\sin(k x+\omega t)$ in the integral (\ref{sol-tilde-int-chi}) at the points of stationary phase may be written
\be
\label{prefactor}
\dfrac{k_{\pm}}{k_{\pm}^2+(m\pi)^2}\,=\,\dfrac{k_{\pm}\omega_{\pm}^2}{(2m\pi)^2}\,=\,\dfrac{x}{\omega_{\pm} t}\,.
\ee
There a routine steepest descent calculation involving $(2\iR)^{-1}\exp[\iR(k x+\omega t)]$ across the saddle-points in the complex $k$-plane  gives the dominant contributions to the integrals, which, when added to their complex conjugates, determine
\be
\label{chit-SD*}
\chit_m(x,t)\,\approx\,\left\{\begin{array}{l}
\ds{\sum_{k=k_{\pm}}}\,\dfrac{\pm 1}{\omega_{\pm}}\biggl(\dfrac{2x^2}{\pm \pi c^{\,\prime}_{\!g\pm}t^3}\biggr)^{\!1/2}\cos(k_{\pm}x+\omega_{\pm} t\mp \pi/4)\,,\\[0.9em]
\ds{\sum_{K=K_{\pm}}}\,\dfrac{\pm 1}{\Upsilon_{\pm}}\biggl(\dfrac{2}{\pm\pi{\ddot{\Upsilon}}_{\pm}X}\biggr)^{\!1/2}
\cos[(K_{\pm}+ \Upsilon_{\pm})X\mp \pi/4]\,.
\end{array}\right.
\ee
In the following subsections we describe the nature of the solution (\ref{chit-SD*}) as $\vartheta$ is increased from zero. 

\subsubsection{$\vartheta\ll 1$\label{vartheta-small}}

The simultaneous limits $X\gg 1$ and $\vartheta\ll 1$ restrict $m\pi x$ to the range
\be
\label{X-large*}
(2t)^{-1}\, \ll\, m\pi x \, \ll \, 2t\,,
\ee
which only exists for $t \gg 1$. With $\vartheta$ small, the roots of (\ref{saddle-point-varphi}$c$) are $\varphi_+\approx 1$ and $\varphi_-\approx \vartheta^2$. Accordingly (\ref{st-ph-1}) determine
\bme
\label{small-stat-phase}
\qie
\begin{align}
K_+\approx\, &\,1\,,  & \Upsilon_+\approx \,&\,1\,,& \ddot{\Upsilon_+}\approx\, &\,2\,,&  k_+\approx\,&\,{m\pi}\big/{\vartheta}\,,& \omega_+\approx\,&\, 2\vartheta\,,\\
K_-\approx \,&\,\vartheta^{3},  & \Upsilon_-\approx\,&\,\vartheta^{-1},& \ddot{\Upsilon_-}\approx\, &\,-\,\vartheta^{-3},&  k_-\approx\, &\,m\pi\vartheta^2,& \omega_-\approx\,& \,2\,,
\end{align}
\eme
with which (\ref{chit-SD*}) becomes
\be
\label{stat-phase-int-small-vartheta}
\chit_m\,\approx\,\dfrac{\cos(2X-\pi/4)}{(\pi X)^{1/2}}\,-
\,\dfrac{\vartheta^{5/2}\cos(2t+\pi/4)}{(\pi X/2)^{1/2}}\,.
\ee

The former $\chit_m^+$-mode is linked to the essential singularity at $p=0$ ($\varphi_+$). In a restricted limit  $t\gg z^2/x \gg x$, the $z$-Fourier series (\ref{FS}$a$) may by summed asymptotically on the basis that the series is dominated by terms with $m=O(xt/z^2)$ large. In this way, we can recover (\ref{Poisson-sum-sol-large-t}$b$) in the limited domain $\xS\ll 1$ ($\xS$ defined by (\ref{Poisson-sum-LT}$b$,$c$)). This suggests that the $\chit_m^+$-modes with large $m$ are responsible for the fine structure visible on figures~\ref{fig7},~\ref{fig8} in the vicinity of the corner $(r,z)=(0,\ell)$. The latter $\chit_m^-$-mode, smaller by a factor $O(\vartheta^{5/2})$, is linked to the cuts at $p=\pm 2 \iR$ ($\varphi_-$). There is no asymptotic regime that the ensuing $z$-Fourier series (of small terms) may by summed over $m$. That said, the very small relative size $O(\vartheta^{5/2})$ of the $\chit_m^-$-modes would seem to render them irrelevant. Of course, proper summation of the large $m$ modes $\chit_m$ is best understood via the analysis of \S\ref{images}.

\subsubsection{$\vartheta=\vartheta_c$: The critical line $x=-\,(c_g)_ct$\label{vtheta-crit}}

Though the $\chit_m^-$-mode is small at small $\vartheta$, on increasing $\vartheta$ its amplitude increases and the contributions from both $\chit_m^+$ and $\chit_m^-$ become comparable when $\vartheta=O(1)$; a trend that continues until $K_+$ and $K_-$ merge. There, ${\ddot{\Upsilon}}_\pm=0$ and so (\ref{st-ph-1}$c$) determines the critical value of $\varphi_c$. That and  $\vartheta_c$ were given previously by (\ref{saddle-point-critical}$a$,$b$):
\bme
\label{critical values*}
\be
\varphi_c=3^{-1/2}  \hskip 20mm \vartheta_c\,=\,2^{1/2}3^{-3/4}\,,
\ee
which on substitution into (\ref{st-ph-1}$a$,$b$) determine
\be\te
K_c\,=\,3^{-3/4}\,, \hskip 10mm    \Upsilon_c\,=\,3^{1/4}\,, \hskip 10mm  K_c+\Upsilon_c\,=\,4\cdot 3^{-3/4}\,.
\ee
\eme
With (\ref{K-Upsilon*})--(\ref{group-velocity}) they yield
\bse
\label{front-line-prelim}
\begin{align}
(m\pi)^{-1}k_c\,=\,&\,{K_c}\big/{\vartheta_c}\,=\,2^{-1/2}\,\doteqdot\,0.707\,,\\
\omega_c\,=\,&\,2\vartheta_c\Upsilon_c\,=\,2^{3/2}\cdot 3^{-1/2}\,\doteqdot\,1.633\,,\\
-\,m\pi(c_g)_c\,=\,&\,2\vartheta_c^2\,=\,4\cdot 3^{-3/2}\,\doteqdot\,0.770\,,\\
(c_p)_c\,=\,&\,-\,(\Upsilon_c/K_c)\,(c_g)_c\,=\,-\,3\,(c_g)_c\,.
\end{align}
\ese

We conclude that the wave-like stationary phase solutions (\ref{chit-SD*}) only exist on increasing $x$ (equivalently $\vartheta$) from zero, at fixed $t$, for $x<x_c(t)$ ($\vartheta<\vartheta_c$), where
\be
\label{front-line}
x_c(t)\,=\,-\,(c_g)_ct\,\doteqdot\,0.245\, t/m\,.
\ee
Of course, the divergence of $\chit_m$ given by the solution (\ref{chit-SD*}) when $x=x_c(t)$, a straight line in the $x$-$t$ plane (see figure~\ref{fig6} below) on which ${\ddot\Upsilon}_c=0$  ($\vartheta=\vartheta_c$), is not real but rather reflects our incorrect implementation of the steepest descent approximation for that limiting case. Following the saddle point mergence $K_+=K_-$, they separate again and move off the real $K$-axis into the complex $K$-plane. The solutions, that they describe for $x>x_c(t)$ ($\vartheta>\vartheta_c$), decay exponentially with increasing $x$ and, for $\vartheta\gg 1$, are given by (\ref{chith-large-vartheta-LT*}) a result that holds for all $t>0$.

%%%%%%%%%%%%%%%%%%%%%%%%%%%%%%
%%%%%%%%%% FIGURE 7 %%%%%%%%%%
%%%%%%%%%%%%%%%%%%%%%%%%%%%%%%

\begin{figure}
\centerline{}
\vskip 3mm
\centerline{
\includegraphics*[width=1.0 \textwidth]{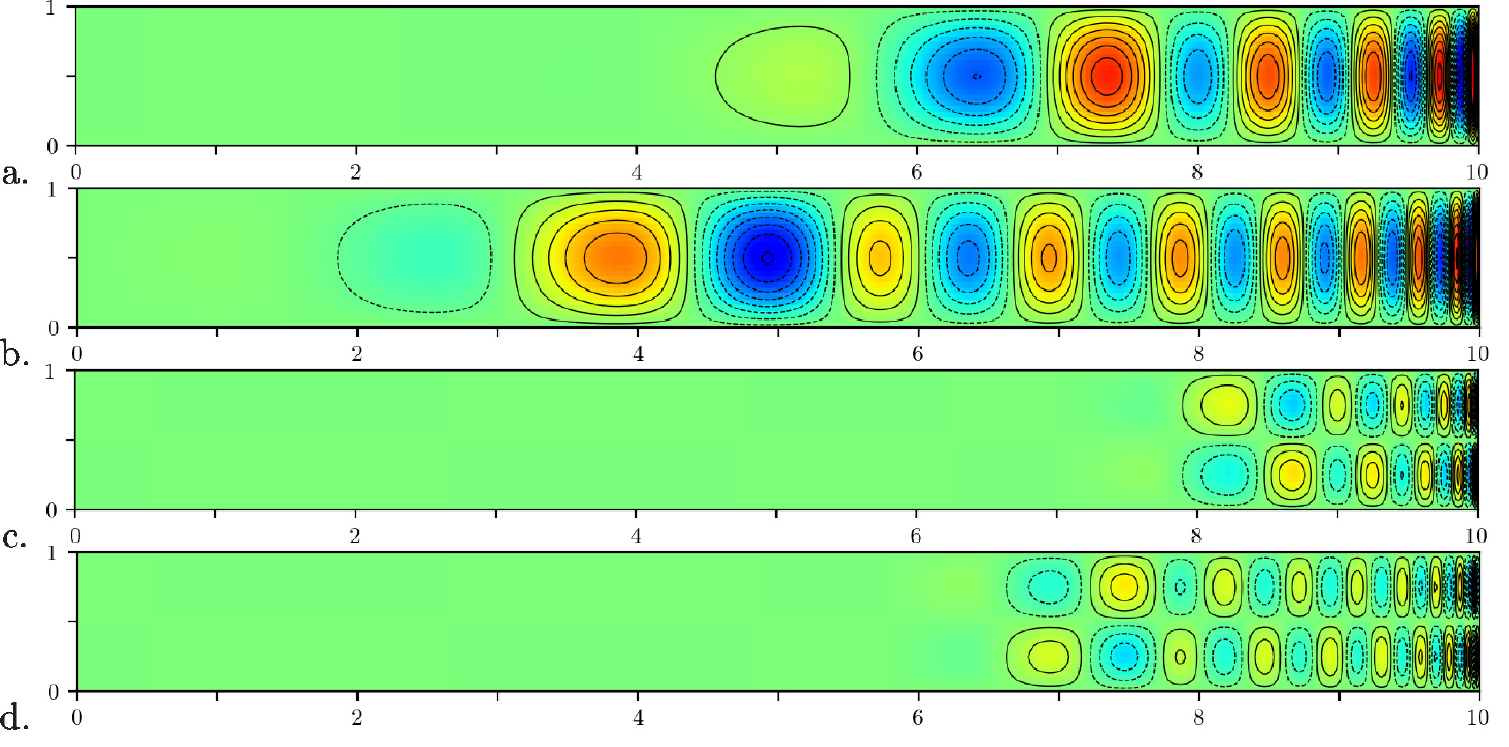}  
}
\caption{(Colour online) $\chit_m^{\,\rm {wave}}$-contours (see (\ref{Fourier-m-modes})) for the cases $m=1$ at ($a$)~$t=15$, ($b$)~$t=25$, and $m=2$ also at ($c$)~$t=15$, ($d$)~$t=25$. The colour scale ranges from $-1$ to $1$.
}
\label{fig5}
\end{figure}

%%%%%%%%%%%%%%%%%%%%%%%%%%%%%%
%%%%%%%%%%%%%%%%%%%%%%%%%%%%%%

\subsection{Comparison with numerics\label{Comparison-with-numerics}}

Since $X$ is large, the termination at $x=x_c(t)$ is essentially abrupt, giving rise to a front, behind which, $0<x \le x_c(t)$, the wave is confined. The property (\ref{front-line}) indicates that $x_c(t)$ is inversely proportional to $m$ implying that the $m=1$ mode penetrates the furthest to the left. We portray the individual Fourier modes $\chit_m^{\,\rm {wave}}(r,z,t)$ (\ref{Fourier-m-modes}) in figure~\ref{fig5}, which contrasts the behaviour of the $m=1$ and $2$ modes. Even though $\chit_m^{\,\rm {wave}}(r,z,t)$ was determined in the true cylindrical geometry, it is remarkable how well the rectangular Cartesian asymptotic formula $\ell-r_c(t)=x_c(t)$,  predicts the front location for $r$ as small as $2$ ($\ell=10$). The fact that $[\ell-r_c(t)]_{m=1}=2[\ell-r_c(t)]_{m=2}$ is clearly illustrated by comparing panels~($a$) with~($c$) and ($b$) with~($d$). Furthermore, the property (\ref{front-line-prelim}$a$) indicates that the half wave length $\pi/k_c$ is also inversely proportional to $m$ implying that the half wave length also decreases with increasing $m$, a trend again confirmed by comparing panels~($a$) with~($c$) and ($b$) with~($d$).

%%%%%%%%%%%%%%%%%%%%%%%%%%%%%%
%%%%%%%%%% FIGURE 8 %%%%%%%%%%
%%%%%%%%%%%%%%%%%%%%%%%%%%%%%%

\begin{figure}
\centerline{}
\vskip 3mm
\centerline{
\includegraphics*[width=1.0 \textwidth]{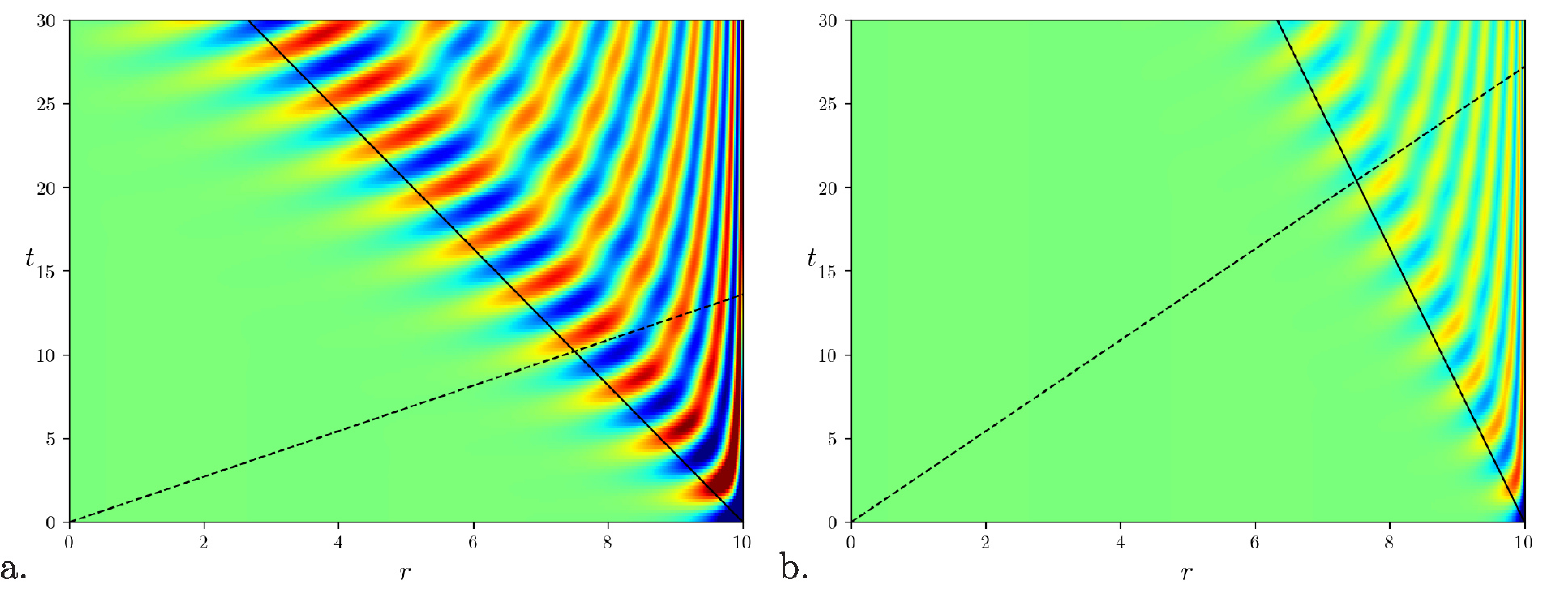}  
}
\caption{(Colour online) $\chi_m^{\,\rm {wave}}$-contours at fixed $z=z_m$, in the $r$-$t$ plane for $\ell=10$. The solid black line is $(c_g)_c t +(\ell-r) =0$ with $(c_g)_c\doteqdot -0.245/m$ (see (\ref{front-line-prelim}$c$)). The dashed black line is $(c_p)_c t=r$ with $(c_p)_c = -3(c_g)_c\doteqdot 0.735/m$ (cf.~(\ref{front-line-prelim}$d$)). ($a$) $m=1$ with $z_1=0.5$; ($b$) $m=2$ with $z_2=0.25$.
}
\label{fig6}
\end{figure}

%%%%%%%%%%%%%%%%%%%%%%%%%%%%%%
%%%%%%%%%%%%%%%%%%%%%%%%%%%
%%%

To test the validity of the formulae (\ref{front-line-prelim}$c$,$d$) for the group and phase velocities at the front $x=x_c(t)$, we consider $\chit_m^{\,\rm {wave}}(r,z,t)$ at fixed $z=z_m$ chosen to maximise $\chit_m^{\,\rm {wave}}$, i.e., $z_1=0.5$, $z_2=0.25$. Then in figure~\ref{fig6}, we consider space-time contour plots in the $r$--$t$ plane. Reassuringly the extent of wave activity is bounded by the line $\ell-r=x_c(t)$. As the maximum amplitude of the wave (i.e., crests) moves at the local phase velocity $c_p$, the tangent to its track has slope $1/c_p$. For that reason we plot the line $t=r/(c_p)_c$ and see that this property is indeed met at the front $x=x_c(t)$, where the line is reasonably parallel to the wave crest tracks. The evolution of $\chit_m^{\,\rm {wave}}(r,z,t)$ is followed on figures~\ref{fig6}($a$,$b$) up until $t=30$. Later, however, by $t\approx 40$ the $m=1$ wave front reaches the axis, after which it is reflected, leading to less well ordered pulsating structures for $t\gtrapprox 40$. The $m=2$ wave is reflected at $t\approx 80$ and so on.

Finally we reassess figures~\ref{fig7}~and~\ref{fig8} in the light of our present findings. Sufficiently far to the left, only the $m=1$ mode is visible. On halving the distance to the right-hand outer boundary, $r=10$, some interference from the $m=2$ mode is visible. Yet further reduction of that distance leads to interference from successive higher harmonics, that complicates the picture more. Note too that, though the waves penetrate further to the left with time ($c_g<0,\,\forall\,m$), the waves themselves propagate to the right ($c_p>0,\,\forall\,m$). As figures~\ref{fig7}~and~\ref{fig8} concern $t\le 25$, our previous remarks about wave reflection at $t\approx 40$ are pertinent here too.

%%%%%%%%%%%%%%%%%%%%%%%%%%%%%%%%%%%%%
%%%%%%%%%%%%%%%%%%%%%%%%%%%%%%%%%%%%%
%%%%%       SECTION 7
%%%%%%%%%%%%%%%%%%%%%%%%%%%%%%%%%%%%%
%%%%%%%%%%%%%%%%%%%%%%%%%%%%%%%%%%%%%

\section{Concluding remarks\label{conclusion}}

The primary feature of any spin-down process is the evolution of the azimuthal QG-flow $v$ on the spin-down time scale, visible for our problem in the DNS ($E=10^{-3}$) results for $E^{-1/2}v_\tDNS$  reported in figures~\ref{fig3},~\ref{fig4} panels ($a$), ($d$), ($g$). The meridional flow, characterised by the streamfunction $r\chi$ and smaller by a factor $E^{1/2}$, needed to provide the vortex line compression, is apparent in the same panels of figures~\ref{fig1},~\ref{fig2} for $E^{-1/2}\chi_\tDNS$. Like the QG meridional flow, all components of the superimposed MF-inertial waves are $O(E^{1/2})$. Consequently they are visible in figures~\ref{fig1},~\ref{fig2} for $E^{-1/2}\chi_\tDNS$ but not in figures~\ref{fig3},~\ref{fig4} for $E^{-1/2}v_\tDNS$, where they are overwhelmed by the dominant QG-part. Being a manifestation of the transient Ekman layer, as discussed in \S\ref{Introduction} \citep[previously identified by][]{GH63}, the MF-waves decay algebraically ($\propto t^{-1/2}$) with time. Outside an expanding boundary layer, width $\Delta(t)=(Et)^{1/2}$, the horizontal components of the MF-waves are $z$-independent (see (\ref{MF-spin-down-prelim})), and in that respect are similar in character to the QG-flow.

The aforementioned characteristics are found in the unbounded layer $\ell\to\infty$. Our objective here has been to identify the extra inertial waves triggered by a boundary at $r=\ell$ large but finite. Like the MF-waves, they are visible in the $E^{-1/2}\chi_\tDNS$ contour plots of figures~\ref{fig1},~\ref{fig2}, but are not clearly identified until consideration of the filtered DNS (\ref{filter-v}) and (\ref{filter-u}$b$) in panels ($b$), ($e$), ($h$) of figures~\ref{fig1}-\ref{fig4}, in which the QG-contribution has been removed.

Since the extra inertial waves are only clearly visible in the DNS when $\ell\gg 1$ (for us $\ell=10$), we considered an asymptotic solution based on $\ell\gg 1$ in \S\ref{mathematical-problem}. There we simply determined the response to the QG-trigger (\ref{triggers}$a$), which reflects the failure of the unbounded QG-flow solution to meet the impermeable boundary condition at $r=\ell$. The results were found to compare well with the filtered DNS at instants when the MF-contributions were absent (cf.~figures~\ref{fig2}~and~\ref{fig3}, their panels ($b$),($e$),($h$) with ($c$),($f$),($i$)). In our sequel Part~II, we consider the additional response when the failure of the unbounded MF-flow solution is properly accounted for, and find the detailed comparison significantly improved. 

In addition to Ekman damping, inertial waves of short length scale suffer significant internal viscous dissipation. To filter out that damping, which is considerable at $E=10^{-3}$, we considered our asymptotic solutions in the zero Ekman number limit in \S\S\ref{No-damping}--\ref{IFm}. The resulting triggered waves, illustrated in figures~\ref{fig7} and~\ref{fig8}, reveal  very detailed structure near the $r=\ell$ boundary, previously hinted at by figures~\ref{fig1}-\ref{fig4} panels ($c$),($f$),($i$). To explain the origins of that structure, we considered analytically the rectangular Cartesian limit, appropriate to $\ell-r =O(1)$ ($\ell\gg 1$) in \S\ref{Cartesian-limit}. Two complimentary approaches were adopted. On the one hand, in \S\ref{images}, we employed the method of images, which revealed the nature of the wave generation, particularly as it pertained to small $\ell-r$. The considerable wave interference identified leads to resonances manifested ultimately by simpler structures at large $\ell-r$. On the other hand, in \S\ref{IFm}, we considered individual $z$-Fourier $m$-modes and used the methods of stationary phase (see  \S\ref{stationary-phase}) and  steepest descent (see \S\ref{X-very-large}) to identify respectively their wave and evanescent wave structure. Together they identify a wave front (see \S\ref{vtheta-crit} and figures \ref{fig5} and \ref{fig6}: panels ($a$) $m=1$, ($b$) $m=2$). As the dominant $m=1$ mode suffers relatively little internal dissipation, it decays slowly and remains visible in the DNS (or more clearly in the filtered DNS) as time proceeds. Moreover, larger $m$-modes propagate a shorter distance from $r=\ell$. For those two reasons, we regard the analysis of \S\ref{IFm} (particularly \S\ref{vtheta-crit}) as the cornerstone of our study with respect to the interpretation of the DNS. We did consider the wave motion triggered in containers with $O(1)$ aspect ratio (particularly $\ell=1$), but for them the inertial wave activity showed little structure and decayed rapidly. There was some evidence of fan-like behaviour near the corner $(r,z)=(\ell,1)$, but none of the other travelling wave or frontal behaviour. That is unsurprising because waves are reflected promptly at the axis with no time available to create the coherent travelling structures like those reported in this paper. As a result of the almost immediate reflection, there is considerable wave interference and a shortening of the length scales leading to enhanced internal dissipation.

There has been a considerable amount of research on spin-up/down \citep[see, e.g.,][and references therein]{Letal12} together with the associated inertial wave activity (but see also the related studies of the linear inertial wave activity in a precessing plane layer, \citealt{MK02}, and linear and nonlinear waves in a container, \citealt{JO14}, \citealt{Betal19}). Much of the recent ongoing research is of the DNS-type, though often with additional physical mechanisms and in containers of more complicated geometry. Indeed, even in our cylindrical geometry, the comprehensive studies of \cite{KB95} and \cite{ZL08} were largely concerned with identifying the free modes and determining their decay rates. They did not address the matter of relative wave amplitude between individual modes during the spin-down process, or for that matter their accumulated structure.  By that we mean that, like \citet{G68} before, they considered a model expansion of the combined $z$-Fourier (\ref{FS}) and $r$-Fourier-Bessel (\ref{solution}) series  type, but unlike in (\ref{solution}) the individual mode amplitudes remained undetermined. Our main thrust has been to gain insight about the structures exhibited by the DNS in a simple geometry via the application of asymptotic methods to solve the initial value problem via the LT-method, albeit we make use of the Ekman damping decay rates for individual modes (see \S\ref{Ekman-layer-damping}), as found by \cite{KB95} and \cite{ZL08}.

%%%%%%%%%%%%%%%%%%%%%%%%%%%%%%%%%%%%%
%%%%%%%%%%%%%%%%%%%%%%%%%%%%%%%%%%%%%
%%%%%            APPENDICES
%%%%%%%%%%%%%%%%%%%%%%%%%%%%%%%%%%%%%
%%%%%%%%%%%%%%%%%%%%%%%%%%%%%%%%%%%%%

\appendix

\section{MF-harmonic expansion\label{MF-harmonic}}

As the MF-boundary-layer, width $\Delta=\sqrt{Et}$, thickens the asymptotic solution (\ref{MF-combined}) becomes unreliable. The only way to properly resolve the solution for
\be
\label{MF-diffusion-range}
                  Et\,=\,O(1)
\ee
is to invoke the inverse of the complete LT-solution to the initial value problem given approximately by eqs.~(3.9), (3.10) of \cite{GH63}. In the inversion, the residue at the pole close to $p=0$ identifies the QG-solution. The remaining poles elsewhere determine
\bse
\label{GH-waves}
\begin{align}
\chi_{\tGH}\,\approx\,& E\,\dfrac{r}{\ell}\cos(2t)\sum_{m=1}^\infty\biggl[(z-1)\,-\,\dfrac{\sin[\xi_m(z-1)]}{\sin\xi_m}\biggr]\exp\bigl(-E\xi_m^2 t\bigr)\,,\\
v_{\tGH}\,\approx\,& E\,\dfrac{r}{\ell}\sin(2t)\sum_{m=1}^\infty\biggl[1\,-\,\dfrac{\cos[\xi_m(z-1)]}{\cos\xi_m}\biggr]\exp\bigl(-E\xi_m^2 t\bigr)\,,
\end{align}
where $\xi_m$ are given by the positive roots of
\be
\tan\xi_m\,=\,\xi_m\,+\,O\bigl(E^{1/2}\bigr)   \qquad\qquad    (m\ge 1).
\ee
To appreciate the nature of (\ref{GH-waves}$a$,$b$) it is helpful to set
\be
\xi_m\,=\bigl(m+\tfrac12\bigr)\pi-\iota_m\qquad\qquad    \bigl(0<\iota_m<\tfrac12 \pi\bigr),
\ee
\ese
in which $\iota_m$ is moderately small: $\iota_1\approx 0.219$, $\iota_2\approx 0.129$, $\iota_3\approx 0.091$ with $\iota_m\downarrow 0$ as $m\to \infty$ (see, e.g., http://mathworld.wolfram.com/TancFunction.html).

The MF-flow is fully $z$-dependent when $Et=O(1)$ and by that time $\chi_{\tGH}$ and $v_{\tGH}$ (\ref{GH-waves}$a$,$b$) are small $O(E)$. Recall too that this is the time scale on which the QG-sidewall shear layer has spread laterally an $O(1)$ distance. That was a central consideration in \cite{OSD17}, in which the QG-solution on that time scale was given by their eq.~(3.8$a$). The first non-trivial zero $\xi_1\approx 4.4934$ of (\ref{GH-waves}$d$) determines the $m=1$ mode with the slowest decay rate that dominates as $Et\to\infty$, a property consistent with our numerical results reported in \S\ref{numerics}.

For $Et\ll 1$, the small amplitude factor $E$ in (\ref{GH-waves}$a$,$b$) is misleading because asymptotic evaluation of the sums determines larger amplitudes. To see this, we note that for sufficiently small $Et$, the factor $\exp\bigl(-E\xi_m^2 t\bigr)$ is approximately unity for $m \ll (Et)^{-1/2}$. So the sums are dominated from the high harmonic contributions with $m=O\bigl((Et)^{-1/2}\bigr)$, for which $\xi_m\approx (m+\tfrac12)\pi$ (see (\ref{GH-waves}$d$)). Accordingly, we may approximate the sums by integrals, so, e.g.,
\bse
\label{sum-to-integral}
\be
\sum_{m=1}^\infty\exp\bigl(-E\xi_m^2 t\bigr)\,\approx \int_0^\infty\exp\bigl(-E(m\pi)^2 t\bigr)\,\dR m\,=\,\dfrac{1}{\sqrt{4\pi Et}}\,,
\ee
which diverges as $Et\downarrow 0$. Accordingly, the first terms in (\ref{GH-waves}$a$,$b$) recover the mainstream \\
\vskip -4.5mm
\noindent
parts of $\chib_{\tGH}$, $\vb_{\tGH}$ (\ref{MF-spin-down-prelim}$b$,$a$) respectively, in which $\UG_{\!\tGH}$ and $\VGxw_{\!\tGH}\approx \VG_{\!\tGH}$ (see (\ref{MF-spin-down-prelim-more}$b$)) are together given by (\ref{outflow-sol-MF}$d$) in the large $t$ limit. Likewise the sum of the second terms in (\ref{GH-waves}$a$) is approximately
\begin{align}
-\,\sum_{m=1}^\infty\dfrac{\sin[\xi_m(z-1)]}{\sin\xi_m}\,\exp\bigl(-E\xi_m^2 t\bigr)
\,\approx \,&\int_0^\infty\cos(m\pi z)\,\exp\bigl(-E(m\pi)^2 t\bigr)\,\dR m  \nonumber\\
=\,&\dfrac{1}{\sqrt{4\pi Et}}\,\exp\biggl(-\dfrac{z^2}{4Et}\biggr)
\end{align}
\ese
\citep[use \S1.4 eq.~(11) of][]{EMOT54I}, while its  $z$-derivative pertains to the sum  of the second terms in (\ref{GH-waves}$b$). Upon recalling that $\ub_{\tGH}=-\partial\chib_{\tGH}/\partial z$ (see (\ref{MF-spin-down-prelim}$b$)), we may recover the leading order boundary layer terms in (\ref{MF-combined}$a$) for $[u_{\tGH}, v_{\tGH}]$. In a similar style to (\ref{sum-to-integral}), \cite{GH63} propose an estimate in their un-numbered equation, p.~389, l.~2, that leads to $v_{\tGH}=O(E^{1/2}t^{-1})$ (our notation). Unfortunately, it fails to capture the mainstream and boundary layer estimates $O(E^{1/2}t^{-1/2})$ and $O(t^{-1})$ predicted by (\ref{MF-combined}$a$) and (\ref{outflow-sol-MF}$d$).

\section{A Fourier-Bessel series\label{Fourier-Bessel}}

We derive the Fourier-Bessel series for $\JR_1(m\pi q r)$ ($q=\,$const.). According to \S18.1, eqs.~(3), (4) of \cite{W66} it is
\bse
\label{Fourier-Bessel-begin}
\be
\JR_1(m\pi q r)\,=\,\dfrac{2}{\ell^2}\sum_{n=1}^\infty\bigl[\JR_0(j_n)\bigr]^{-2}\biggl[\int_0^\ell r\JR_1(m\pi q r)\JR_1(j_nr/\ell)\,\dR r\biggr]\JR_1(j_nr/\ell)\,,
\ee
where $j_n$ denotes the $n^{\rm{th}}$ zero ($>0$) of $\JR_1(x)$ with the consequence that
\be
\JR_2(j_n)\,=\,\JR_0(j_n)\,=\,\JR'_1(j_n)\,.
\ee
\ese
With $\DS_m$ defined by (\ref{F-Series-eqs}$e$) and $q_{mn}=j_n/(m\pi\ell)$ (\ref{Fourier-Bessel-series}$e$), the identities
\bse
\label{Fourier-Bessel-middle}
\begin{align}
\DS_m\,\JR_1(m\pi q r)\,=&\,-\bigl(q^2+1\bigr)(m\pi)^2\JR_1(m\pi q r)\,,\\
\DS_m\,\JR_1(j_nr/\ell)\,=&\,-\,\bigl(q_{mn}^2+1\bigr)(m\pi)^2\JR_1(j_nr/\ell)
\end{align}
follow. Their use in a routine integration by parts leads to
\begin{align}
-\,\bigl(q_{mn}^2-q^2\bigr)&(m\pi)^2\int_0^\ell r\JR_1(m\pi q r)\JR_1(j_nr/\ell)\,\dR r\,\nonumber\\
&=\,\int_0^\ell r[\JR_1(m\pi q r)\DS_m\,\JR_1(j_nr/\ell)\,-\,\JR_1(j_nr/\ell)\DS_m\,\JR_1(m\pi q r)]\,\dR r\qquad\qquad\nonumber\\
&=\,j_n\JR_1(m\pi q \ell)\JR_0(j_n)\,.
\end{align}
\ese
Substitution into (\ref{Fourier-Bessel-begin}$a$) determines
\be
\label{Fourier-Bessel-end}
\dfrac{\JR_1(m\pi q r)}{\JR_1(m\pi q \ell)}\,=\,-\,\sum_{n=1}^\infty\dfrac{2q_{mn}^2}{q_{mn} ^2-q^2}\,\dfrac{\JR_1(j_nr/\ell)}{j_n\JR_0(j_n)} \hskip 10mm \mbox{on} \hskip 10mm  0\,\le\, r\,<\,\ell\,.
\ee
The representation fails at $r=\ell$, where $\JR_1(j_n)=0$ and each term vanishes.

\section{The transient boundary layer flow $[\ubr,\vbr]_{bl}$\label{esoteric}}

We investigate the nature of $[\ubr,\vbr]_{bl}$ (\ref{Poisson-sum-sol-large-t}$c$) for $t\gg 1$. We anticipate that it is localised in a boundary layer of $z$-thickness $\Delta_{bl}=x t^{-1/2}$, near $z=0$, for which convenient co-ordinates are $(\zeta, \zS)$:
\bme
\label{tS-def}
\be\te
\zeta\,=\,2(1-\xS)t\,=\,O(1)\,, \hskip 12mm  2(1-\xS)\,=\,\zS^2\,+\,O(\zS^4)\,,\hskip 8mm \zS\,\ll\,1\,.
\ee
\eme

The evaluation of (\ref{Poisson-sum-sol-large-t}$c$) for large $t$ is helped by writing
\bme
\label{bl-large-t}
\be
[\,\ubr\, ,\,\vbr\,]_{bl}\,=\,(\pi x)^{-1}[\,\xS\FS_i\,,\,\FS_r+\FS_0\,]\,,
   \hskip 10mm    \FS_0(t)\,=\,\int_t^{\infty}\dfrac{\JR_1(2\tau)}{\tau}\,\dR\tau
\ee
with
\be\se
\FS_r\,+\iR\FS_i\,\equiv\,\FS(\xS,t)\,=\,-\,\int_t^{\infty}\dfrac{\JR_1(2\tau)}{\tau}\,\exp\bigl(\iR 2\xS(t-\tau)\bigr)\,\dR\tau\,.
\ee
\eme
Since $\JR_1(2\tau)= (\pi \tau)^{-1/2}\cos (2\tau-3\pi/4)+O\bigl(\tau^{-3/2}\bigr)$, we have
\be
\label{F0}
\FS_0 \,\approx\,\tfrac12 \pi^{-1/2} t^{-3/2}\cos (2t-\pi/4) \,=\,O\bigl(t^{-3/2}\bigr),
\ee
which is small compared to the resonant contribution to $\FS$:
\[
\FS\,=\,-\,\dfrac{\exp[\iR(2t-3\pi/4)]}{\pi^{1/2}}\int_t^\infty\dfrac{\exp\bigl[\iR 2(1-\xS)(\tau-t)\bigr]}{2\tau^{3/2}}\,\dR\tau\,+\,O\bigl(t^{-3/2}\bigr),
\]
obtained on the basis that $\xS$ is close to unity. It may be expressed as
\bse
\label{F-large-t}
\be
\FS\,=\,\iR\,\dfrac{\exp[\iR(2t-\pi/4)]}{(\pi t)^{1/2}}\,\GS(\zeta)\,+\,O\bigl(t^{-3/2}\bigr),
\ee
where
\begin{align}
  \GS(\zeta)\,=\,&\,\dfrac{\zeta^{1/2}}{2}\int_0^\infty \dfrac{\eR^{\iR\zeta'}}{(\zeta+\zeta')^{3/2}}\,\dR \zeta'\,=\,1-(-\iR\pi\zeta)^{1/2} \,\eR^{-\iR\zeta}\,\erfc\!\bigl((-\iR \zeta)^{1/2}\bigr) \\[0.3em]
=\,&\,-\,\zeta^{1/2}\,\od{\,}{\zeta}\biggl[\int_0^\infty \dfrac{\eR^{\iR\zeta'}}{(\zeta+\zeta')^{1/2}}\biggr]\,\dR\zeta'\,=\,-\,(\iR\pi\zeta)^{1/2}\od{\,}{\zeta}\bigl[\eR^{-\iR\zeta}\,\erfc\!\bigl((-\iR \zeta)^{1/2}\bigr)\bigr], 
\end{align}
\ese
is a function of the similarity variable $\zeta$ (\ref{tS-def}$a$). Use of (\ref{F-large-t}$c$) shows that
\be
\label{int-G}
\int_0^\infty\dfrac{\GS(\zeta)}{\zeta^{1/2}}\,\dR\zeta\,=\,(\iR\pi)^{1/2}\,.
\ee

With the help of (http://dlmf.nist.gov/7.6.E2), we may express (\ref{F-large-t}$b$) in the form
\bse
\label{point-sink-solution-later-time-asym-more-1}
\be
\GS(\zeta)\,=\,-(-\iR\pi\zeta)^{1/2} \,\eR^{-\iR\zeta}\,+\, \PS(\zeta)\,,
\ee
where $\PS(\zeta)$ is an entire function with the power series expansion
\begin{align}
\PS(\zeta)\,\equiv\,\PS_{\!r}(\zeta)\,+\,\iR\PS_{\!i}(\zeta)\,=\,&\,1\,+\,(-\iR \pi\zeta)^{1/2}\eR^{-\iR\zeta}\,\erf\!\bigl((-\iR \zeta)^{1/2}\bigr)\nonumber\\
=\,&\,1+\sum_{n=1}^\infty\,\dfrac{(-\iR 2\zeta)^n}{1\cdot3\cdots(2n-1)}\,.
\end{align}
Explicitly the real and imaginary parts are
\be
\PS_{\!r}(\zeta)\,=\,1\,-\,{(2\zeta)^2}\big/{3}\,+\,\cdots\,, \hskip 15mm \PS_{\!i}(\zeta)\,=\,-\,2\zeta\,+\,{(2\zeta)^3}\big/{15}\,+\,\cdots\,.
\ee
\ese
The value of $\FS$ determined by substitution of only the first term $-(-\iR\pi\zeta)^{1/2} \,\eR^{-\iR\zeta}$ of  $\GS$ into (\ref{F-large-t}$a$) is $-\sqrt{2(1-\xS)}\exp(2\iR\xS t)\approx -|\zS|\exp(2\iR\xS t)$. It  defines the  contribution $-(\pi x)^{-1}|\zS|[\xS\sin(2\xS t),\cos(2\xS t)]$ to the flow $[\ubr,\vbr]_{bl}$ (\ref{bl-large-t}$a$), which exactly cancels the wave part of $[\ubr,\vbr]_{ms}$ (\ref{Poisson-sum-sol-large-t}$b$) so that their sum is simply $[0,\vbr_\tG]$. Hence the remaining second term $\PS(\zeta)$ determines the complete boundary layer flow $[\ubr,\vbr]_{ms+bl}$:
\bse
\label{point-sink-PS-bl}
\be
\left[\begin{array}{c}  \!\!(\pi x)\ubr_{ms+bl}  \!\!\\[0.2em]
    \!\! (\pi x)\vbr_{ms+bl}+ 1\!\!  \end{array}\right]\,=\,
\dfrac{1}{(\pi t)^{1/2}}
\left[\begin{array}{cc}  \!\!\xS\PS_{\!r}(\zeta) &-\, \xS\PS_{\!i}(\zeta)\!\!\\[0.2em]
    \!\!\,-\,\PS_{\!i}(\zeta) &-\, \PS_{\!r}(\zeta)\!\!    \end{array}\right]
\left[\begin{array}{c}  \!\! \cos(2t\,-\,\pi/4) \!\!\\[0.2em]
    \!\!\sin(2t\,-\,\pi/4) \!\!  \end{array}\right]+O(t^{-3/2})\,,
\ee
in which (see (\ref{tS-def}))
\be
\zeta=\zS^2t(1+O(t^{-1}))  \hskip 10mm \xS = 1+O(t^{-1}) \hskip 10mm \mbox{when}\hskip 10mm \zeta=O(1)\,.
\ee
\ese
Note that the contribution from $\FS_0$ is $O(t^{-3/2})$ and contained in the error estimate. The $\zeta=0$ values of (\ref{point-sink-PS-bl}$a$) agree with the asymptotic ($t\gg 1$) values of (\ref{Poisson-sum-sol-small-t-zero-z}) at $z=0$.

For large $\zeta$, rather than (\ref{point-sink-solution-later-time-asym-more-1}),  we use the asymptotic form 
\be
\label{point-sink-solution-later-time-asym-more-2}
\GS(\zeta)\,=\,\tfrac12 \iR \zeta^{-1}\,+\,O\bigl(\zeta^{-2}\bigr)  \hskip 10mm \mbox{for} \hskip 10mm |\zeta|\gg 1
\ee
of (\ref{F-large-t}$b$). To evaluate $[\ubr,\vbr]_{bl}$ from  (\ref{bl-large-t}) in that limit, we find it tidier, though not essential, to reinstate the the asymptotic value (\ref{F0}) of $\FS_0$. Then substitution of (\ref{point-sink-solution-later-time-asym-more-2}) into (\ref{F-large-t}$a$) determines
\begin{align}
\FS+\FS_0\,=\,&\,(\pi t)^{-1/2}\zeta^{-1}\bigl[-\,\exp\bigl(\iR(2t-\pi/4)\bigr)\,+(1-\xS)\cos(2t-\pi/4)\bigl]+O(t^{-3/2})\nonumber\\
=\,&\,-\,(\pi t)^{-1/2}\zeta^{-1}\bigl[\xS\cos(2t-\pi/4)+\iR\sin(2t-\pi/4)\bigl]+O(t^{-3/2})\,.\label{FS-large-tS}
\end{align}
In turn substitution into  (\ref{bl-large-t}$a$) yields
\be
\label{uv-bl-large-tS}
[\,\ubr\, ,\,\vbr\,]_{bl}\,=\,-\,(\pi \varpi\zeta)^{-1}(\pi t)^{-1/2}\bigl[\,\sin(2t-\pi/4)\,,\,\cos(2t-\pi/4)\,]+O(t^{-3/2})\,,
\ee
which tends to zero at fixed $x$ as $z\to\infty$.

We make the approximation $\xS\approx 1$ in (\ref{bl-large-t}$a$), continue to neglect $\FS_0$ and evaluate the mean value $(\pi x)^{-1}\langle\FS\rangle$, using (\ref{F-large-t}$a$), to obtain
\bse
\label{bl-mean-values-large-t}
\be
\langle \vbr_{bl}\rangle\,+\,\iR\langle \ubr_{bl}\rangle\,\approx\, \iR\, (\pi x)^{-1}(\pi t)^{-1/2}\exp\bigl(\iR(2t-\pi/4)\bigr)\int_0^1\GS(\zeta)\,\dR z\,,
\ee
which under the further approximation $\zS\approx (\zeta/t)^{1/2}$ (see (\ref{point-sink-PS-bl}$b$)), implying $x^{-1}\dR z\approx\tfrac12 (t\,\zeta)^{-1/2}\dR\zeta$, yields
\be
\langle \vbr_{bl}\rangle\,+\,\iR\langle \ubr_{bl}\rangle\,\approx\,\iR\, \dfrac{\exp\bigl(\iR(2t-\pi/4)\bigr)}{2\pi t}\,
\dfrac{1}{\pi^{1/2}}\!\int_0^{t/\varpi^2} \dfrac{\GS(\zeta)}{\zeta^{1/2}}\,\dR \zeta\,.
\ee
Then in the limit $t/\varpi^2\to \infty$, use of (\ref{int-G}) determines
\be
\bigl[\,\langle \ubr_{bl}\rangle\,,\,\langle \vbr_{bl}\rangle\,\bigr]\,=\,
\,(2\pi t)^{-1}[\,\cos(2t)\, ,\,-\,\sin(2t) \,]\,+\,O(t^{-3/2})\,.
\ee
\ese
Many approximations have been made in obtaining this result, but significantly it shows that the boundary layer volume flux $[\langle \ubr_{bl}\rangle,\langle \vbr_{bl}\rangle]$ from the $m=0$ sink alone, without consideration of any possible far field contributions from the  $m\not=0$ sinks, accounts for the flux deficit of the mainstream flow determined by (\ref{mean-values-large-t-more}) and (\ref{mean-values-large-t-yet-more}).

\end{document}